\documentclass[twocolumn]{aastex631}

\usepackage[T1]{fontenc}
\usepackage[utf8]{inputenc}

\usepackage{xcolor}
\usepackage{amsmath}
\usepackage{natbib}
\usepackage{bm}
\usepackage{multirow}
\usepackage{comment}

\newcommand\St{\,{\rm St}}
\newcommand{\bmu}{\bm{u}}
\newcommand{\bmw}{\bm{w}}
\newcommand{\bmv}{\bm{v}}
\newcommand{\p}{\partial}
\newcommand{\xhat}{\hat{\bm{x}}}
\newcommand{\yhat}{\hat{\bm{y}}}
\newcommand{\zhat}{\hat{\bm{z}}}
\newcommand{\bmB}{\bm{B}}
\newcommand{\bmA}{\bm{A}}
\newcommand{\Bhat}{\widehat{\bm{B}}}
\newcommand{\rhod}{\rho_\mathrm{d}}
\newcommand{\rhog}{\rho_\mathrm{g}}
\newcommand{\taus}{\tau_\mathrm{s}}
\newcommand{\fdust}{f_\mathrm{d}}
\newcommand{\fgas}{f_\mathrm{g}}
\newcommand{\st}{\mathrm{St}}
\newcommand{\dd}{\delta}
\newcommand{\ii}{\mathrm{i}}
\newcommand{\ik}{\mathrm{i}\bm{k}}
\newcommand{\ikx}{\mathrm{i}k_x}
\newcommand{\ikz}{\mathrm{i}k_z}
\newcommand{\etahall}{\eta_\mathrm{H}}
\newcommand{\va}{V_\mathrm{A}}
\newcommand{\re}{\operatorname{Re}}
\newcommand{\etahat}{\hat{\eta}}
\newcommand{\Ha}{\operatorname{Ha}}

\newcommand{\vext}{v_\mathrm{ext}}
\newcommand{\Hg}{H_\mathrm{g}}
\newcommand{\Hd}{H_\mathrm{d}}

\begin{document}

\title{Dust Dynamics in Hall-effected Protoplanetary Disks.\\ I. Background Drift Hall Instability}
\author[0000-0003-3728-8231]{Yinhao Wu}
\affiliation{School of Physics and Astronomy, University of Leicester, Leicester LE1 7RH, UK; \href{mailto:email@domain}{yw505@leicester.ac.uk}}
\affiliation{Institute of Astronomy and Astrophysics, Academia Sinica, Taipei 10617, Taiwan, R.O.C.; \href{mailto:email@domain}{mklin@asiaa.sinica.edu.tw}}

\author[0000-0002-8597-4386]{Min-Kai Lin}
\affiliation{Institute of Astronomy and Astrophysics, Academia Sinica, Taipei 10617, Taiwan, R.O.C.; \href{mailto:email@domain}{mklin@asiaa.sinica.edu.tw}}
\affiliation{Physics Division, National Center for Theoretical Sciences, Taipei 10617, Taiwan, R.O.C.}

\author[0000-0003-3180-0038]{Can Cui}
\affiliation{Department of Astronomy and Astrophysics, University of Toronto, Toronto, ON M5S 3H4, Canada}
\affiliation{DAMTP, University of Cambridge, Wilberforce Road, Cambridge CB3 0WA, UK}

\author[0000-0001-7671-9992]{Leonardo Krapp}
\affiliation{Department of Astronomy and Steward Observatory, University of Arizona, Tucson, AZ 85721, USA}

\author[0000-0003-3497-2329]{Yueh-Ning Lee}
\affiliation{Department of Earth Sciences, National Taiwan Normal University, Taipei 116059, Taiwan, R.O.C.}
\affiliation{Center of Astronomy and Gravitation, National Taiwan Normal University, Taipei 116059, Taiwan, R.O.C.}
\affiliation{Physics Division, National Center for Theoretical Sciences, Taipei 10617, Taiwan, R.O.C.}

\author[0000-0002-3644-8726]{Andrew N. Youdin}
\affiliation{Department of Astronomy and Steward Observatory, University of Arizona, Tucson, AZ 85721, USA}
\affiliation{The Lunar and Planetary Laboratory, University of Arizona, Tucson, AZ 85721, USA}

\begin{abstract}
Recent studies have shown that the large-scale gas dynamics of protoplanetary disks (PPDs) are controlled by non-ideal magneto-hydrodynamics (MHD), but how this influences dust dynamics is not fully understood. To this end, we investigate the stability of dusty, magnetized disks subject to the Hall effect, which applies to planet-forming regions of PPDs. We find a novel Background Drift Hall Instability (BDHI) that may facilitate planetesimal formation in Hall-effected disk regions. Through a combination of linear analysis and nonlinear simulations, we demonstrate the viability and characteristics of BDHI. We find it can potentially dominate over the classical streaming instability (SI) and standard MHD instabilities at low dust-to-gas ratios and weak magnetic fields. We also identify magnetized versions of the classic SI, but these are usually subdominant.  We highlight the complex interplay between magnetic fields and dust-gas dynamics in PPDs, underscoring the need to consider non-ideal MHD like the Hall effect in the broader narrative of planet formation.
\end{abstract}


\section{Introduction} \label{sec:intro}

Understanding the processes involved in planetesimal and planet formation is essential for unraveling the origins and evolution of our own Solar System and those around other stars \citep[e.g.,][]{eDiskI,eDiskII}. In the core accretion scenario, the formation of 1-100 km sized planetesimals ---the building blocks of planets---is one of the most critical steps for planet formation \citep[e.g.,][]{Chiang_Youdin_2010,Johansen_2014,Nayakshin_2022,Drazkowska-PPVII}.

Specifically, the growth of solid particles from micron-sized grains via sticking is limited to pebbles ranging in size from millimeters to centimeters. Once the particles reach this size range, collisions often lead to either bouncing or fragmentation rather than further growth (see \citealp{Blum2008ARAA} for a review). Moreover, the presence of gas drag can induce a swift inward migration of solid particles in the process \citep{Whipple1972,Weidenschilling1977}, creating a radial drift barrier \citep{Birnstiel2010,Birnstiel2012}.

One way to overcome these growth barriers is through the collective, self-gravitational collapse of a swarm of particles directly into planetesimals \citep{Goldreich1973,Youdin_Shu_2002}. The particle swarm must reach a dust-to-gas mass density ratio, $\epsilon$, significantly higher than unity \citep{Shi2013}. This contrasts the typical value of $\epsilon\sim0.01$ found in the interstellar medium but can be reasonably expected in protoplanetary disks \citep[PPDs,][]{Testi_PPVI}, especially in their initial stages of evolution. Consequently, an additional mechanism is required to enhance the local dust-to-gas ratio. This mechanism may be facilitated through dust settling, radial drift, dust trapping by pressure bumps or vortices, and additional dust-gas instabilities \citep{Chiang_Youdin_2010,Johansen_2014}.

One prominent mechanism is the streaming instability \citep[SI;][]{Youdin-Goodman2005,johansen+07,Youdin-Johansen2007,bs2010a,bs2010b,Yang_Johansen_2014,Yang2017,Li2018,Jaupart2020, Li2021}. This phenomenon arises from the relative drift between the gas and dust components in PPDs \citep{Jacquet2011, Lin-Youdin_2017}. In the limit of $\epsilon\ll 1$, the SI is a member of a broader class of `resonant drag instabilities' that results from the resonance between neutral waves in the gas and the relative dust-gas drift \citep[RDIs][]{Squire-Hopkins-2018a,Squire-Hopkins2020}. While the basic SI only requires mutually interacting dust and gas in rotation, realistic PPD conditions need to be considered to fully assess planetesimal formation through the SI. To this end, researchers have extended the classical SI of \citeauthor{Youdin-Goodman2005} to account for additional complexities. These include: particle self-gravity \citep{Simon2016,Simon2017,Schafer2017,Abod2019,Li2019}, gas turbulence  \citep{Chen_Lin_2020,Gole2020,Schafer2020,Umurhan2020}, accretion flows \citep{Hsu_Lin_2023}, vertical disk structure \citep{lin21}, interactions between multiple grain sizes \citep{Schaffer2018,Schaffer2021,FARGO3D_multifluid,Krapp2019,Paardekooper2020,McNally2021,Zhu_Yang_2021}, etc. These studies are conducted in the purely hydrodynamic limit. 

On the other hand, magnetic fields are known to play a leading role in driving gas dynamics in PPDs \citep[e.g.,][]{Jonathan_2011_ARAA,Crutcher_2012_ARAA}. The radial transport of angular momentum is mediated by magneto-rotational instability (MRI) induced turbulence \citep{bh91}. The vertical transport of angular momentum is through magnetized disk winds \citep{bs13,bethune+17,bai17,wang19,cui2021}. Moreover, non-ideal MHD effects --- Ohmic resistivity, Hall effect, and ambipolar diffusion --- have significant impact due to the extremely weak ionization in PPDs \citep{gammie96,armitage11}. It is, therefore, imperative to understand how planetesimals form in non-ideal MHD.

Dust-gas dynamics under the influence of magnetic fields have been investigated in the literature. Early works focused on the dust radial migration, vertical diffusion, and settling in MRI turbulent disks \citep{fromang+05,fromang+06,johansen+05,johansen+06}. More recent numerical simulations found dust clumping was common in ideal and non-ideal MHD-dominated disks, where the clumping is attributed to the SI \citep{johansen+07,balsara09,tilley10}, weak radial diffusion in Ohmic resistive disks \citep{yang+18}, or pressure bumps and zonal flows induced by the MRI turbulence \citep{johansen+09,dittrich+13}, ambipolar diffusion \citep{riols20,xb22}, and the Hall effect \citep{krapp18}.

The Hall effect corresponds to the electron-ion drift. The effect of Hall drift on disk dynamics depends on the magnetic polarity that co- or counter-aligns with the angular momentum of the disk. The Hall effect suppresses linear MRI modes when co-aligned \citep{wardle99,bt01}, while diffusive MRI may operate when counter-aligned \citep{DMRI,Latter-Kunz2022}. Local simulations demonstrate the non-linear saturation of the MRI modified by Hall \citep{ss22a,ss22b,kunz13,bai15}. Global simulation reveals its importance on the evolution of disk magnetic flux transport \citep{bs17}. Analytical studies of MHD winds-driven accretion taken into account the Hall effect are also conducted \citep{wardle93,konigl+10,salmeron+12}. 
The Hall effect is applicable over the bulk disk from one to several tens of AU \citep{armitage15,Lesur-review}, i.e. the planet-forming regions of PPDs. 

In this work, we study dust-gas interaction in a Hall-dominated disk. In addition to the classical SI, we are also motivated by recent work on 
magnetic RDIs related to MHD waves \citep{Squire-Hopkins-2018a,Squire-Hopkins-2018b,hs18}. In the ideal-MHD limit, the fast magnetosonic, slow magnetosonic, and Alfv\'{e}n wave can be in resonance with dust-gas drift \citep[see][for the `Alfv\'{e}n wave' RDI in PPDs]{Lin-Hsu-2022}. Incorporating the Hall effect enables two more waves, the electron-cyclotron (whistler) wave and the ion-cyclotron wave \citep{Lesur-review}, which can lead to new branches of RDIs. We indeed find such `Hall RDIs', but these are usually sub-dominant. However, we also find a novel instability that is associated with the transport of magnetic fields by dust-induced gas flows. This non-RDI can dominate over other instabilities in dust-poor disks and could thus be important for planetesimal formation.

The paper is structured as follows. In $\S$ \ref{sec:eqs}, we present the basic equations governing a dusty, magnetized PPD and specify the physical setups that are being considered. Additionally, we provide an overview of the linear problem and discuss selected analytic results in $\S$ \ref{subsec:linear}. 
In $\S$ \ref{sec:3}, we present numerical solutions to the linear stability problem. In $\S$ \ref{sec:BG drift model}, we discuss the newly discovered instability mentioned above and present a toy model for it. In $\S$ \ref{simulations}, we present illustrative non-linear simulations of this instability.  In $\S$ \ref{sec:discuss}, we discuss our results in the context of PPDs and planetesimal formation. We summarize in $\S$ \ref{sec:8}. 

\section{Basic Physics} \label{sec:eqs}

We consider a three-dimensional (3D) disk composed of gas and {dust orbiting a central star of mass} $M_{\star}$. {Cylindrical coordinates $(r,\phi, z)$ are centered on the star. The gas has} density $\rhog$, velocity $\textbf{\emph{V}}$, and pressure $p$. {We assume an isothermal gas with} $p=C_{s}^{2}\rhog$, where $C_{s}=H_\mathrm{g}\Omega_\mathrm{K}$ is the strictly constant sound-speed, $H_\mathrm{g}$ is gas pressure scale height, $\Omega_\mathrm{K} = \sqrt{GM_*/r^3}$ is the Keplerian frequency, and $G$ is the gravitational constant. The gas is threaded by a magnetic field $\bmB$ and we assume the Hall effect is the dominant non-ideal MHD term. 

We consider a single species of dust grains that interacts with the gas through a drag force characterized by a stopping time $\tau_s$. For small grains with small $\taus$, dust grains are tightly --- but not necessarily perfectly --- coupled to the gas. In this limit, the dust grains can be modeled as a second, pressureless fluid with density $\rhod$ and velocity $\textbf{\emph{W}}$ \citep{Jacquet2011}.

{For the most part, we neglect gas viscosity, ohmic diffusion, and dust diffusion. However, we do include these dissipative effects in a few selected simulations for numerical stability.}

\subsection{Shearing Box Approximation}\label{subsec:Shearing-Box}

We consider the stability of the magnetized, dusty-gas on scales much smaller than the typical disk radius $r$. As such, we adopt the shearing box approximation \citep{Goldreich1965} to model a local patch of the disk that rotates around the star at the Keplerian frequency $\Omega_{0} = \Omega_\mathrm{K}(r_0)$, where $r_0$ is the radius of the patch. Cartesian coordinates $(x, y, z)$ in the box correspond to the $(r,\phi, z)$ directions in the global disk. 

In the shearing box, Keplerian rotation appears as the linear shear flow $\bm{U}_K = -(3/2)\Omega_0 x \yhat$. Defining the gas and dust velocities $\bmv$ and $\bmw$ relative to this shear flow, assuming axisymmetry ($\p_y\equiv 0$), and focusing on regions close to the disk midplane ($z=0$) and neglecting the vertical component of stellar gravity, the equations that govern the magnetized gas and dust are: 
\begin{equation}\label{shear-eq-density-gas}
    \frac{\partial\rhog}{\partial t} + \nabla\cdot\left(\rhog\textbf{\emph{v}}\right) = 0,
\end{equation}
\begin{align}\label{shear-eq-velocity-gas}
\frac{\partial\textbf{\emph{v}}}{\partial t} + \textbf{\emph{v}}\cdot\nabla\textbf{\emph{v}} = 
    & 2v_{y}\Omega_0\hat{\textbf{\emph{x}}} - v_{x}\frac{\Omega_0}{2}\hat{\textbf{\emph{y}}} - \frac{1}{\rhog}\nabla p \notag\\
    &+ 2\eta r_0 \Omega_0^2 \xhat +\frac{1}{\mu_0\rhog}\left(\nabla\times\bmB\right)\times\bmB \notag\\
    &- \frac{\epsilon}{\taus}\left(\bmv - \bmw\right),
\end{align}
\begin{align}\label{shear-eq-B} \frac{\partial\textbf{\emph{B}}}{\partial t} = &\nabla\times\left(\textbf{\emph{v}}\times\textbf{\emph{B}}\right) + \nabla\times\left(\bm{U}_{K}\times\textbf{\emph{B}}\right) \notag\\ &-\eta_H\nabla\times\left[\left(\nabla\times\bmB\right)\times\Bhat\right],
\end{align}
\begin{equation}\label{shear-eq-density-dust}
    \frac{\partial\rhod}{\partial t} + \nabla\cdot\left(\rhod\textbf{\emph{w}}\right) = 0,
\end{equation}
\begin{equation}\label{shear-eq-velocity-dust}  \frac{\partial\textbf{\emph{w}}}{\partial t} + \textbf{\emph{w}}\cdot\nabla\textbf{\emph{w}} = 2w_{y}\Omega_0\hat{\textbf{\emph{w}}}
    - w_{x}\frac{\Omega_0}{2}\hat{\textbf{\emph{y}}} - \frac{1}{\tau_{s}}\left(\textbf{\emph{w}}-\textbf{\emph{v}}\right),
\end{equation}
where
\begin{align}
    \eta \equiv \left.-\frac{1}{2 r\Omega_\mathrm{K}^2\rhog}\frac{\p p}{\p r}\right|_{r=r_0}
\end{align}
is a dimensionless measure of the radial pressure gradient from the background global disk, $\etahall$ is the constant Hall diffusivity, $\widehat{\bmB} = \bmB/|\bmB|$ is the unit vector along the magnetic field, and $\mu_0$ is the magnetic permeability. For clarity, hereafter we drop the subscript $0$ on $r_0$ and $\Omega_0$. 

\subsection{{Disk equilibrium and parameters}}

Our shearing box admits a steady-state equilibrium with constant densities, velocities, and a uniform vertical field $B_{z0}$. The velocities are given by
\begin{align}
    v_x &= \frac{2\epsilon\st}{\Delta^2}\eta r \Omega,\label{eqm_vx}\\
    v_y & = -\frac{1+\epsilon+\st^2}{\Delta^2}\eta r \Omega,\label{eqm_vy}\\
    w_x & = -\frac{2\st}{\Delta^2}\eta r \Omega,\\
    w_y & = -\frac{1+\epsilon}{\Delta^2}\eta r \Omega,
\end{align}
where $\Delta^2 = (1+\epsilon)^2 + \st^2$, and $v_z = w_z = 0$. Here, 
\begin{align}
    \st \equiv \taus \Omega
\end{align}
is the Stokes number, which is assumed to be constant. Note that the presence of a constant vertical field has no impact on the equilibrium flow. 

In smooth disks, $\eta$ is of order $h_\mathrm{g}^2$, where $h_\mathrm{g} \equiv H_\mathrm{g}/r$ is the aspect-ratio. For convenience, we define the reduced pressure gradient
\begin{align}
\widehat{\eta} \equiv \frac{\eta}{h_\mathrm{g}},
\end{align}
which is the relevant parameter in linear theory when velocities are normalized by $C_s$. 

The magnetic field strength $B_{z0}$ is defined through the Alfv\'{e}n speed $V_\mathrm{A}$ 
\begin{align}
    V_\mathrm{A}^2 \equiv \frac{B_{z0}^2}{\mu_0 \rhog},    
\end{align}
itself characterized through the plasma beta parameter, which measures the ratio between thermal and magnetic pressure:
\begin{align}
\beta_z \equiv \frac{2C_s^2}{V_\mathrm{A}^2}.
\end{align}
To parameterize the Hall effect, we follow \cite{Latter-Kunz2022} and define the Els\"asser number 
\begin{align}
    \Ha \equiv \frac{V_\mathrm{A}^2}{2\etahall\Omega} = \frac{C_s^2}{\beta_z \etahall \Omega},\label{Ha_def}
\end{align}
which is used to set the Hall diffusivity $\etahall$. Note that $\Ha$ is positive (negative) when the equilibrium field is aligned (anti-aligned) with the disk rotation. In discussions, we also refer to the Hall length $l_\mathrm{H}$,
\begin{align}
    l_\mathrm{H} \equiv \frac{\etahall}{V_\mathrm{A}}.\label{lH_def}
\end{align}

\subsection{Single-fluid models}
Most of our analyses are based on the above explicit, two-fluid model equations. However, we also make use of a modified, single-fluid formulation that is more amenable to the numerical simulations that we present later. In this approach, the gas is approximated to be incompressible (so $\rhog$ is a constant); we solve for the relative dust-gas velocity 
\begin{align}
    \bmu \equiv \bmw - \bmv  
\end{align}
instead of solving for $\bmw$ directly; the Hall term is slightly augmented (which has no effect on linear stability); and dissipation is included for numerical stability. This single-fluid model is detailed in Appendix \ref{full_single_fluid}. 

Furthermore, in the limit of tightly-coupled dust grains with $\st\ll1$, it is possible to simplify the single-fluid model by approximating $\bmu$ with the `terminal velocity approximation' (TVA). We describe the TVA in Appendix \ref{TVA}. 

\section{Linear Theory}\label{subsec:linear}

We consider axisymmetric Eulerian perturbations to a variable $f$ of the form 
\begin{align}
    f \to f + \re\left[\dd f \exp{\left(\sigma t + k_x x + k_z z\right)}\right],
\end{align}
where $\delta f$ is a complex amplitude, $\sigma$ is the complex growth rate, and $k_{x,z}$ are real wavenumbers. We take $k_{x,z}>0$ without loss of generality. 

Linearizing Eqs. \ref{shear-eq-density-gas}---\ref{shear-eq-velocity-dust} gives: 
\begin{equation}\label{eq-gas-rho}
    \sigma W = - ik_{x}v_{x}W - ik_{x}\delta v_{x} - ik_{z}\delta v_{z},
\end{equation}
\begin{equation}\label{eq-dust-rho}
    \sigma Q = - ik_{x}w_{x}W - ik_{x}\delta w_{x} - ik_{z}\delta w_{z},
\end{equation}
\begin{align}\label{eq-vx}
    \sigma\delta v_{x} =
    &- ik_{x}v_{x}\delta v_{x} + 2\Omega\delta v_{y} - ik_{x}C_{s}^{2}W \notag\\
    &- \frac{\epsilon}{\tau_{s}}\left(w_{x} - v_{x}\right)\left(W - Q\right) + \frac{\epsilon}{\tau_{s}}\left(\delta w_{x} - \delta v_{x}\right)\notag\\ 
    &+ iV_\mathrm{A}\left(k_{z}\delta b_{x} - k_{x}\delta b_{z}\right),
\end{align}
\begin{align}\label{eq-vy}
    \sigma\delta v_{y} =
    &- ik_{x}v_{x}\delta v_{y} - \frac{\Omega}{2}\delta v_{x} - \frac{\epsilon}{\tau_{s}}\left(w_{y} - v_{y}\right)\left(W - Q\right)\notag\\ 
    &+ \frac{\epsilon}{\tau_{s}}\left(\delta w_{y} - \delta v_{y}\right) + iV_\mathrm{A}k_{z}\delta b_{y},
\end{align}
\begin{equation}\label{eq-vz}
    \sigma\delta v_{z} =
    - ik_{x}v_{x}\delta v_{z} + \frac{\epsilon}{\tau_{s}}\left(\delta w_{z} - \delta v_{z}\right) - ik_{z}C_{s}^{2}W,
\end{equation}
\begin{equation}\label{eq-wx}
    \sigma\delta w_{x} =
    - ik_{x}w_{x}\delta w_{x} + 2\Omega\delta w_{y} - \frac{1}{\tau_{s}}\left(\delta w_{x} - \delta v_{x}\right),
\end{equation}
\begin{equation}\label{eq-wy}
    \sigma\delta w_{y} =
    - ik_{x}w_{x}\delta w_{y} - \frac{\Omega}{2}\delta w_{x} - \frac{1}{\tau_{s}}\left(\delta w_{y} - \delta v_{y}\right),
\end{equation}
\begin{equation}\label{eq-wz}
    \sigma\delta w_{z} =
    - ik_{x}w_{x}\delta w_{z} - \frac{1}{\tau_{s}}\left(\delta w_{z} - \delta v_{z}\right),
\end{equation}
\begin{equation}\label{eq-bx}
    \sigma\delta b_{x} =
    - ik_{x}v_{x}\delta b_{x} + ik_{z}V_\mathrm{A}\delta v_{x} - \eta_{H}k_{z}^{2}\delta b_{y},
\end{equation}
\begin{align}\label{eq-by}
    \sigma\delta b_{y} =
    &- ik_{x}v_{x}\delta b_{y} + ik_{z}V_\mathrm{A}\delta v_{y} - \frac{3}{2}\Omega\delta b_{x}\notag\\ 
    &+ \eta_{H}k_{z}^{2}\delta b_{x} - \eta_{H}k_{x}k_{z}\delta b_{z},
\end{align}
\begin{equation}\label{eq-bz}
    \sigma\delta b_{z} =
    - ik_{x}v_{x}\delta b_{z} + ik_{z}V_\mathrm{A}\delta v_{z} + \eta_{H}k_{x}k_{z}\delta b_{y},
\end{equation}
where $W \equiv \delta\rhog/\rhog$, $Q \equiv \delta\rhod/\rhod$, and $\dd\bm{b} = \delta \bmB /\sqrt{\mu_0\rhog}$. See \cite{Lin-Hsu-2022} for a similar set of equations for the case with ohmic resistivity instead of the Hall effect. 

The linearized system can be regarded as an eigenvalue problem 
\begin{equation}\label{matrix}
    \bm{M}\bm{X} = \sigma \bm{X},
\end{equation}
which $\bm{M}$ is the $11$-dimensional matrix representation of the right-hand-side of Eqs.~\ref{eq-gas-rho}---\ref{eq-bz}, and $\bm{X} = \left[W, Q, \delta\textbf{\emph{v}}, \delta\textbf{\emph{w}}, \delta\textbf{\emph{b}}\right]^{T}$ is the eigenvector of complex amplitudes. To solve Eq. \ref{matrix}, we use standard numerical methods in the \textsc{eigvals} package for \textsc{PYTHON}.

\subsection{Dust-free Dispersion Relation}\label{subsec:dispersion relation}

In the dust-free limit and approximating the gas to be incompressible (thus filtering out sound waves), the linearized equations yield the quartic dispersion relation 
\begin{equation}\label{Lesur-dispersion-relation}
    \sigma^{4}+\mathcal C_{3}\sigma^{3}+\mathcal C_{2}\sigma^{2}+\mathcal C_{1}\sigma+\mathcal C_{0}=0,
\end{equation}
with $\mathcal{C}_3 = \mathcal{C}_1 = 0$, see \cite{Lesur-review}. If ohmic resistivity and ambipolar diffusion are included, $\mathcal{C}_3$ and $\mathcal{C}_1$ become non-zero but are real. In this work, we only consider the Hall effect, in which case Eq. \ref{Lesur-dispersion-relation} is a bi-quadratic equation for $\sigma^2$ with 
coefficients $\mathcal{C}_{2,0}$ given by 
\begin{align}
   \mathcal C_{2}=&~\frac{k_{z}^{2}}{k^2}\kappa^{2}+2\omega_\mathrm{A}^{2}+l_\mathrm{H}k_{z}\omega_\mathrm{A}\left(\frac{k^2}{k_{z}^{2}}l_\mathrm{H}k_{z}\omega_\mathrm{A}-q\Omega\right),\\
\mathcal{C}_{0}=&~\omega_\mathrm{A}^{2}\left(\omega_\mathrm{A}^{2}-2q\Omega^{2}\frac{k_{z}^{2}}{k^2}\right)\notag\\&~+l_\mathrm{H}\omega_\mathrm{A}k_{z}\left[\left(4-q\right)\Omega\omega_\mathrm{A}^{2}-q\Omega\frac{k_{z}^{2}}{k^2}\kappa^{2}+l_\mathrm{H}\omega_\mathrm{A}k_{z}\kappa^{2}\right],\label{lesur_C0}
\end{align}
where $k^2=k_{x}^{2}+k_{z}^{2}$, $\kappa^2\equiv2\Omega^{2}\left(2-q\right)$ is the squared epicycle frequency with $q \equiv -\p\ln{\Omega}/\p \ln{r}$ being the local shear rate, and the Alfv\'en frequency $\omega_\mathrm{A}\equiv k_z V_\mathrm{A}$. Unless otherwise stated, we set $q=3/2$ as appropriate for Keplerian flow, in which case $\kappa=\Omega$. 

According to Eq. \ref{Lesur-dispersion-relation}, a sufficient condition for instability is $\mathcal{C}_0<0$, which implies at least one positive real root. To understand the MHD effects brought about by different negative $\Ha$, \cite{Latter-Kunz2022} considered three different regimes: Regime 1 corresponds to $\Ha<-1$, i.e. $\Ha$ is sufficiently large and negative, we associate instability with the standard MRI, slightly modified by the Hall effect. Regime 2 corresponds to $-1\leq\Ha<-0.25$, where instability is associated with the diffusive MRI \citep[DMRI,][]{DMRI}. Regime 3 corresponds to $-0.25\leq\Ha<0$, i.e. small negative $\Ha$, where MHD instabilities are suppressed. When $\Ha > 0$, the situation becomes more straightforward: the MRI/HSI is deactivated when $\beta_{z}$ is sufficiently small.

\subsection{Resonant Drag Instabilities}\label{subsec:RDI}

The presence of dust can drive drag-related instabilities when the mutual friction between dust and gas is accounted for. \cite{Squire-Hopkins-2018a,Squire-Hopkins-2018b} proposed the RDI theory as a robust framework for characterizing such instabilities. RDIs stem from the resonance between waves in the gas and the equilibrium relative drift between the gas and dust. This occurs when the condition $\omega_\mathrm{gas}\left(\textbf{\emph{k}}\right)=\textbf{\emph{k}}\cdot\left(\textbf{\emph{w}}-\textbf{\emph{v}}\right)$ is met. Here, $\omega_\mathrm{gas}$ is the neutral frequency of a wave mode with wavevector $\bm{k}$ in the gas when no dust is present. 

The resonance condition in our unstratified, axisymmetric shearing box can be expressed as:
\begin{equation}\label{eq: omega_gas}
    \omega_\mathrm{gas}\left(k_{x}, k_{z}\right) = k_{x}(w_{x} - v_{x}).
\end{equation}
Several candidates for $\omega_\mathrm{gas}$ may be considered, depending on the physics involved in the pure gas. Once $\omega_\mathrm{gas}$ is specified, Eq. \ref{eq: omega_gas} can be used to find the relation between $k_x$ and $k_z$ of resonant or most unstable modes. 

\subsubsection{RDIs with inertial waves: classical SI}

In a hydrodynamic disk with $\epsilon\ll1$, the classic SI of \cite{Youdin-Goodman2005} is an RDI with inertial waves \citep{Squire-Hopkins-2018a}. Inertial waves are most easily recovered by setting $\omega_\mathrm{A}\to 0$ and $\sigma = -\ii \omega_\mathrm{gas}$ in Eqs. \ref{Lesur-dispersion-relation}---\ref{lesur_C0}. We find 
\begin{equation}\label{eq: Classic-RDI}
    \omega_\mathrm{gas}^{2} = \frac{k_{z}^{2}}{k_{x}^{2}+k_{z}^{2}}\Omega^{2}.
\end{equation}
By utilizing Eq. \ref{eq: Classic-RDI} and writing $u_x = w_x - v_x$ (which is given explicitly by Eq. \ref{eqm_ux}), the resonance condition stated in Eq. \ref{eq: omega_gas} can be rewritten as:
\begin{equation}\label{eq: classic-kz}
k_z^2 = \frac{u_x^2 k_x^4}{\Omega^2 - u_x^2 k_x^2}.
\end{equation}
Notice as $k_x^2 \to \Omega^2/u_x^2$, $k_z^2\to\infty$. That is, without dissipation, there is no minimum scale to the SI. 

\subsubsection{RDIs with Hall-MHD: No rotation}\label{3.2.2}

In the presence of an active magnetic field, one can consider a number of MHD waves for $\omega_\mathrm{gas}$ \citep{Squire-Hopkins-2018a}. Differential rotation (such as Keplerian shear) enables MHD instabilities such as the MRI, which complicates the application of the RDI recipe. For simplicity, we first neglect rotation, which is expected to be appropriate on small scales (large $k_{x,z}$). This limit also conveniently filters out inertial waves and allows us to focus on pure MHD effects. 

Setting $\Omega = \kappa = 0$ in Eq. \ref{Lesur-dispersion-relation}, we find 
\begin{equation}\label{eq: Hall-RDI-no-rotation}
\omega_{\mathrm{gas}} = \omega_\mathrm{A}\left(\pm\frac{l_\mathrm{H}k}{2}+\sqrt{\frac{l_\mathrm{H}^{2}k^{2}}{4}+1}\right).
\end{equation}
In ideal MHD ($l_\mathrm{H}\to 0$), we recover Alfv\'{e}n waves with $\omega_{\mathrm{gas}} = \omega_\mathrm{A}$. \cite{Lin-Hsu-2022} indeed found RDIs associated with Alfv\'{e}n waves on small scales, but they are unlikely relevant to PPDs as they only operate under nearly ideal conditions. 

The positive and negative solutions to Eq. \ref{eq: Hall-RDI-no-rotation} are known as whistlers (electron-cyclotron) modes, and ion-cyclotron modes, respectively \citep{Lesur-review}. Inserting Eq. \ref{eq: Hall-RDI-no-rotation} into the RDI condition Eq. \ref{eq: omega_gas}, one can then solve for the resonant $k_z$ as a function of $k_x$ for these Hall RDIs. The explicit solutions are unwieldy and are thus relegated to Appendix \ref{RDI-NoRotation}. 

Here, we consider the strong Hall regime such that $\left|l_\mathrm{H} k\right|\gg 1$. Then
\begin{align}
  \omega_{\mathrm{gas}} \simeq \omega_\mathrm{A}\left(l_\mathrm{H} k\right)^{\pm 1}. \label{strong_hall_no_rotation}
\end{align}
Inserting these into the RDI condition Eq. \ref{eq: omega_gas} and rearranging, we find for whistler waves:
\begin{align}
    k_x^2 = \frac{V_\mathrm{A}^2 l_\mathrm{H}^2 k_z^4}{u_x^2 - V_\mathrm{A}^2 l_\mathrm{H}^2 k_z^2},
    \label{whislter_RDI}
\end{align}
which shows that $k_x^2 \to \infty$ as $k_z^2$ tends to the critical value $k_{z,c}^2$
\begin{align}
    k^2_{z,c} \equiv \left(\frac{u_x}{V_\mathrm{A} l_\mathrm{H}}\right)^2 =\left(\frac{u_x \beta_z \Ha \Omega }{C_s^2}\right)^2,
\end{align}
where in the second equality we utilized Eqs. \ref{Ha_def}---\ref{lH_def} to eliminate $V_\mathrm{A}$ and $l_\mathrm{H}$ in favor of $\Ha$ and $\beta_z$. 

For ion-cyclotron waves, we find
\begin{align}
    k_z^2 = \frac{l_\mathrm{H}^2u_x^2 k_x^4}{V_\mathrm{A}^2 - l_\mathrm{H}^2 u_x^2 k_x^2},
    \label{ion_RDI}
\end{align}
which has the same functional form as for the classic SI, Eq. \ref{eq: classic-kz}, since in both cases $\omega_{\mathrm{gas}}\propto k_z/k$. Indeed, like the SI, Eq. \ref{ion_RDI} shows that $k_z^2\to\infty$ as $k_x^2$ approaches the critical value $k_{x,c}^2$
\begin{align}
k_{x,c}^2 \equiv \left(\frac{V_\mathrm{A}}{l_\mathrm{H} u_x}\right)^2 = 4\Omega^2\left(\frac{\Ha}{u_x}\right)^2.
\end{align}
For $|\Ha| = 0.5$, this $k_{x,c}$ coincides with that of the classic SI, whence the two RDIs overlap. For $|\Ha|$ smaller (larger) than $0.5$, the ion-cyclotron RDI occurs at smaller (larger) $k_x$ than the classical SI. 

\subsubsection{RDIs with Hall-MHD: Keplerian rotation}
For Hall-only MHD, the general solution to Eq. \ref{Lesur-dispersion-relation} is given, in terms of the wave frequency, by 
\begin{equation}
\omega_{\mathrm{gas}}^2=\frac{1}{2}\left(\mathcal C_{2}\pm\sqrt{\mathcal C_{2}^2-4\mathcal C_{0}}\right). \label{eq: normation-rotation}
\end{equation} 
Since RDI theory is based on neutral waves in the dust-free problem, we apply Eq. \ref{eq: normation-rotation} to the RDI condition Eq. \ref{eq: omega_gas} to obtain the relation between $k_x$ and $k_z$ for resonant modes only when $\omega_\mathrm{gas}^2>0$. This is met for most of the parameter space we explore, except when $|\Ha|$ is large and $\omega_\mathrm{gas}^2<0$ for the MRI.

\section{Numerical Results} \label{sec:3}

In this section, we present numerical solutions to the eigenvalue problem specified by Eq. \ref{matrix}. In the first section, we vary $\Ha$ and fix other parameters to study the impact of the Hall effect on stability (Figure \ref{fig:1}, $\S$\ref{sec:3.1}). As alluded to above, we indeed find whistler and ion-cyclotron RDIs, but they generally do not dominate over the classical SI or the MRI. However, we shall also find a non-RDI, novel ``background drift Hall instability" (BDHI) that can dominate in certain regimes, which we further explore in $\S$\ref{sec:3.2} (Figures \ref{growth_rates_vary_betaz} and \ref{growth_rates_vary_drag}). We quote the dimensionless wavenumbers $K_{x,z}\equiv k_{x,z}H_\mathrm{g}$ and normalize growth rates by $\Omega$. 

In Table \ref{tab:1}, we summarize the parameters space and comment on key results of our calculations. With names of each mode reflecting one or more major model assumptions, e.g., mode ``e02s01b4h-025'' indicates its $\epsilon=0.2$, $\st=0.1$, $\beta_{z}=10^{4}$ and $\Ha=-0.25$. Here we fixed the value of $\widehat{\eta}=0.05$ in all of our calculations. The table is broken into two horizontal blocks separated by single solid lines for $\S$\ref{sec:3.1} and $\S$\ref{sec:3.2}, respectively.

\begin{deluxetable*}{lccccc}
\tabletypesize{\scriptsize}
\tablewidth{0pt} 
\tablecaption{Selected unstable modes in Hall-dusty protoplanetary disks. List of parts modes adopted in the numerical calculation. The models in first part is presented in $\S$\ref{sec:3.1}, the models in second are selected from our parameters space study (see in $\S$\ref{sec:3.2}). The main parameters are listed in Columns 2 to 5: the dust to gas ratio $\epsilon$, the Stokes number $\st$, the plasma beta parameters $\beta_{z}$ for the vertical magnetic field, and the Els\"asser number $\Ha$ to represent the Hall effect. The last column uses a comment to characterise main results of the calculations. \label{tab:1}}
\tablehead{
\colhead{Mode} & \colhead{$\epsilon$} & \colhead{$\St$} & \colhead{$\beta_{z}$} &
\colhead{$\Ha$} & \colhead{Comments} \\
} 
\colnumbers
\startdata 
e02s01b4h-001 & 0.2 & 0.1 & $10^{4}$ & -0.01 & the case of limit of strong Hall, instabilities dominated by classic SI\\
e02s01b4h-025 & 0.2 & 0.1 & $10^{4}$ & -0.25 & instabilities composed of classic SI, Hall-RDI and BDHI\\
e02s01b4h-05 & 0.2 & 0.1 & $10^{4}$ & -0.5 & instabilities dominated by DMRI\\
e02s01b4h-2 & 0.2 & 0.1 & $10^{4}$ & -2 & the case of weak Hall, instabilities dominated by classic MRI\\
e02s01b4h+0015 & 0.2 & 0.1 & $10^{4}$ & 0.01 & instabilities dominated by MRI\\
e02s01b4h+025 & 0.2 & 0.1 & $10^{4}$ & 0.25 & instabilities dominated by MRI\\
e02s01b4h+05 & 0.2 & 0.1 & $10^{4}$ & 0.5 & instabilities dominated by MRI\\
e02s01b4h+2 & 0.2 & 0.1 & $10^{4}$ & 2.0 & instabilities dominated by MRI\\\hline
e02s01b2h-025 & 0.2 & 0.1 & $10^{2}$ & -0.25 & instabilities composed of classic SI and weak BDHI\\
e02s01b3h-025 & 0.2 & 0.1 & $10^{3}$ & -0.25 & instabilities composed of classic SI and weak BDHI\\
e02s01b5h-025 & 0.2 & 0.1 & $10^{5}$ & -0.25 & instabilities dominated by BDHI\\
e02s01b6h-025 & 0.2 & 0.1 & $10^{6}$ & -0.25 & instabilities dominated by BDHI\\
e002s01b4h-025 & 0.02 & 0.1 & $10^{4}$ & -0.25 & instabilities composed of classic SI, Hall-RDI and BDHI\\
e3s01b4h-025 & 3.0 & 0.1 & $10^{4}$ & -0.25 & instabilities dominated by classic SI\\
e02s001b4h-025 & 0.2 & 0.01 & $10^{4}$ & -0.25 & very weak instabilities \\
e02s1b4h-025 & 0.2 & 1.0 & $10^{4}$ & -0.25 & instabilities composed of classic SI, Hall-RDI and BDHI\\
\enddata
\end{deluxetable*}

\subsection{Hall effect on the Streaming Instability}\label{sec:3.1}

The Hall Els\"asser number $\Ha$ sets the direction and strength of the Hall effect with a given magnetic field intensity, here measured by the plasma beta parameter $\beta_{z}$. We begin with $\beta_{z}=10^{4}$, which is a relatively mild field strength \citep{cui2021}. Following \citet[][see their Figure 4]{Latter-Kunz2022}, we vary the Els\"asser number $\Ha$, including its sign, to adjust the possible impact of the Hall effect. For simplicity, here we fix grain size to $\st=0.1$ and consider a dust-poor disk with $\epsilon=0.2$, as considered in the `AA' setup for the classic SI by \cite{johansen07}.  

\begin{figure*}
\centering
\includegraphics[width=0.245\hsize]{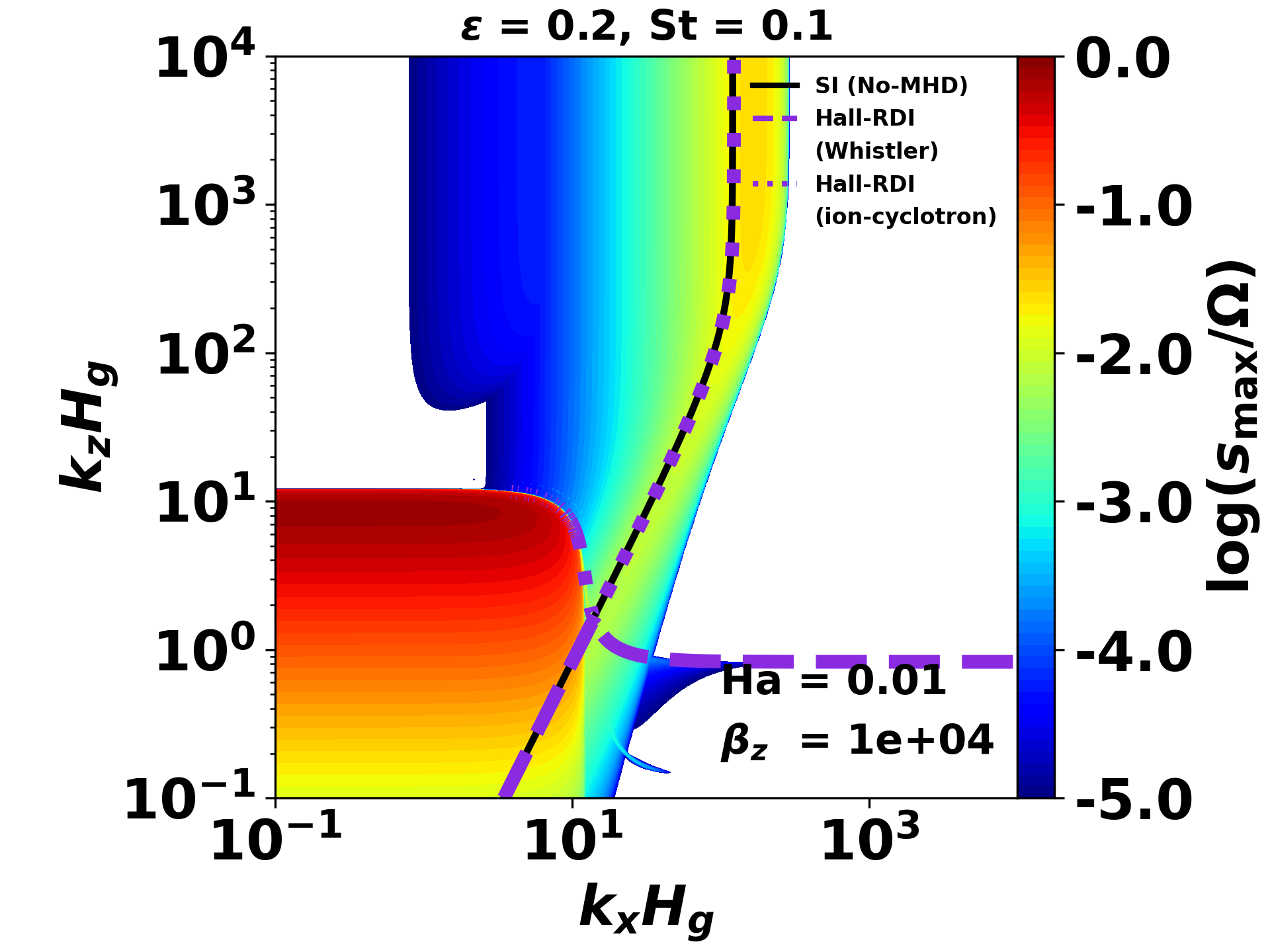}
\includegraphics[width=0.245\hsize]{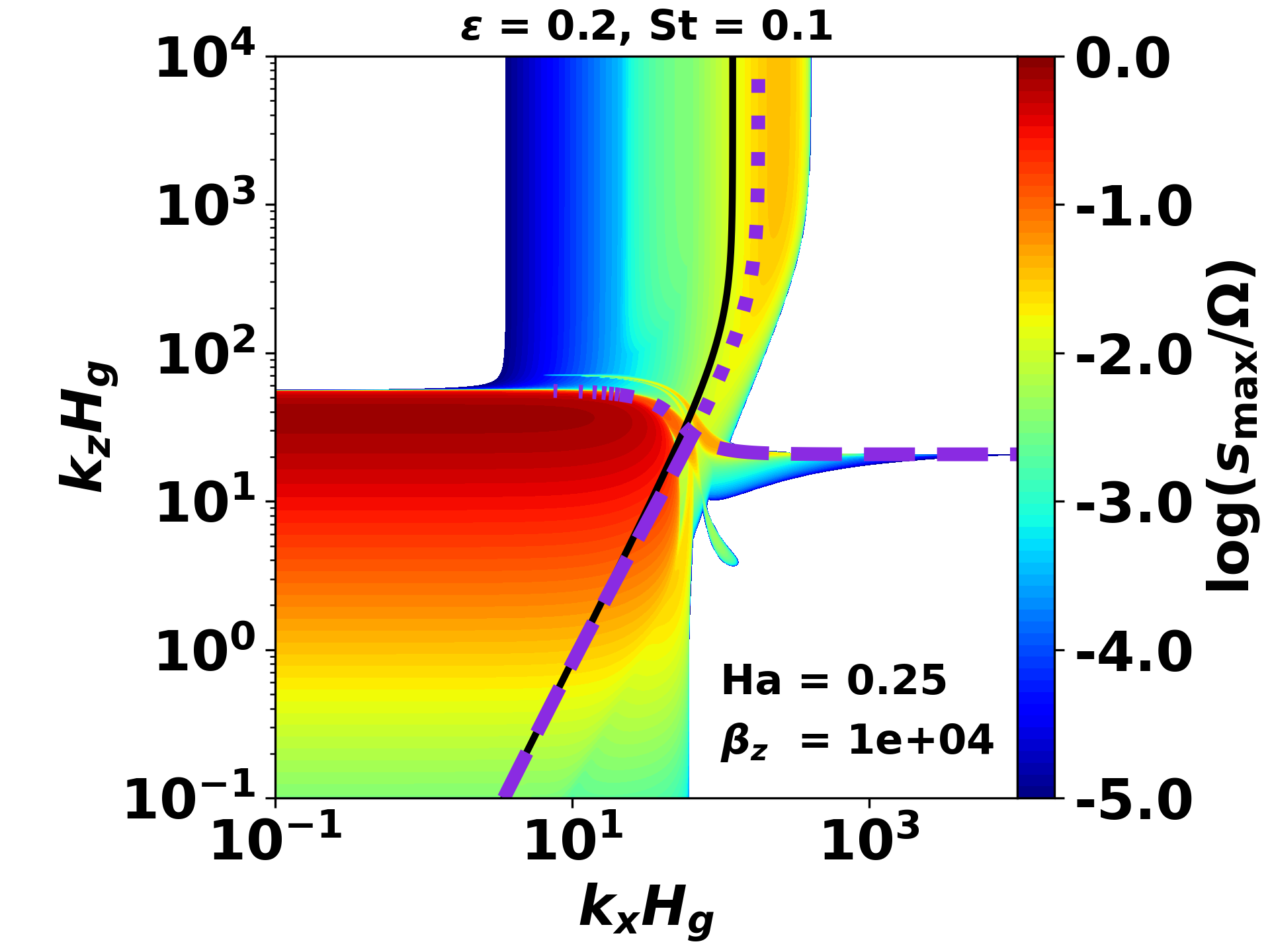}
\includegraphics[width=0.245\hsize]{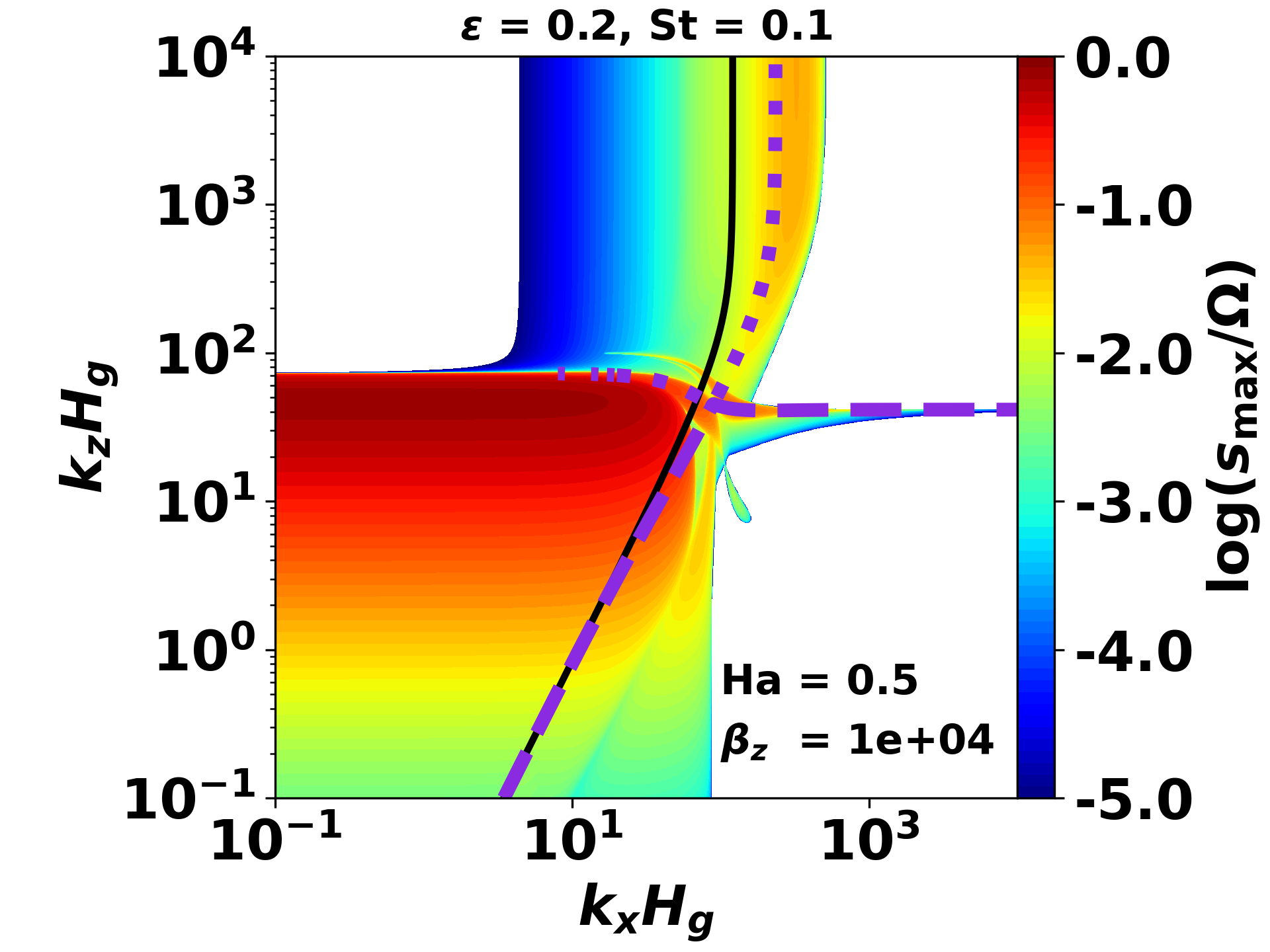}
\includegraphics[width=0.245\hsize]{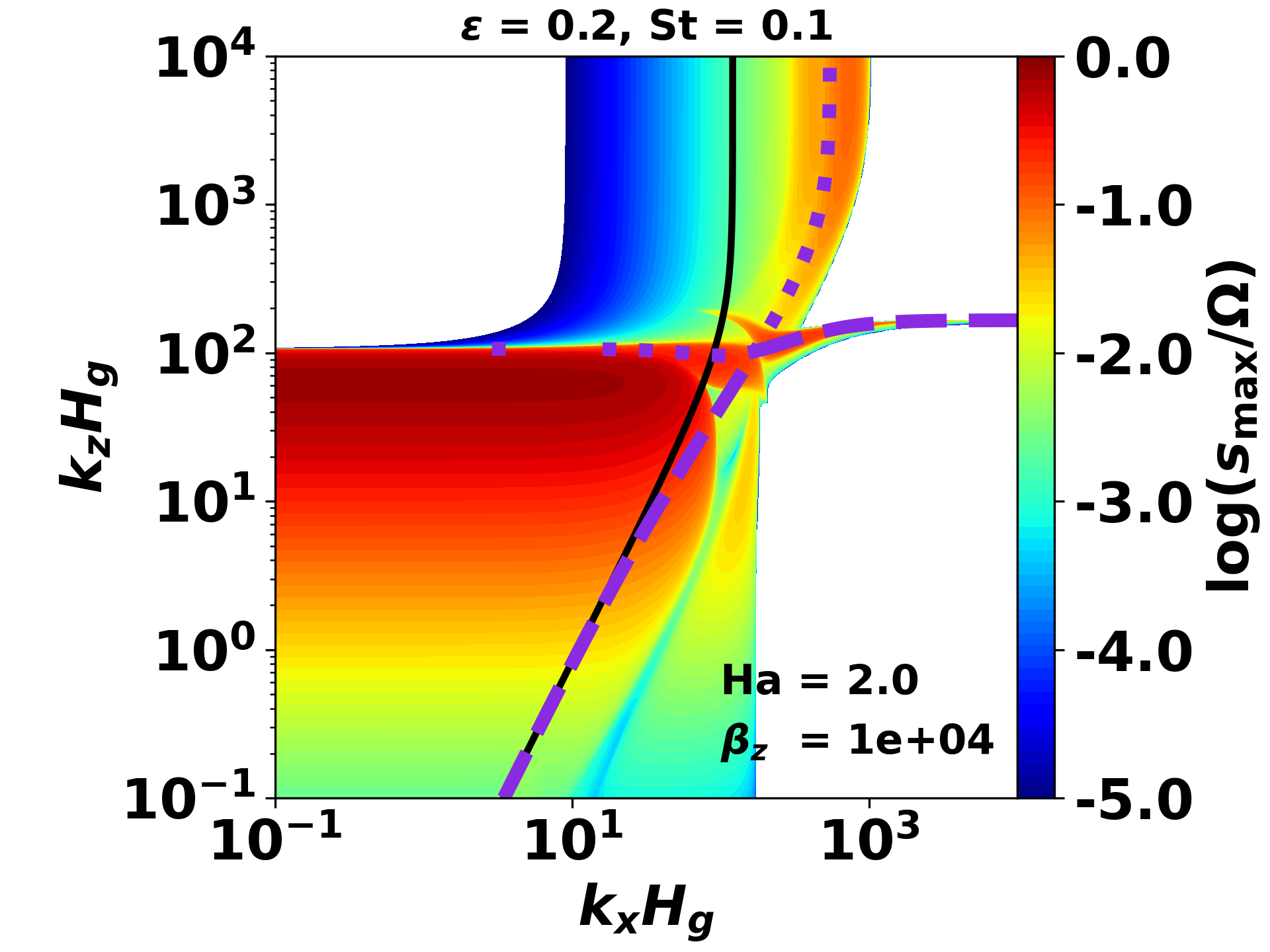}
\includegraphics[width=0.245\hsize]{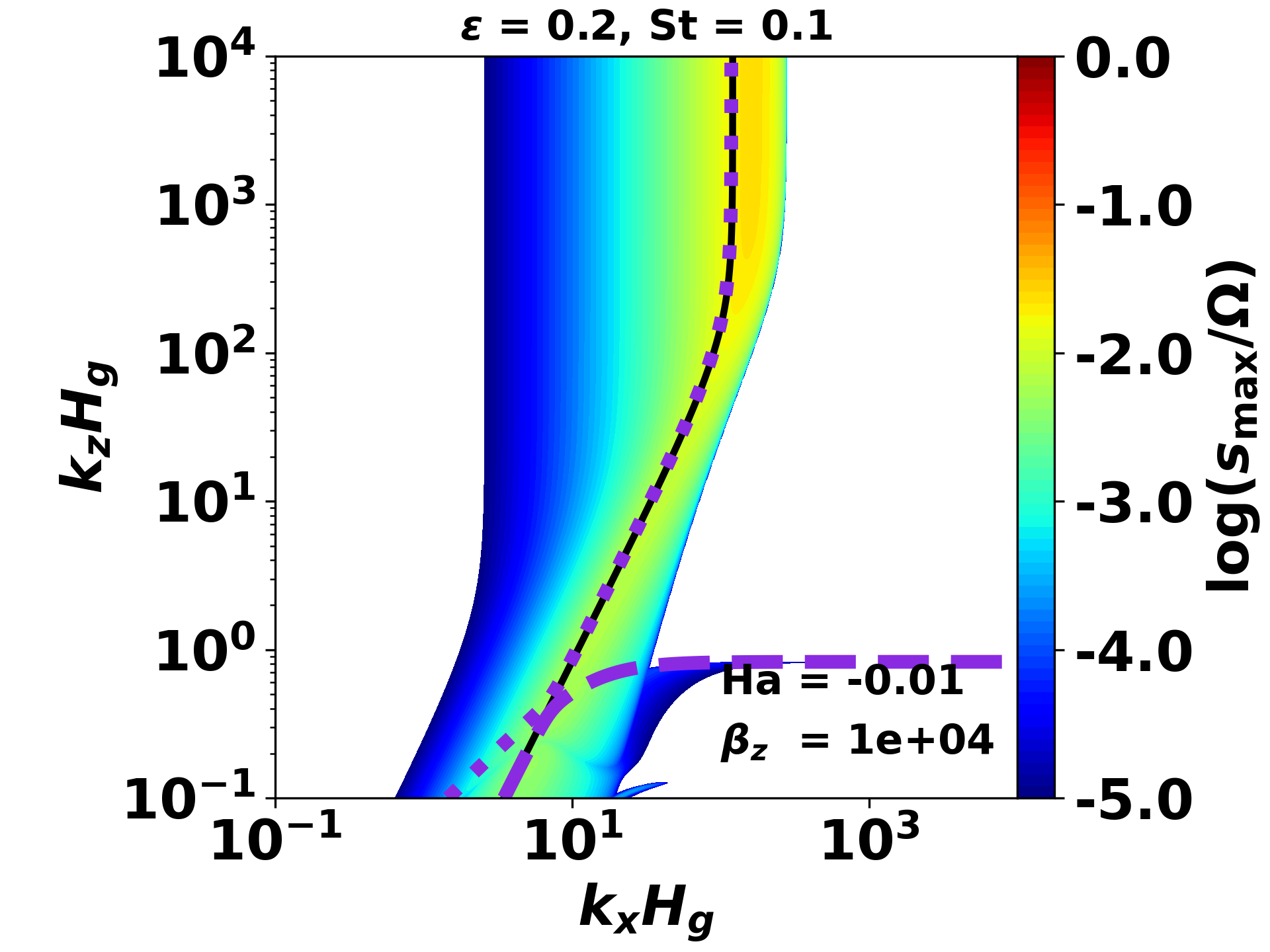}
\includegraphics[width=0.245\hsize]{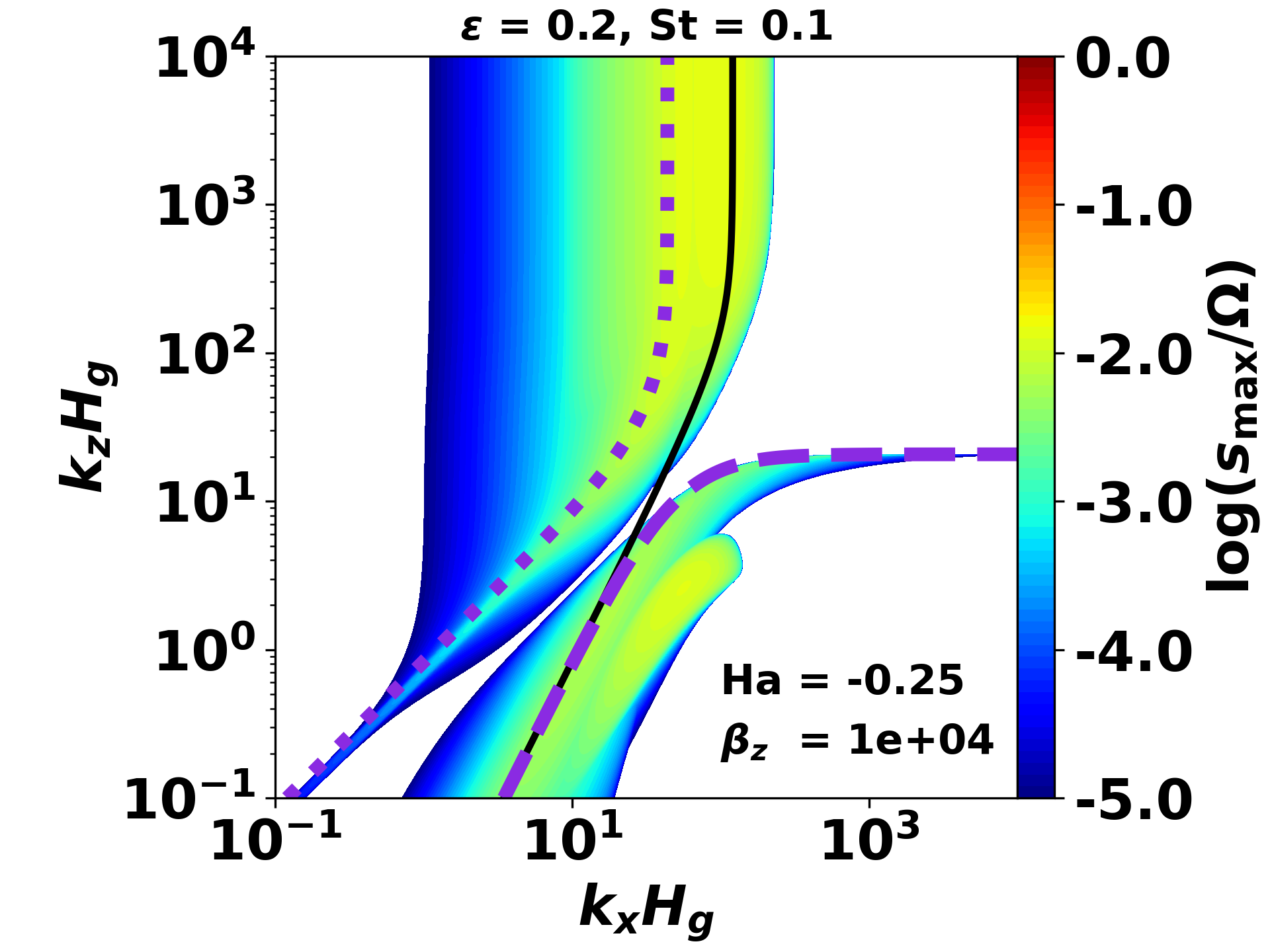}
\includegraphics[width=0.245\hsize]{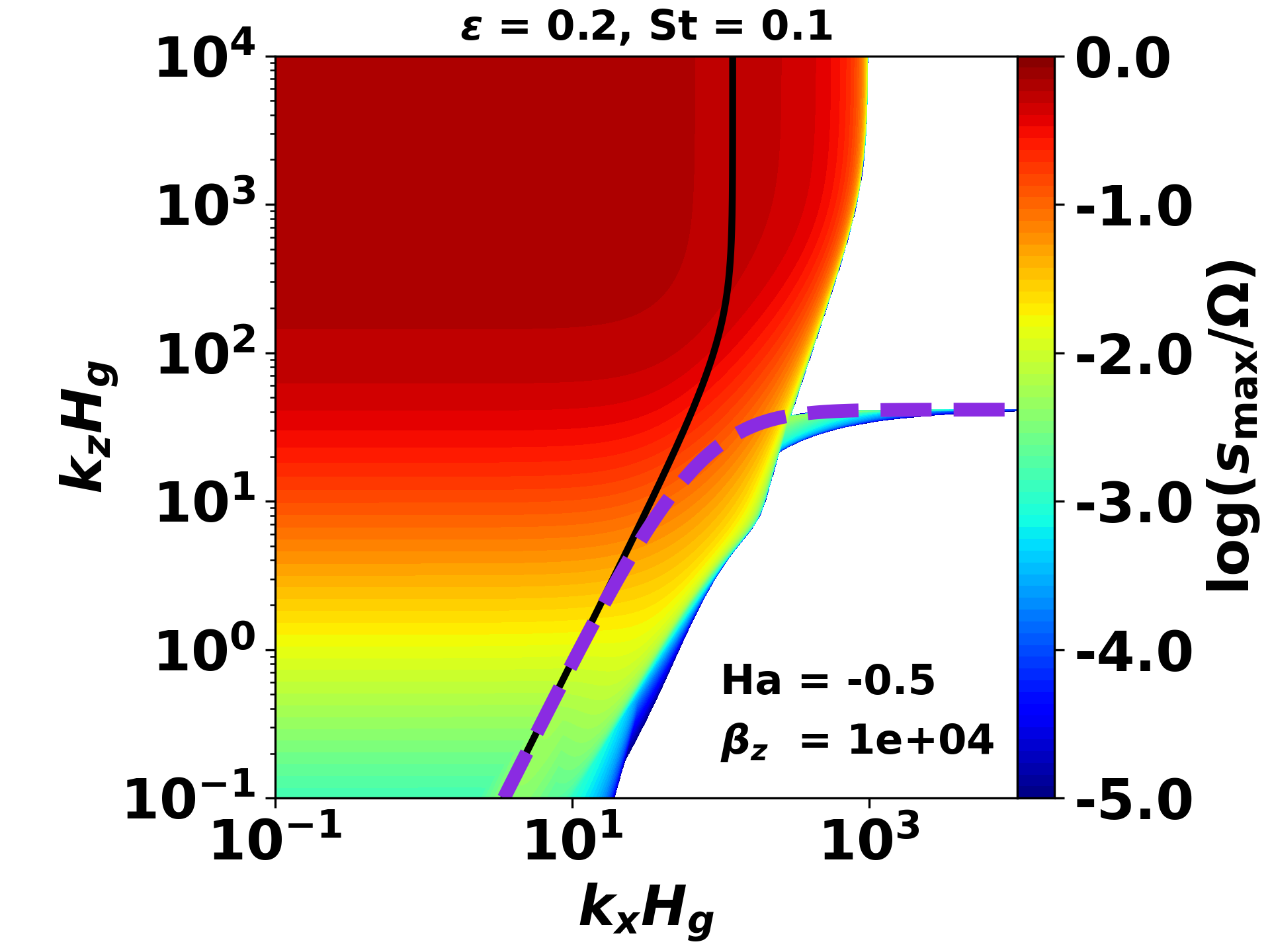}
\includegraphics[width=0.245\hsize]{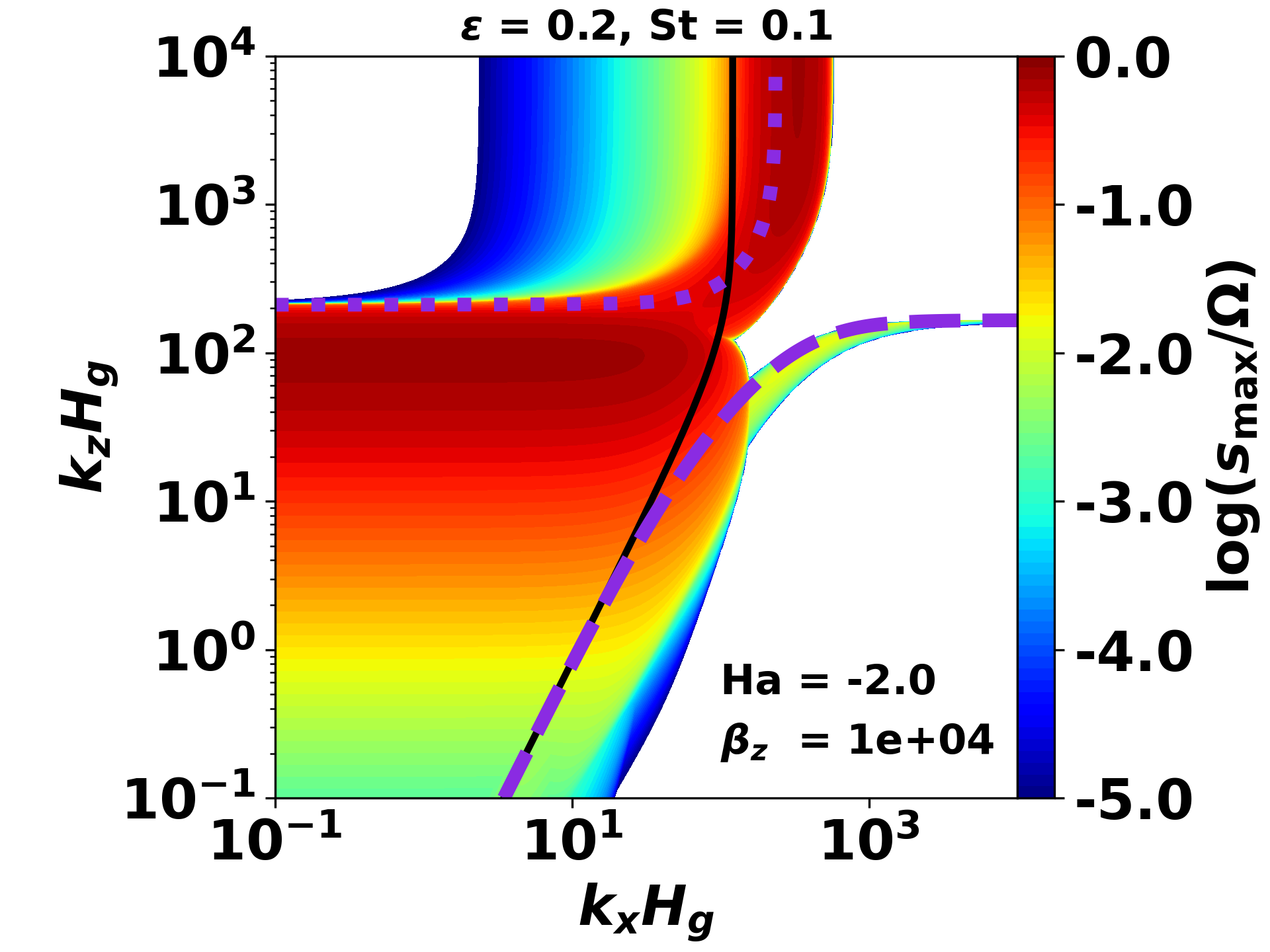}
\caption{\label{fig:1}The growth rates of SI in an actively magnetized dusty-disk initialized with a vertical field in the limit of ideal MHD. The upper row is for $\Ha>0$, and the lower row is for $\Ha<0$. The black solid line is the RDI condition for the classic streaming instability without MHD effects (Eq. \ref{eq: classic-kz}). The violet dashed line and the violet dotted line are two solutions of Eq.~\ref{eq: normation-rotation}. The above three lines are also marked in the upper right corner of the first panel in the upper row. The title is the parameters of dust in the model. The parameters of the magnetic field are marked in the lower right part of each panel.}
\end{figure*}

Figure \ref{fig:1} shows growth rates of unstable modes as a function of wavenumbers. The absolute value of $\Ha$ increases from the left to right panels, signifying a weakening of the Hall effect. The upper and bottom lower rows display cases for $\Ha>0$ and $\Ha<0$, respectively, i.e. for fields aligned and anti-aligned with the disk rotation. The RDI condition for the classic SI, i.e., without MHD effects, is depicted by the solid line (Eq. \ref{eq: classic-kz}). We also plot the two solutions to Eq.~\ref{eq: normation-rotation} for RDIs associated with whistler waves (dashed) and ion-cyclotron modes (dotted) 
in the absence of rotation. 

For cases with $\Ha>0$,  we identify the MRI as the block of `red-yellow' modes in the lower left corners. For $\Ha=0.01$, we also identify the classic SI modes associated with the solid curve; as well as the weakly growing, ion-cyclotron RDIs associated with the upper end of the dashed curve (vertical blue strip). As $\Ha$ increases, the ion-cyclotron modes shift to larger $k_x$, as expected from Eq. \ref{ion_RDI}; and by $\Ha=2$ their growth rates exceed the classic SI modes in the hydrodynamic limit. A narrow, horizontal band of whistler RDIs can also be identified along the dashed curve. However, regardless of the value of $\Ha$, when $\Ha>0$, the system is dominated by the MRI so we do not expect of any the above dust-gas instabilities to be relevant.

For cases with $\Ha<0$, the situation becomes more interesting. When $|\Ha|\ll 1$ and the impact of the Hall effect is more pronounced (e.g. $\Ha=-0.01$), the classic SI dominates as the system is close to the hydrodynamic limit. As $|\Ha|$ increases to a certain extent ($\Ha=-0.25$) and the system moves slightly closer to the ideal limit, the classic SI weakens while the whistler RDI (the dashed curve) becomes pronounced. Furthermore, a new region of instability arises below it\footnote{The branch can also be just seen in the $\Ha=-0.01$ case with $K_x \sim 10$ and $K_z\sim 10^{-1}$.}, with intensity on par with that of the classic SI. We will refer to it as the "Background Drift Hall Instability" (BDHI) for the reasons given below, where we further explore it. As we increase $|\Ha|$ further ($\Ha=-0.5$ and $\Ha=-2$), the influence of MRI re-emerges and dominates the system. However, note that for $\Ha=-2$ the ion-cyclotron RDI (dashed curve) attains a similar growth rate as the MRI, in which case one expects the two instabilities to operate simultaneously. 

Concluding this section, the majority of the new instability branches align well with the RDI theory. Moreover, at significantly negative values of $\Ha$, MRI and ion-cyclotron RDI exhibit comparable magnitudes. However, in cases with moderate and negative values of $\Ha$, a novel instability, the BDHI, may prevail. 

\subsection{Parameter study} \label{sec:3.2}

\begin{figure*}
\centering
\includegraphics[width=0.49\hsize]{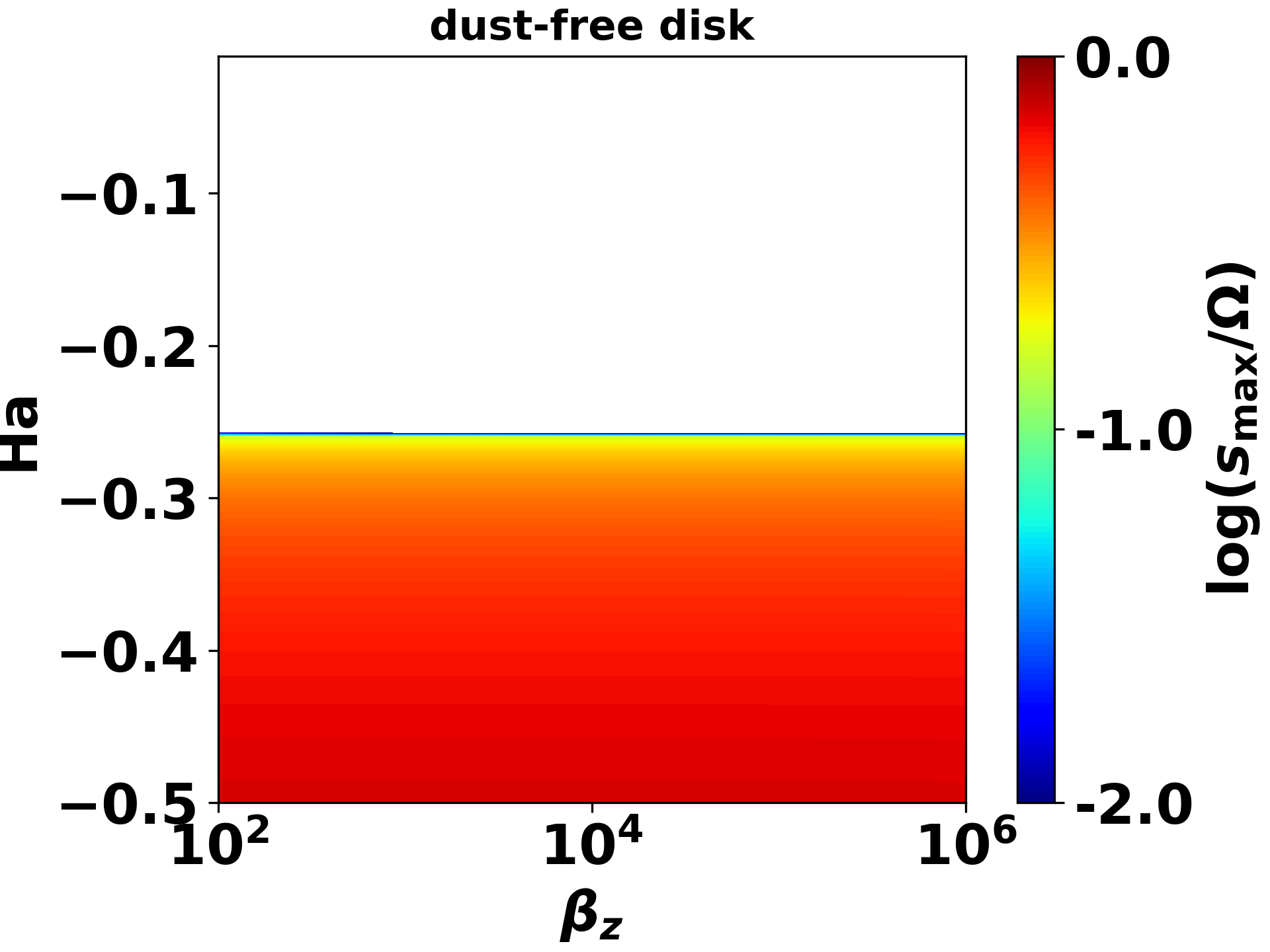}
\includegraphics[width=0.49\hsize]{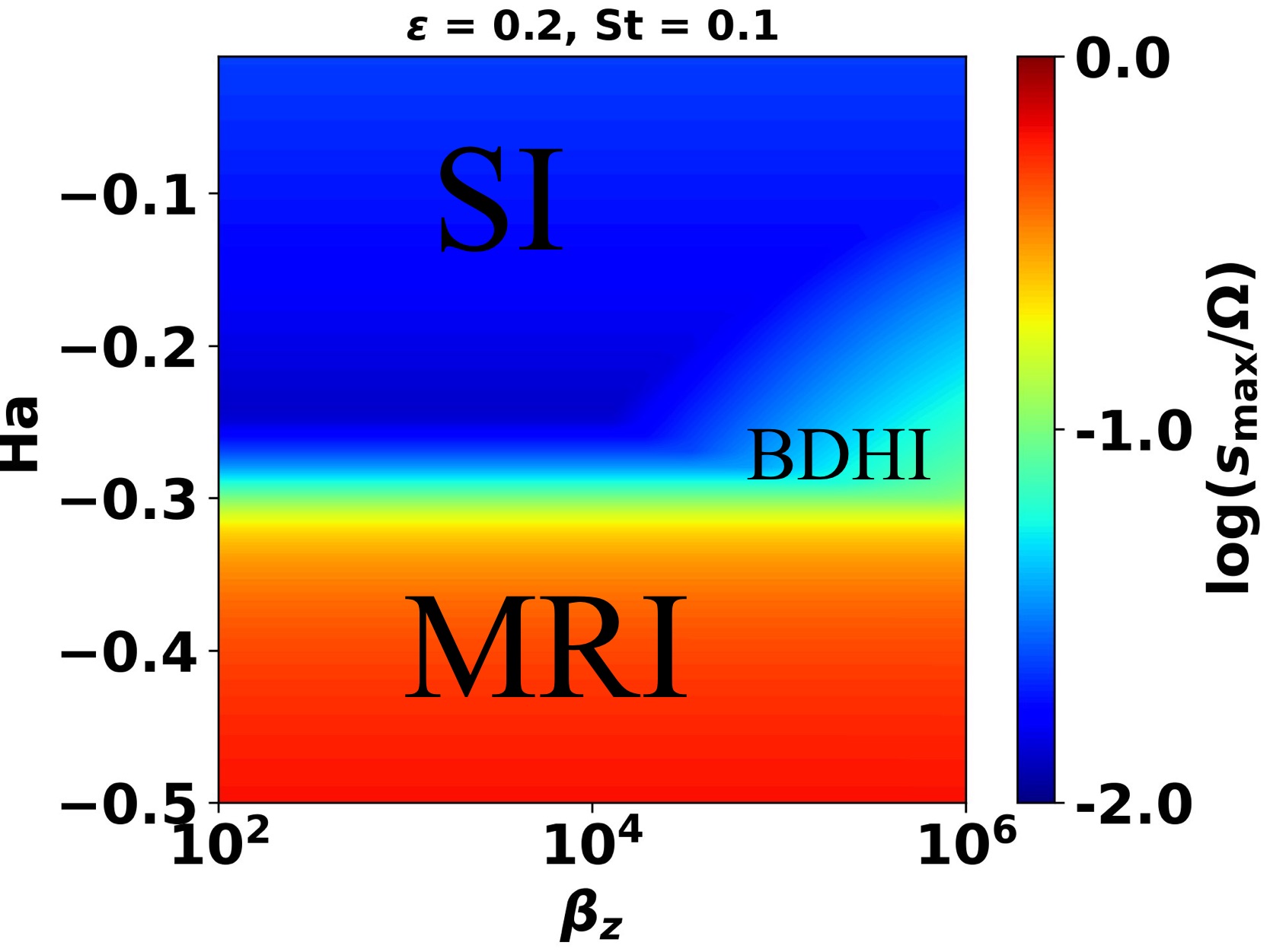}
\caption{The maximum growth rate as a function of $\beta_z$ (x-axis) and $\Ha$ (y-axis). For the left panel, we set $\epsilon=0$, i.e., dust-free and MRI-only disk, as a reference case. For the right panel, we fixed $\epsilon=0.2$ and $\St=0.1$ for each model, as shown in the title. The instability modules occupying the dominant position in each region are labeled in the corresponding regions in the right panel.}
\label{growth_rates_vary_betaz}
\end{figure*}

Here, we explore how the above instabilities depend on the disk parameters. In Figure \ref{growth_rates_vary_betaz}, we plot the maximum growth rates as a function of $\beta_z\in[10^2,10^6]$ and $\Ha\in[-0.5,-0.01]$. The left panel of Figure \ref{growth_rates_vary_betaz} is a dust-free reference case ($\epsilon=0$), in which only the MRI operates for sufficiently negative $\Ha$ ($\lesssim -0.25$). In the right panel of Figure \ref{growth_rates_vary_betaz}, we set $\epsilon=0.2$ and $\St=0.1$. The figure shows a sharp transition in the growth rate at $\Ha\simeq -0.3$ with the MRI pushed to more negative values of $\Ha$ but remains dominant in that regime. 

The classical SI manifests in the `blue' region where $\Ha \gtrsim -0.3$. We conclude this from the fact that the SI is fundamentally a non-magnetic phenomenon, so its growth should remain constant when varying magnetic parameters, and should equal to that for $|\Ha|\to0$. This also simplifies the identification of parameter ranges where the BDHI dominates, namely the cyan region 
where $\beta_z\gtrsim 10^5$.

\begin{figure*}
\centering
\includegraphics[width=0.49\hsize]{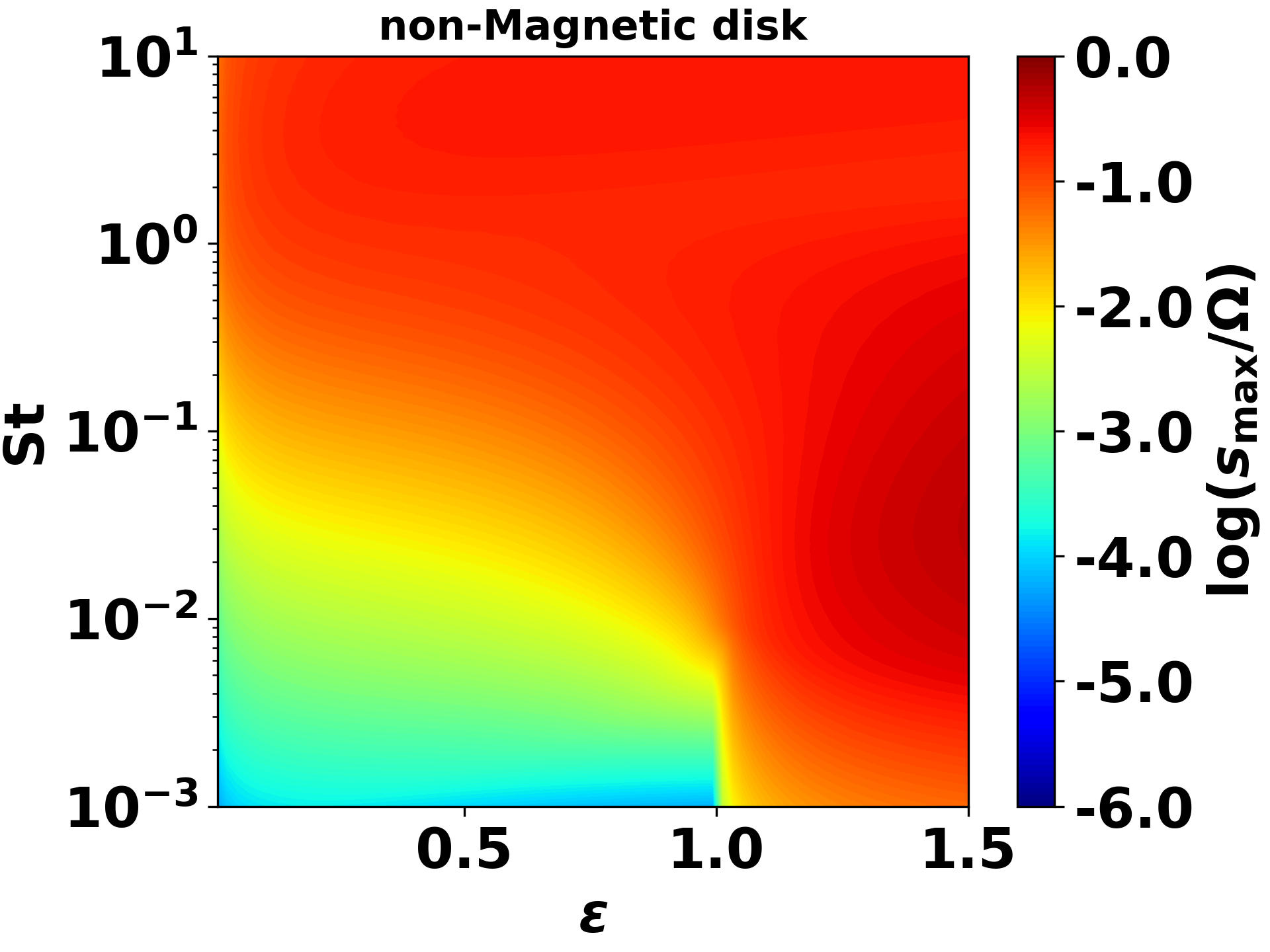}
\includegraphics[width=0.49\hsize]{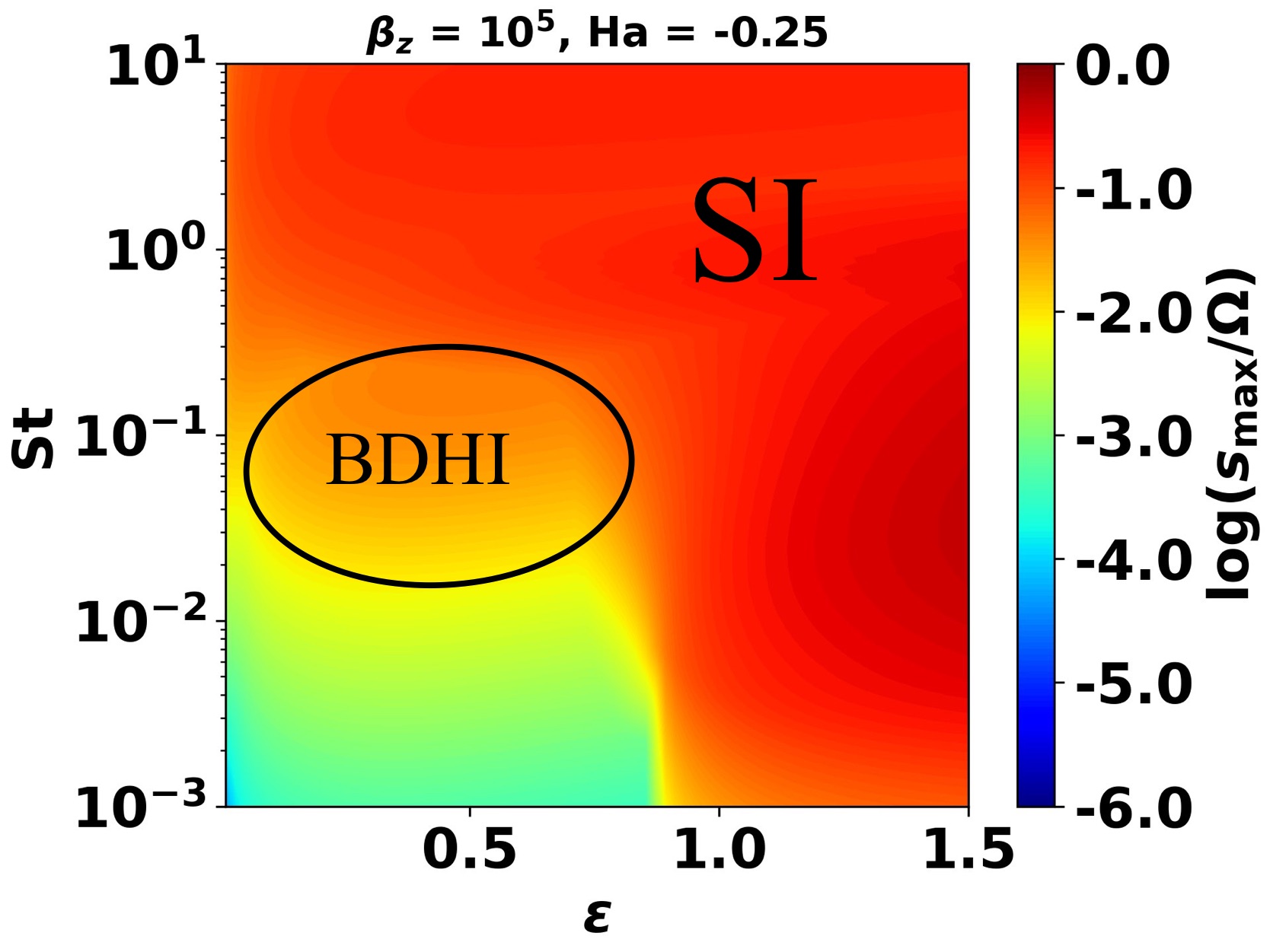}
\caption{The maximum growth rate as a function of $\epsilon$ (x-axis) and $\St$ (y-axis). Similar to Figure \ref{growth_rates_vary_betaz}, we also set a reference case in the left panel, with $\beta_z=\infty$, i.e., non-magnetic disk. In the right panel, we fixed $\beta_z=10^{5}$ and $\Ha=-0.25$ for each model, as shown in 
the title. Same as Figure \ref{growth_rates_vary_betaz}, we labeled the SI occupying the dominant position in the corresponding regions in the right panel, and we marked the rough outline of BDHI using black solid lines.}
\label{growth_rates_vary_drag}
\end{figure*}

We next investigate the impact of dust properties. In Figure \ref{growth_rates_vary_drag}, we plot growth rates as a function of $\epsilon$ and $\St$ with fixed $\beta_z=10^5$ and $\Ha=-0.25$. Similar to Figure \ref{growth_rates_vary_betaz}, we also prepared a reference case for comparison. In the left panel of Figure \ref{growth_rates_vary_drag}, we set $\beta_z=\infty$ as non-magnetic disk models. In such a disk, the source of instability should be the classical SI. Then, we vary $\epsilon$ from 0.01 to 3, and $\St$ from $10^-3$ to 10, depend on the right panel of Figure \ref{growth_rates_vary_betaz}. Because BDHI could potentially assume a dominant position within this parameter space. 

In a dust-rich disk with $\epsilon\gtrsim 1$ or with large grains of $\St\gtrsim 0.3$, we find the classical SI is always the dominant instability with high growth rates ($s_\mathrm{max}\sim 0.1\Omega$). By contrast, in a dust-poor disk with $\epsilon\lesssim 1$ and $0.01\lesssim \St \lesssim 0.3$, we find the BDHI dominates (orange-yellow). However, for tightly coupled grains with $\St\lesssim 10^{-2}$ at low dust-to-gas ratios, both the classical SI and BDHI are weak, making it indiscernible which is dominant.  

\section{Background Drift Hall Instability} \label{sec:BG drift model}

In $\S$\ref{sec:3}, we identified a new branch of unstable modes --- unrelated to any of the RDIs in \S\ref{subsec:RDI} --- that appear at low dust-to-gas ratios ($\epsilon< 1$) and a weak field ($\beta_z\sim 10^5$) anti-aligned with the disk rotation with moderate Hall strength ($-0.25\lesssim \Ha <0$). See, for example, the dominant modes in the left panel of Figure  \ref{growth_rates_vary_betaz}. In this section, we reproduce this new instability in a toy model, which shows that it results from the advection of magnetic field perturbations relative to the dusty gas' center of mass. 

\subsection{Destabilization by dust-induced advection of magnetic perturbations}\label{sec:4.1}

\begin{figure}
\centering
\includegraphics[width=\linewidth]{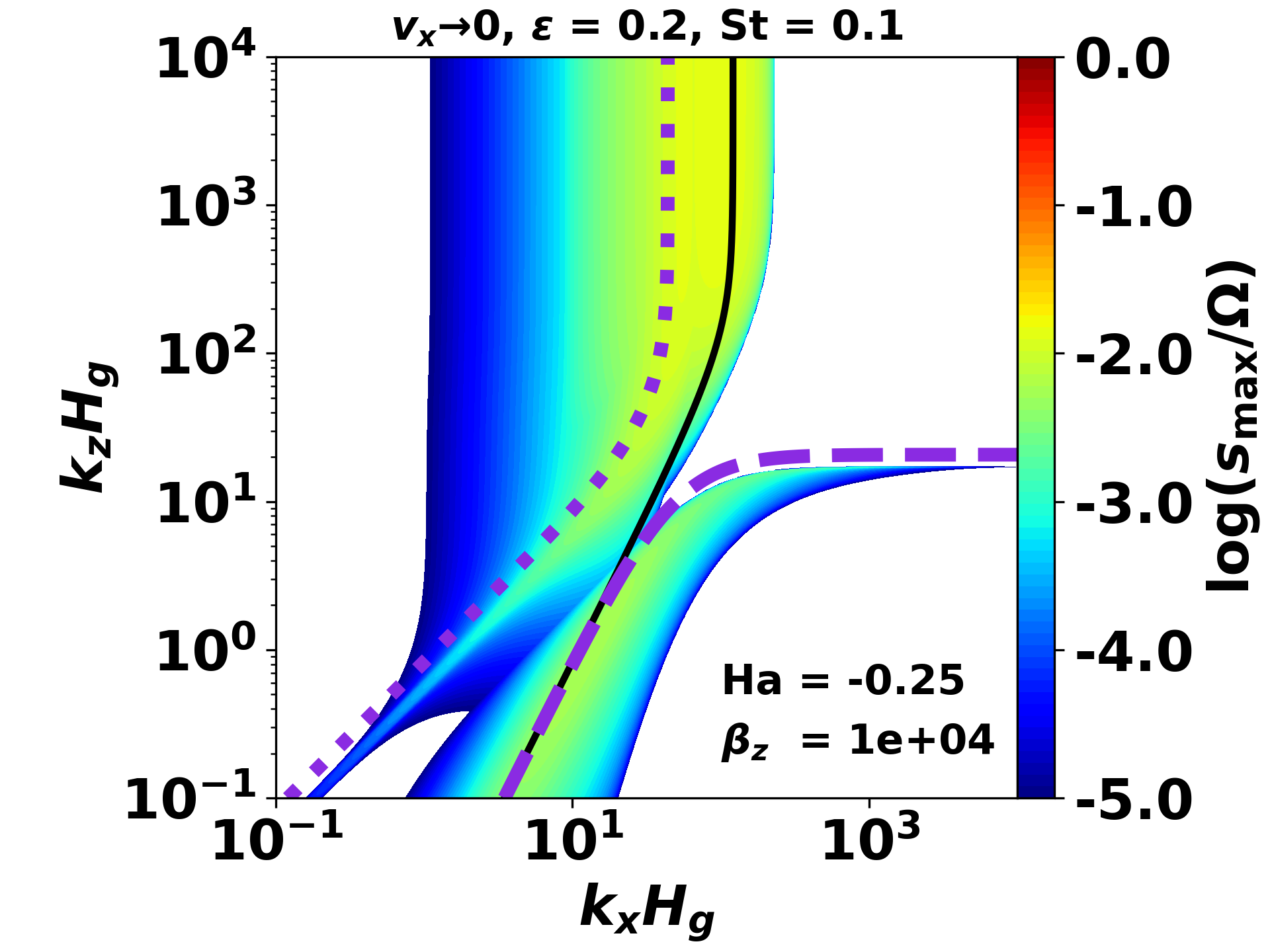}
\caption{Same as the case with $\Ha=-0.25$ in Figure \ref{fig:1} but set the equilibrium gas radial velocity $v_x\to0$ in Eq. \ref{eq-bx}-\ref{eq-bz}.}
\label{BDHI_full_novx}
\end{figure}

The fact that this new instability is not an RDI indicates that dust does not play an active role, specifically, that is through the drag forces in the linearized momentum equations. However, dust-gas drag also manifests through the equilibrium radial velocities. Here, we show that the new instability is related to the advection of magnetic fields by such an equilibrium flow. To this end, we repeat the standard calculation in Figure \ref{fig:1} (the second panel from the left in the bottom row, with $\Ha = -0.25$) but artificially set the equilibrium gas radial velocity $v_x\to0$ in the linearized induction equations Eq. \ref{eq-bx}-\ref{eq-bz}. The result is shown in Figure \ref{BDHI_full_novx} and we see that the new instability has disappeared. We, therefore, refer to it as the ``background-drift Hall instability" (BDHI). 

\subsection{Toy model} \label{bg_drift_hallSI_toy}

As shown above, the BDHI arises from the advection of magnetic perturbations by the dust-induced background flow. This effect can be captured in the standard, pure gas incompressible Hall-MHD equations by adding a source term in the induction equation to transport magnetic field perturbations. The governing equations of such a toy model are: 
\begin{align}
    \nabla\cdot \bm{v} = \nabla\cdot \bm{B} = 0\label{toy_model_solenoidal},
\end{align}
\begin{align}
    \frac{\partial\textbf{\emph{v}}}{\partial t} + \textbf{\emph{v}}\cdot\nabla\textbf{\emph{v}} =&     2v_{y}\Omega\hat{\textbf{\emph{x}}} - v_{x}\frac{\Omega}{2}\hat{\textbf{\emph{y}}} \notag\\&-\frac{\fgas}{\rhog}\nabla \Pi 
    +\frac{\fgas}{\mu_0\rhog}\bmB\cdot\nabla\bmB,\label{toy_model_vg}
\end{align}
where $\Pi$ is the total pressure, 
\begin{align}
      \frac{\partial\textbf{\emph{B}}}{\partial t} =& \nabla\times\left(\bm{v}\times\bmB\right) + \nabla\times\left(\bm{v}_\mathrm{ext}\times\Delta\bmB\right)
      -\frac{3}{2}\Omega B_x\yhat  \notag\\& - \left(\frac{\eta_H}{B_{z0}}\right)\nabla\times\left[\left(\nabla\times\bmB\right)\times\bmB\right],\label{toy_model_induction}
\end{align}
$\bm{v}_\mathrm{ext}$ is a velocity field prescribed below, and 
\begin{align}
    \Delta\bmB = \bmB - B_{z0}\zhat
\end{align}
is the magnetic field deviation (not necessarily small) from the equilibrium, constant vertical field $B_{z0}$. In the toy induction equation Eq. \ref{toy_model_induction}, we augmented the Hall term slightly to facilitate our nonlinear simulations in \S\ref{simulations} (see also Appendix \ref{full_single_fluid}), but this does not affect the system's linear stability. 

Our toy model should be viewed in conjunction with a corresponding dusty problem. We thus set 
$\bmv_\mathrm{ext} = v_\mathrm{ext}\xhat$, where $v_\mathrm{ext}$ is the dust-induced, radial gas drift  given by Eq. \ref{eqm_vx}. Furthermore, in Eq. \ref{toy_model_vg}, we reduce the pressure and magnetic tension forces by a constant gas fraction $\fgas$. This reduction manifests in the dusty problem for tightly-coupled grains because dust-loading increases the gas' inertia (see Appendix \ref{TVA} and Eq. \ref{gas_based_mom_tva}). 

One can regard the $\bmv$ in our toy model to represent the corresponding dusty-gas' center of mass velocity (which experiences no equilibrium drift). However, since the magnetic field is physically carried by the gas component only, there exists a drift between the magnetic field and the system's center of mass. This is the spirit of our toy model. 

\subsection{Linearized equations and dispersion relation}
Linearizing the toy model about $\bmv = 0$ and a constant vertical field gives:
\begin{align}
    0 = \ikx \delta v_x + \ikz \dd v_z = \ikx \dd B_x + \ikz \dd B_z,\label{toy_linear_solenoidal}
\end{align}
\begin{align}
    \sigma\dd\bmv =& 2\Omega\dd v_y\xhat - \frac{\Omega}{2}\dd v_x \yhat - \fgas\left(\ik\frac{\dd\Pi}{\rhog} - \frac{\ikz B_z}{\mu_0\rhog}\dd\bmB\right) \label{toy_linear_mom},
\end{align}
\begin{align}
    \sigma \dd\bmB =& \ikz B_z\dd\bmv - \ikx \vext \dd \bmB - \frac{3}{2}\Omega\dd B_x\yhat -\etahall k_z^2\dd B_y\xhat +\notag\\& \etahall k^2 \dd B_x \yhat +\etahall k_x k_z \dd B_y\zhat,\label{toy_linear_induction}
\end{align}
which may also be obtained from the linearized single-fluid formulation of the dusty problem (Eqs. \ref{gas_based_incompress}---\ref{gas_based_linear_induction}) by setting $\delta\epsilon=\st=0$ where they appear explicitly. 

These equations yield the dispersion relation 
\begin{align}
    \sigma^4 + \widetilde{\mathcal{C}}_3\sigma^3 + \widetilde{\mathcal{C}}_2\sigma^2 + \widetilde{\mathcal{C}}_1 \sigma + \widetilde{\mathcal{C}}_0 = 0\label{toy_dispersion},
\end{align}
with
\begin{align}
    \widetilde{\mathcal{C}}_3 =& 2 \ikx \vext,\\
    \widetilde{\mathcal{C}}_2 =& \frac{k_z^2}{k^2}\Omega^2 + 2 \fgas k_z^2\va^2 
    + \etahall k_z^2\left(\etahall k^2 - \frac{3}{2}\Omega\right)
    - k_x^2 \vext^2\\
   \widetilde{\mathcal{C}}_1 =& 2 \ikx \vext \frac{k_z^2}{k^2}\left(\Omega^2 + \fgas  k^2 \va^2 \right),\\
     \widetilde{\mathcal{C}}_0 =& \left( \fgas k_z^2 \va^2 + 2\etahall k^2_z \Omega - 3\frac{k^2_z}{k^2} \Omega^2\right)\times\notag\\&\left( \fgas k^2_z\va^2 + \frac{1}{2}\etahall k^2_z\Omega \right) - \frac{k_x^2k^2_z}{k^2}\Omega^2_0\vext^2.
\end{align}

Without the imposed drift, $\vext=\widetilde{\mathcal{C}}_3=\widetilde{\mathcal{C}}_1=0$, we recover the bi-quadratic dispersion relation for Hall-only MHD, see e.g. \cite{Lesur-review} and  \S\ref{subsec:dispersion relation}. In the presence of the drift, the dispersion relation is a full quartic and must generally be solved numerically. Notice the Alfv\'{e}n speed is reduced by dynamical coupling to dust. 

\subsection{An RDI-like interpretation}

As the toy model is comprised only of a single fluid, the background drift Hall instability cannot be an RDI as described by \cite{Squire-Hopkins-2018a}. However, it turns out that we can estimate the most unstable wavenumbers based on similar ideas, as follows. This is because our toy model also exhibits a drift: between the fluid and the magnetic field. 

We consider the strong Hall regime with $|\Ha|\ll 1$. We first show that without the imposed magnetic drift, the system admits neutral waves. This is akin to the first step of the RDI recipe, where one identifies neutral waves in the corresponding pure gas system. 

In this limit, $\widetilde{\mathcal{C}}_3=\widetilde{\mathcal{C}}_1=0$, and 
\begin{align}
    \widetilde{\mathcal{C}}_2 \simeq & k_z^2\left[ \frac{\Omega^2}{k^2} 
    + \etahall\left(\etahall k^2 - \frac{3}{2}\Omega\right)\right] > 0, \\
     \widetilde{\mathcal{C}}_0 \simeq &  \frac{k_z^4}{k^2}\Omega^2 \etahall \left(\etahall k^2 - \frac{3}{2} \Omega\right).
\end{align}
Solving the dispersion relation Eq. \ref{toy_dispersion} in the above limit, we find the gas wave frequencies $\omega_\mathrm{gas} = \ii \sigma$ satisfy $\omega_\mathrm{gas}^2 = k_z^2\Omega^2/k^2$, corresponding to inertial waves, or  
\begin{align}
    \omega_\mathrm{gas}^2 = \etahall k_z^2\left(\etahall k^2 - \frac{3}{2}\Omega  \right). \label{whislter_strong_hall_rotation}
\end{align}
In the case of interest with $\Ha, \etahall < 0$ and thus $\omega_\mathrm{gas}^2>0$, these correspond to whistler waves modified by Keplerian rotation. Note that in the absence of rotation, Eq. \ref{whislter_strong_hall_rotation} reduces to the positive solution of Eq. \ref{strong_hall_no_rotation}. 

We now suppose magnetic perturbations with radial wavenumber $k_x$ are transported at a radial velocity $\vext$. We may then expect a resonance to occur if this transport velocity matches the phase velocity of neutral whistler waves, i.e. if $\omega_\mathrm{gas}/k_x = \vext$. This is akin to the RDI condition where the dust-gas drift velocity matches the phase speed of neutral waves in the gas. 

The above RDI-like condition can be written in dimensionless form 
\begin{align}
    K_z^2 = \frac{\beta_z|\Ha|}{\left(1+\theta^2\right)}\left(\beta_z |\Ha|\theta^2\frac{\vext^2}{c_s^2} - \frac{3}{2}\right) \label{bg_hall_rdi},
\end{align}
where $\theta \equiv K_x/K_z$. Eq. \ref{bg_hall_rdi} can be regarded as a quadratic equation for $K_z^2$ as a function of $K_x$. Defining $\zeta_x \equiv \vext/C_s$, for $|\theta|\to \infty$ we find $K_z^2 \to (\beta_z \Ha\zeta_x)^2 $. We also require $\theta^2 > 3/\left(2\beta_z|\Ha|\zeta_x^2\right)$ for real vertical wavenumbers. That is, modes must be sufficiently narrow in the radial direction compared to their vertical length scale.   

\subsection{Example solution}\label{toy_model_example}

\begin{figure}
\centering
\includegraphics[width=1\hsize]{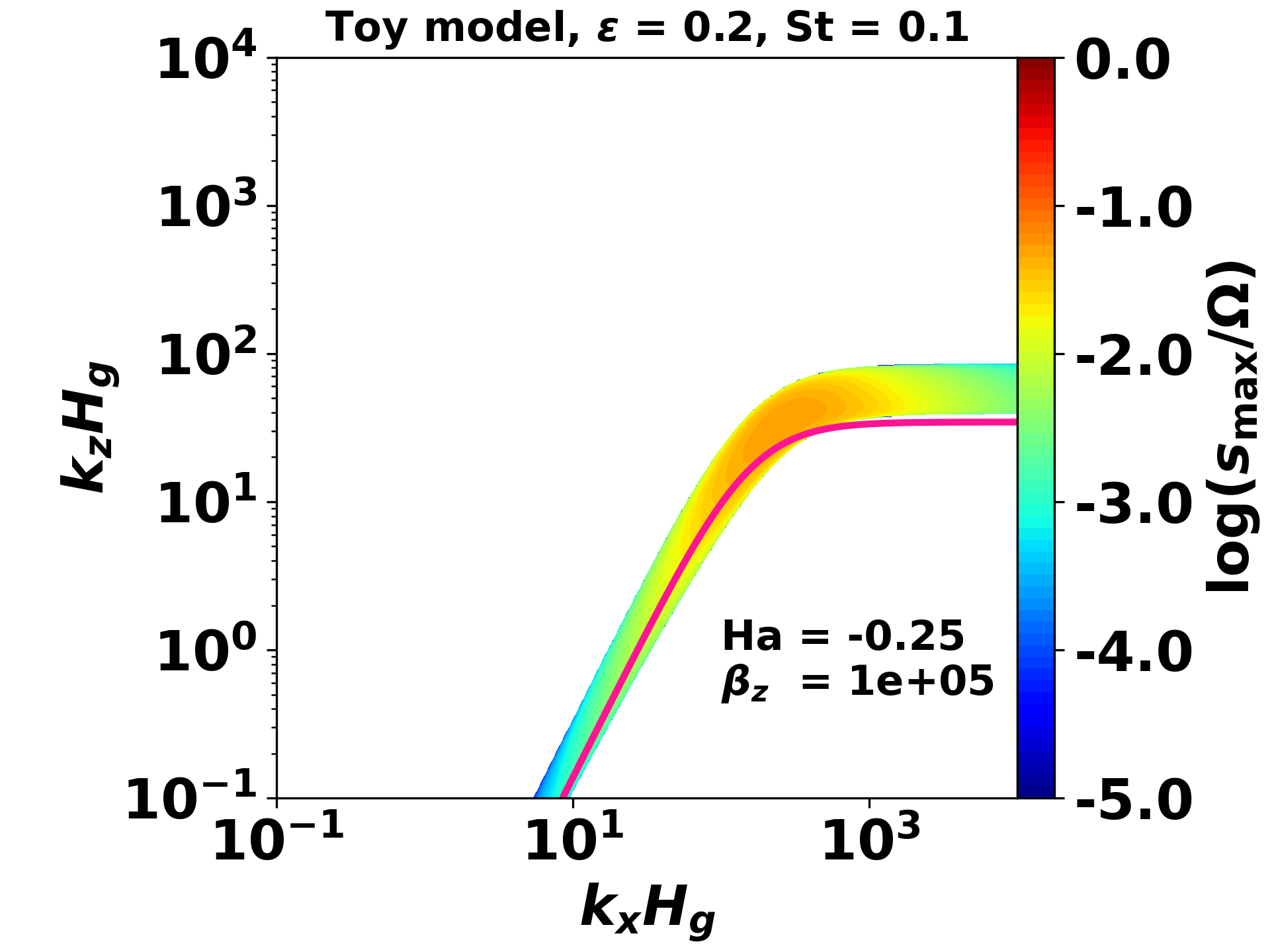}
\includegraphics[width=1\hsize]{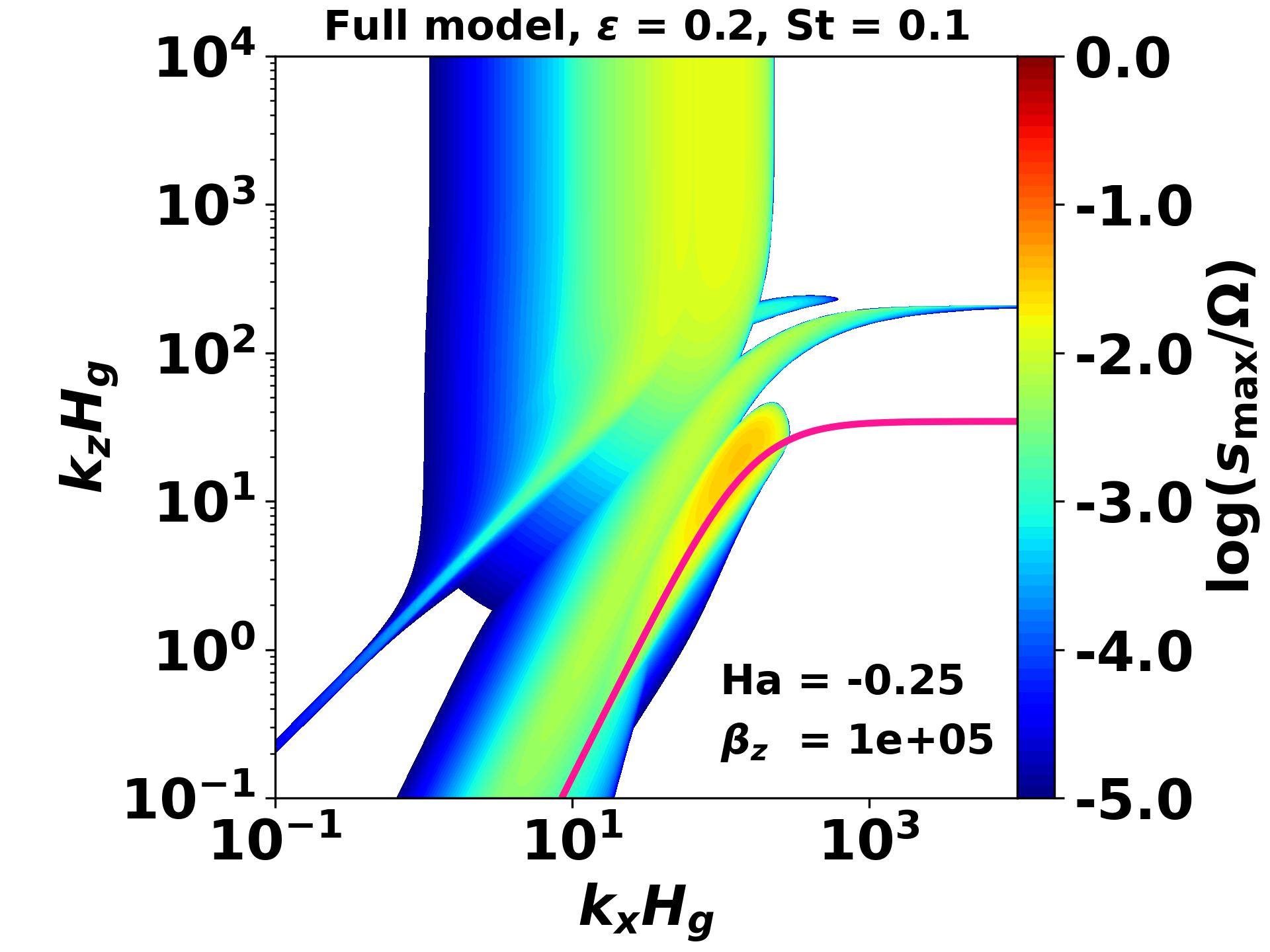}
\caption{Comparison between toy model of the BDHI (the upper panel) and the full model (the lower panel), with $\beta_z=10^5$, $\Ha=-0.25$,$\epsilon=0.2$ and $\St=0.1$. The pink solids line are calculated by Eq. \ref{bg_hall_rdi}.\label{fig:full vs toy}}
\end{figure}

In Figure \ref{fig:full vs toy} we plot growth rates obtained from the dispersion relation of the toy model (Eq. \ref{toy_dispersion}) and compare it with that obtained from the full model. Here, we set  $\epsilon=0.2$, $\st=0.1$, $\etahat=0.1$; $\Ha=-0.25$, and $\beta=10^5$, for which the BDHI dominates. We also over-plot the RDI-like condition given by Eq. \ref{bg_hall_rdi}. The toy model indeed reproduces the BDHI, although with slightly higher growth rates. The toy model also predicts significant growth for $K_x\gtrsim 10^3$; while in the full model, this region has negligible growth. Nevertheless, the coincidence between the BDHI region with the curves supports an RDI-like interpretation of the instability. 

\section{Direct Simulations
} \label{simulations}

In this section, we demonstrate the background drift Hall instability using direct numerical simulations. We use the \textsc{dedalus} spectral code \citep{burns20} to evolve both the single-fluid model described in Appendix \S \ref{full_single_fluid} and the toy model described above. In both models, we solve for the magnetic vector potential $\bmA$ such that $\Delta \bmB = \nabla \times \bmA$, which ensures that $\nabla\cdot\bmB = 0$. 

Explicit dissipation is needed in spectral simulations for numerical stability. In the single-fluid model, we include viscosity and resistivity characterized by a constant kinematic viscosity coefficient $\nu_1$ and constant resistivity coefficient $\chi$, respectively. As this model also evolves the relative dust-gas drift $\bm{u} = \bmw - \bmv$ and the dust-to-gas ratio $\epsilon$, we further include a viscous term  $\nu_2\nabla^2 \bm{u}$ in the equation for $\bmu$ and dust diffusion $D\nabla^2\epsilon$ in the dust mass equation; where $\nu_2$ and $D$ are constant viscosity and diffusion coefficients, respectively. We solve for $Q_\epsilon\equiv \epsilon_0\ln{\left(\epsilon/\epsilon_0\right)}$, where $\epsilon_0$ is the initial dust-to-gas ratio, to ensure that $\epsilon>0$. Our implementation of the single-fluid model in \textsc{dedalus} is described in Appendix \ref{full_model_dedalus}. 

In the toy model, we only include gas viscosity ($\nu_1$) and ohmic resistivity ($\chi$) since this model does not involve $\bmu$ or $\epsilon$. The dissipative toy model is described in Appendix \ref{hallSI_diss_toy_model} for ease of reference. More specifically, we evolve Eqs. \ref{toy_model_vg_diss} and \ref{vector_potential_eqn_toy} subject to $\nabla\cdot\bmv = 0$. 

We consider a periodic domain of size $x\in[-L_x/2,L_x/2]$ and $z\in[-L_z/2,L_z/2]$. We decompose all fields into $N_x\times N_z$ Fourier modes and evolve them using a second-order Runge-Kutta time integration. A standard dealiasing factor of 3/2 is employed.
For simplicity, we set 
\begin{align}
\nu_1=\nu_2 = \chi = D = \frac{H_\mathrm{g}^2\Omega}{\re}, 
\end{align}
where the $\re$ is the Reynolds number. We initialize the simulations in a steady state and then add either an eigenmode perturbation or random perturbations. 

\subsection{Code test}

We first test our code implementation against linear theory. The linearized full and toy model equations with dissipation are given by Eqs. \ref{gas_based_incompress}---\ref{gas_based_linear_dg} and Eqs. \ref{toy_linear_mom_diss}---\ref{toy_linear_induction_diss} (subject to  $\nabla\cdot\delta\bmv = 0$), respectively. We consider a system with an equilibrium $\epsilon=0.2$, $\st=0.1$, $\etahat=0.1$; $\Ha=-0.3$, and $\beta_z=10^5$. The Reynolds number $\re = 2\times 10^5$. 
 Recall the imposed radial flow in the toy model is set to $\vext =  2\epsilon\st\etahat c_s/\Delta^2$; which is the dust-induced, equilibrium gas radial flow in the full model. 

\begin{figure*}
    \centering
\includegraphics[scale=0.5,clip=true,trim=0cm 0cm 3.2cm 0cm]{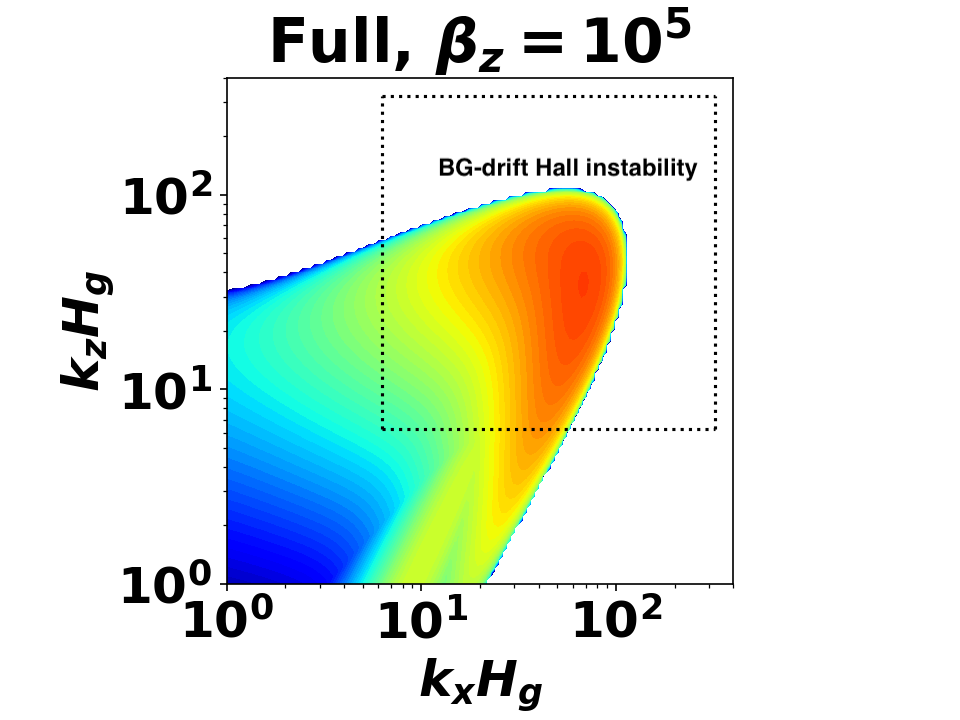}\includegraphics[scale=0.5,clip=true,trim=3.2cm 0cm 3.2cm 0cm]{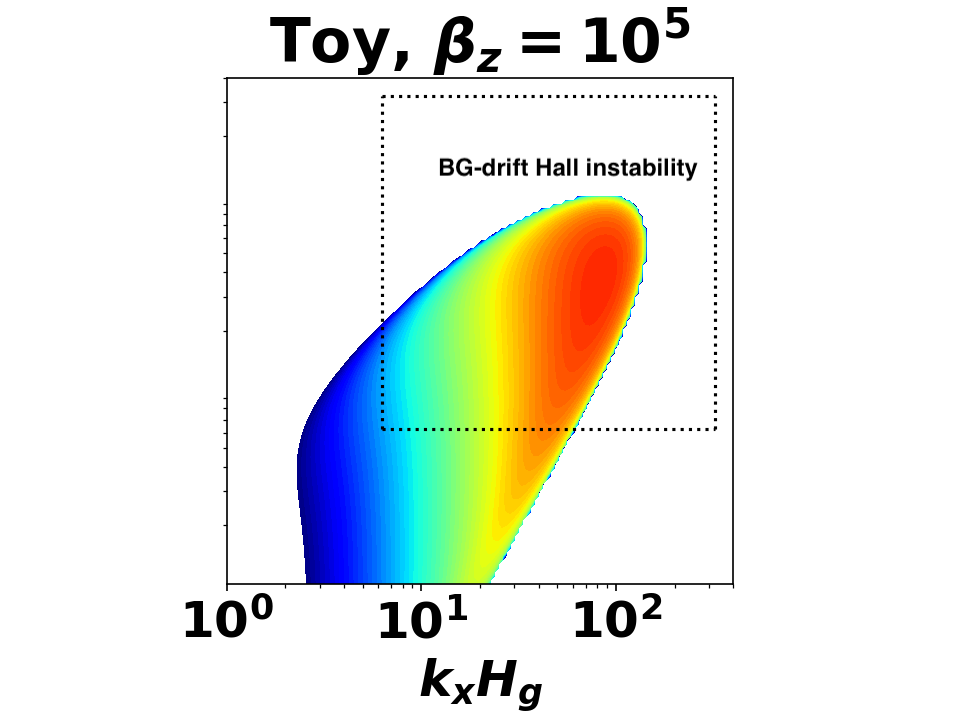}\includegraphics[scale=0.5,clip=true,trim=3.4cm 0cm 0cm 0cm]{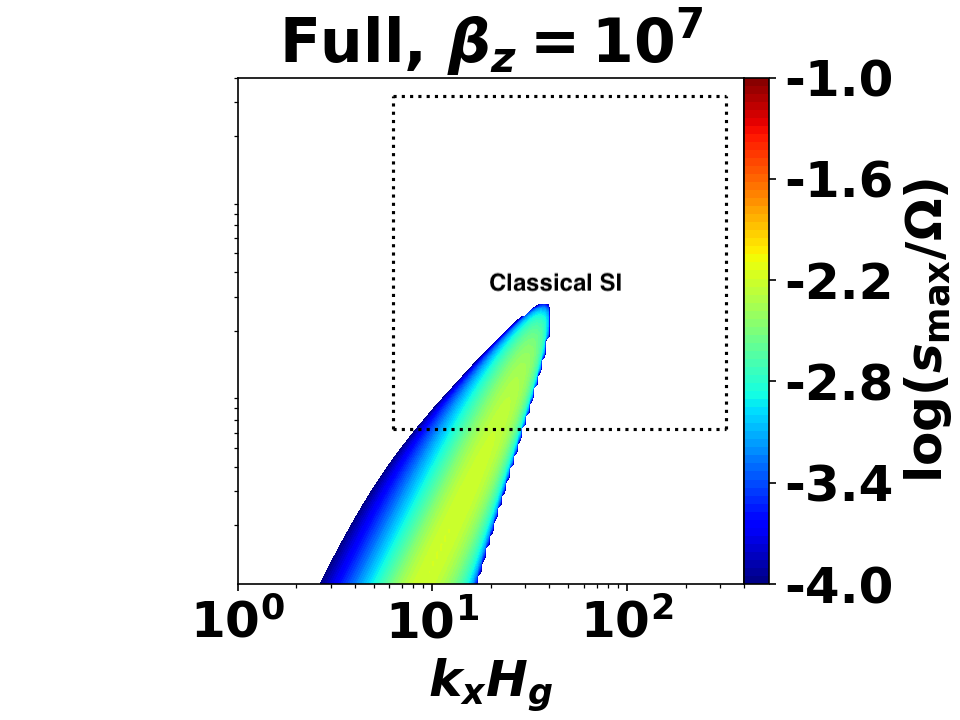}
    \caption{Growth rates as a function of wavenumber in a dusty, magnetized disk with $\beta_z=10^5$ (left) and a corresponding toy model based on a modified, pure gas Hall-MHD equations (middle). The right panel is a full model with $\beta_z=10^7$, for which MHD effects are negligible. 
    The dotted box corresponds to the resolved wavenumbers in the \textsc{dedalus} simulation presented in \S\ref{simulations}. }
    \label{hallSI_examples}
\end{figure*}

In Figure \ref{hallSI_examples}, we solve the linearized equations numerically and show growth rates as a function of $k_{x,z}$ for the full model (left panel) and the toy model (middle panel). For comparison, we also plot growth rates in a full model with $\beta_z=10^7$ in the right panel, for which the background drift Hall instability is negligible and only the classical SI remains. 

The toy model correctly isolates the background drift Hall instability from the corresponding full model, although the former slightly overestimates growth rates. The modes in the full model that extend to $K_x\sim 1$ and $K_z\sim 10$ are RDIs associated with Hall MHD. These are not captured by the toy model. The full model also captures the classical SI as a faint branch of modes below the BG-drift Hall instability. These classical SI modes survive in the $\beta_z=10^7$ case (right panel), wherein MHD effects are effectively absent.

For the \textsc{dedalus} runs, we consider the full and toy models with $\beta_z=10^5$. We add a clean eigenmode solution from the linearized problem. We choose $K_x=69.6$ and $K_z=34.3$, which is approximately the most unstable mode in the full model and corresponds to the background drift Hall instability. The perturbation amplitude is normalized such that $|\delta v_y| = 10^{-4}C_s$. We set the simulation domain to one wavelength in each direction, i.e. $L_{x,z} = \lambda_{x,z}\equiv 2\pi/k_{x,z}$, and use a resolution of $N_x\times N_z=64\times 64$. We note that the Reynolds number associated with the box size $\operatorname{Re}_\mathrm{box}\equiv \left(4\pi^2/K^2\right)\re \simeq 1300$. 

\begin{figure}
    \centering
    \includegraphics[width=\linewidth]{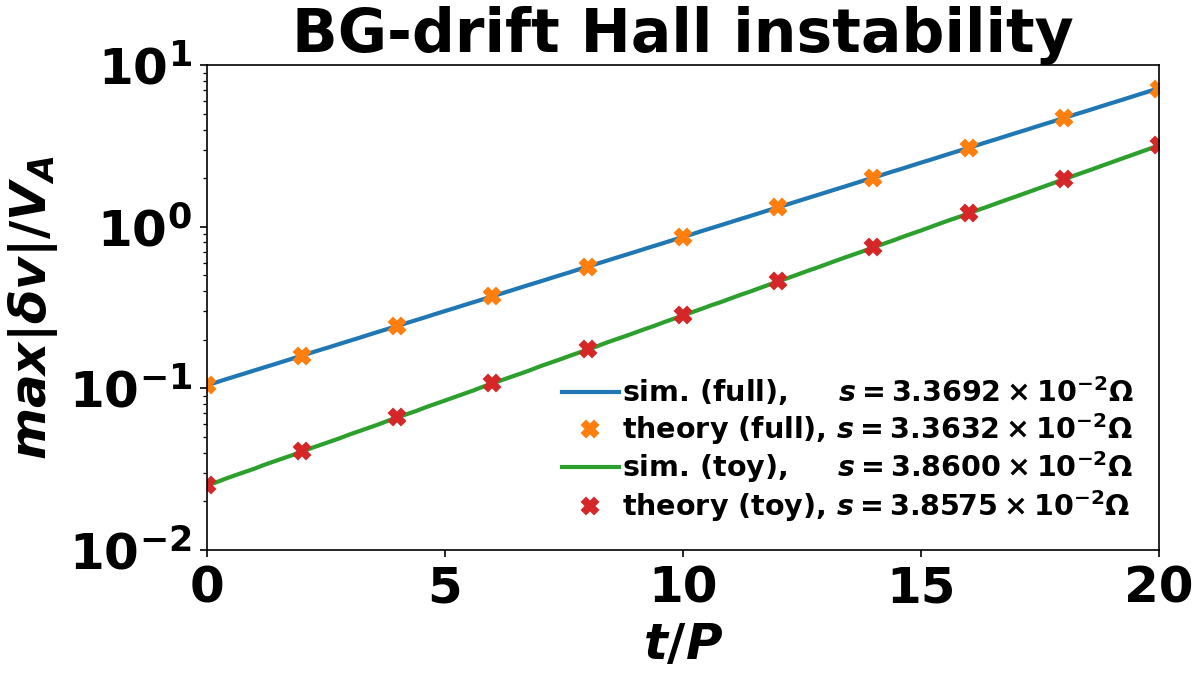}
    \caption{\textsc{dedalus} simulations of the background-drift Hall instability based on the full, single-fluid model (blue solid) and the toy model (green solid). The simulations are initialized with a pure eigenmode with theoretical growth curves shown as crosses for the toy model (orange) and full model (red).}
    \label{hallSI_toy_sim_growth}
\end{figure}

Figure \ref{hallSI_toy_sim_growth} shows the evolution of the maximum velocity perturbation normalized by the Alfv\'{e}n speed $V_\mathrm{A}$. We also plot  in asterisks the expected evolution based on theoretical growth rates and find an excellent match with the simulation in the linear regime.  Growth rates are also shown in the legend, which demonstrate a relative error of $O(10^{-3})$.

\subsection{Nonlinear evolution} \label{dedalus_random}

We now present simulations of the three models shown in Figure \ref{hallSI_examples}. Here, we set the box size to $L_x=L_z=H_\mathrm{g}$ and use a resolution of  $N_x\times N_z = 512 \times 512$. The range of relevant wavenumbers is delineated by the dotted boxes in Figure \ref{hallSI_toy_sim_growth}, which is obtained by assuming the maximum allowed wavelength is the box size and that a minimum of 10 collocation points are needed to resolve a wavelength. We destabilize the system by adding random perturbations to $v_y$ with a maximum amplitude of $10^{-4}C_s$.

\begin{figure}
    \centering
    \includegraphics[width=\linewidth]{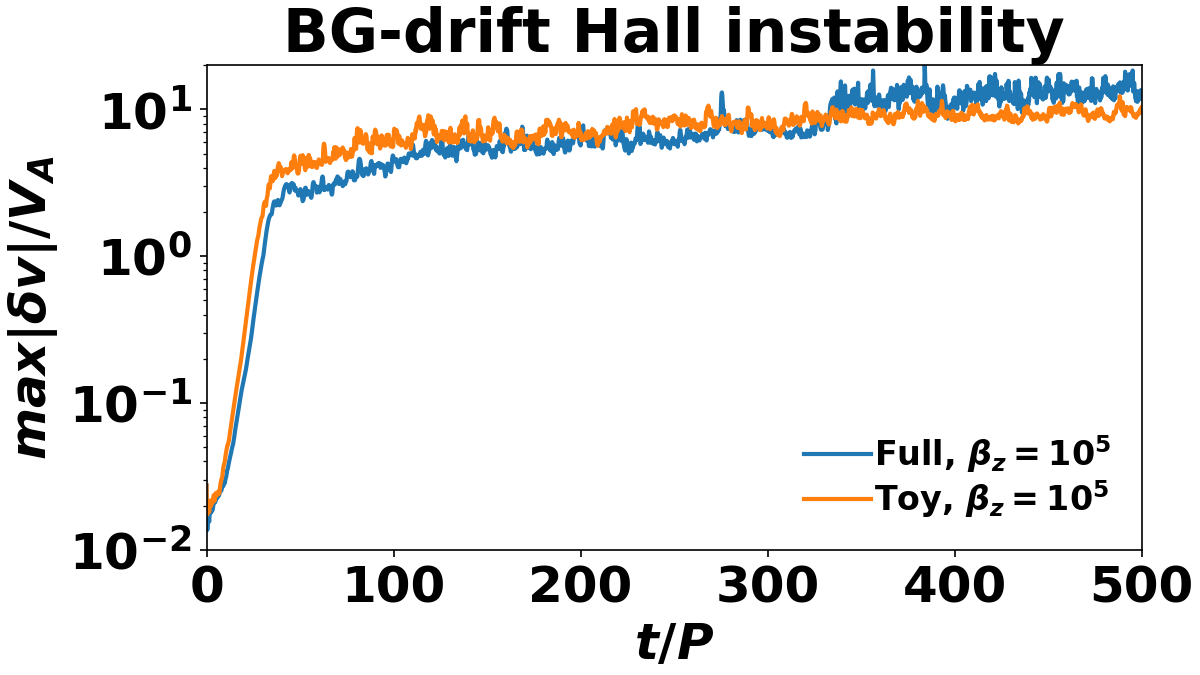}
    \caption{Evolution of the vertical velocity, normalized by the Alf\'{e}n speed, in \textsc{dedalus} simulations of the full model (blue) and the toy model (orange).
    }
    \label{hallSI_vg_evol}
\end{figure}

\begin{figure}
    \centering
    \includegraphics[scale=0.3,clip=true, trim=0cm 0cm 1cm 0cm]{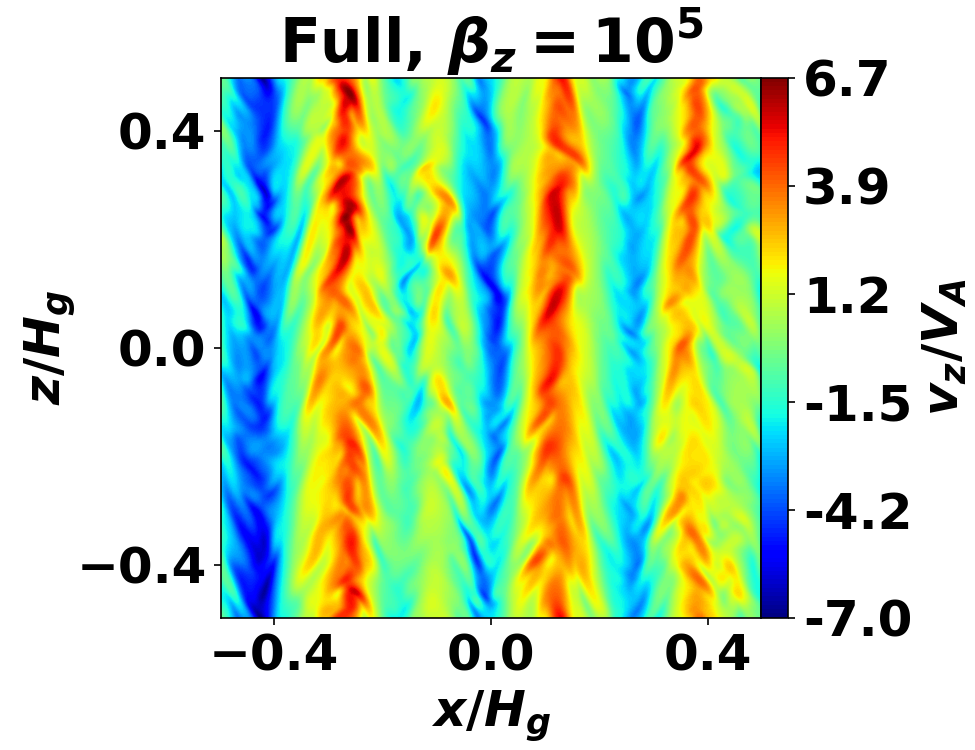}\includegraphics[scale=0.3,clip=true,trim=3.5cm 0cm 0cm 0cm]{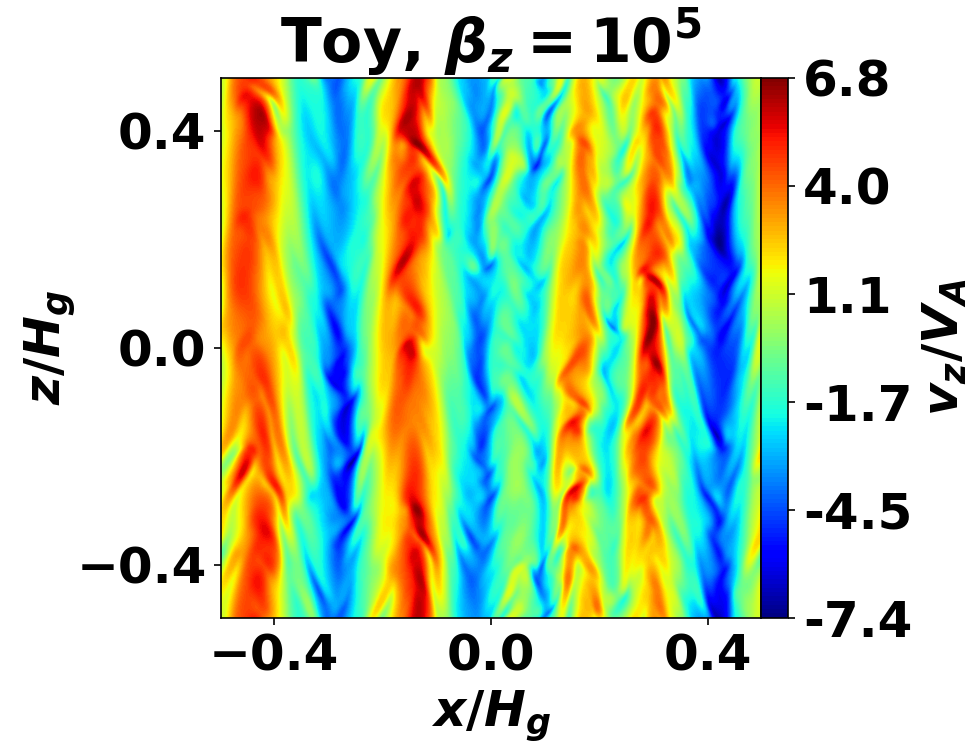}
    \caption{ Vertical gas velocities in the full model (left) and the toy model (right) at $t=300P$ for the simulations shown in Figure \protect\ref{hallSI_vg_evol}.}
    \label{hallSI_vgz_2D}
\end{figure}

In Figure \ref{hallSI_vg_evol}, we compare the evolution of the vertical velocity between the full model and the toy model, both with $\beta_z=10^5$. Both models yield exponential growth with growth rates of $3.15\times 10^{-2}\Omega$, which is somewhat smaller than the theoretical maximum growth rates of $3.36\times 10^{-2}\Omega$ ($4.10\times10^{-2}\Omega$) in the full (toy) models. This is likely due to the reduced resolution compared to the code tests. The models saturate at similar amplitudes with $\left|v_z\right|\sim 6V_\mathrm{A}$. Figure \ref{hallSI_vgz_2D} shows that both runs exhibit large-scale upwelling and downwellings, with small-scale substructures. The similarity between the two models indicates that the background drift Hall instability dominates the system up to the initial saturation.

However, the full model undergoes a secondary, albeit minor, growth around $\sim 330P$, where $\left|v_z\right|$ reaches $\sim 10V_\mathrm{A}$. We find this is associated with the development of small-scale, enhanced dust-density filaments surrounding low dust-density voids (see below). We suspect this is related to classical streaming-type instabilities operating. 

\begin{figure}
    \centering
    \includegraphics[width=\linewidth]{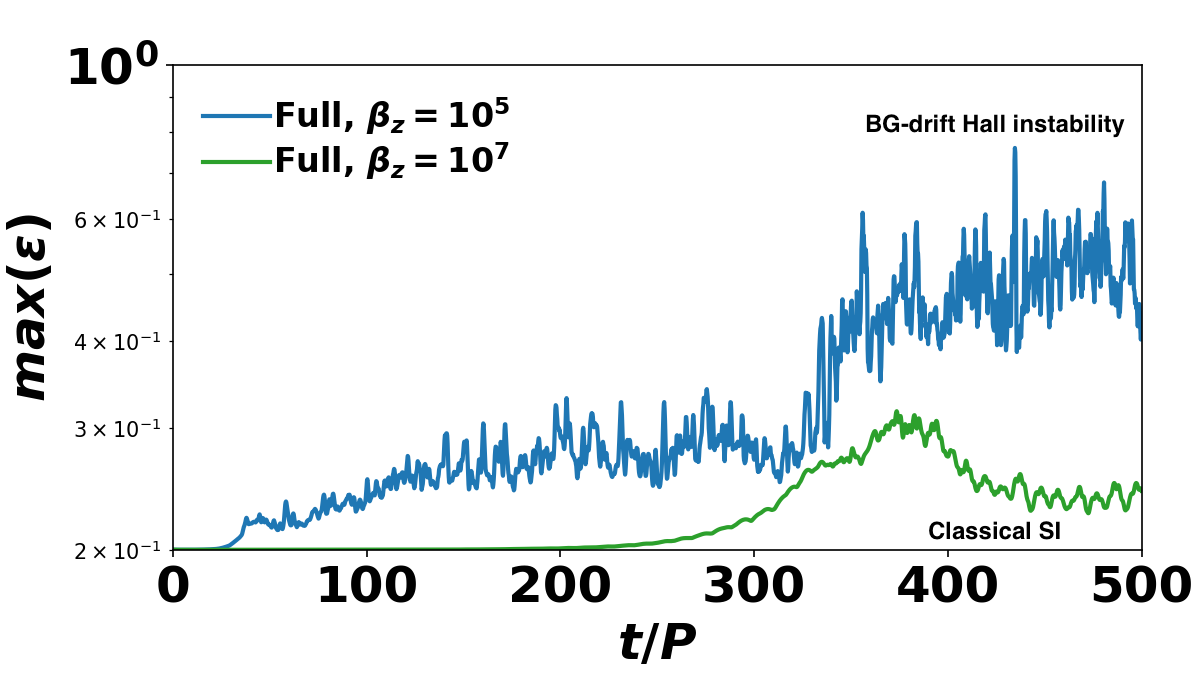}
    \caption{Evolution of the maximum dust-to-gas ratio in \textsc{dedalus} simulations of the full model with $\beta_z=10^5$ (blue; dominated by the background drift Hall instability) and $\beta_z=10^7$ (green; only the classical SI develops). 
    }
    \label{hallSI_epsilon_evol}
\end{figure}

\begin{figure*}
    \centering
    \includegraphics[scale=0.6,clip=true, trim=0cm 0cm 1cm 0cm]{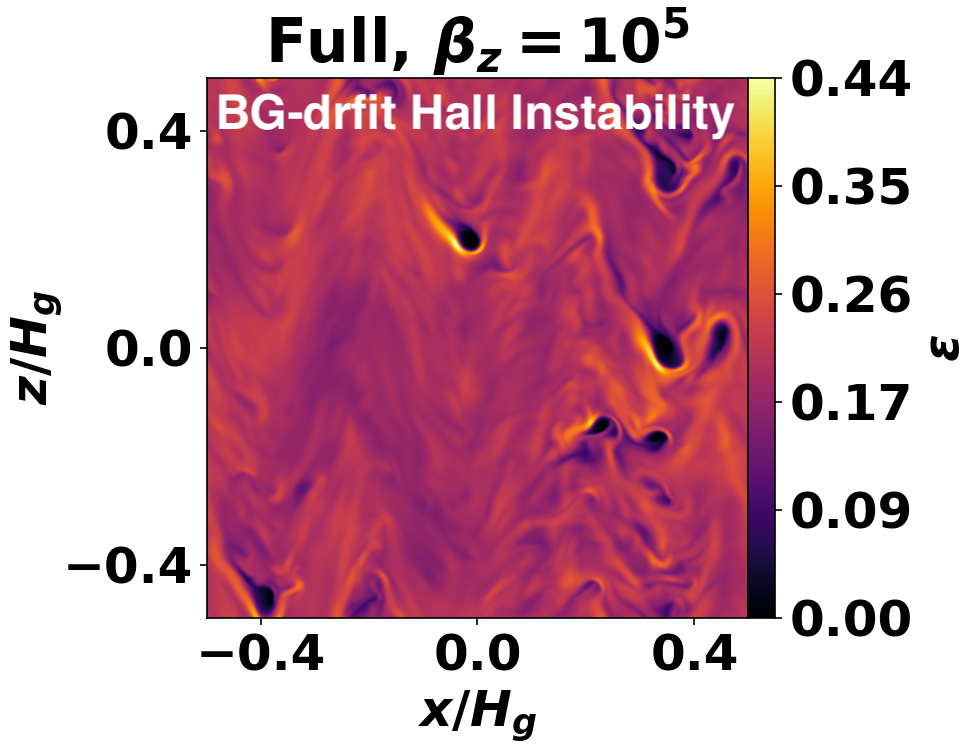}\includegraphics[scale=0.6,clip=true,trim=3.25cm 0cm 0cm 0cm]{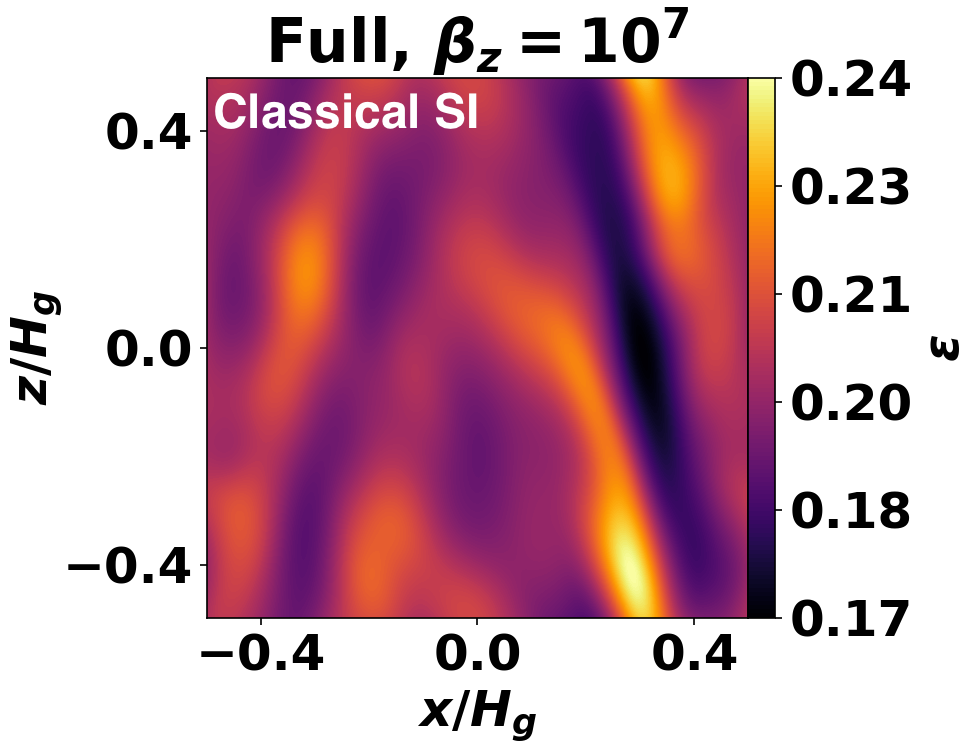}
    \caption{Dust-to-gas ratios at the end of the \textsc{dedalus} runs of the full model with $\beta_z=10^5$ (left) and $\beta_z=10^7$ (right).} 
    \label{hallSI_epsilon_2D}
\end{figure*}

We next compare the full models with $\beta_z = 10^5$ and $\beta_z=10^7$. Figure \ref{hallSI_epsilon_evol} shows the evolution of the maximum dust-to-gas ratio in the box and Figure \ref{hallSI_epsilon_2D} shows $\epsilon$ at the end of the simulations. For $\beta_z=10^7$, where only the classic SI operates, we find the development of weak dust clumps around $t\sim370P$ associated with the break up of initial SI filaments. However, these dust enhancements are not sustained and the system decays to a quasi-steady state with box-scale filaments and a negligible increase in $\epsilon$. This case proceeds similarly to the \cite{johansen+07}'s `AA' run of the SI at low $\epsilon$, which also exhibits transient growth. 

On the other hand, the $\beta_z=10^5$ case evolves quite differently. The system first saturates to a state with $\mathrm{max}(\epsilon)\sim 0.3$, which is sustained for $\sim 100$ orbits. Since the toy model behaves similarly to the full model up to this point, it suggests that dust concentration is passively induced by the background drift Hall instability. However, unlike the $\beta_z=10^7$ case, dust then undergoes a secondary concentration and the system sustains $\mathrm{max}(\epsilon)\sim 0.5$, or about twice the initial value, until the end of the simulation. The left panel of Figure \ref{hallSI_epsilon_2D} shows that the dust enhancements develop around small dust `voids'. This again differs from the classical SI, where dust typically concentrates into filaments instead. 



\section{Discussion} \label{sec:discuss}

We find the classical SI dominates dust-rich systems with $\epsilon\gtrsim 1$. Similarly, for $\Ha > 0$, the MRI always dominates. We generally find Hall RDIs, for which Hall MHD and dust-gas interaction act in concert, to be negligible. The only exception is when $\Ha$ is large and negative, in which case the ion-cyclotron RDI co-dominates the system with the MRI, and they are distinctly associated with high and low wavenumber, respectively. \cite{Squire-Hopkins-2018b} also suggested that MHD RDIs are likely unimportant, at least in the disk midplane, because ohmic resistivity and ambipolar diffusion are both expected to dampen MHD waves (but see below). 

On the other hand, the newly-discovered BDHI can dominate systems with weak fields ($\beta\gtrsim 10^5$), pebble-sized grains with $10^{-2}\lesssim \st\lesssim 0.3$, and low dust abundance ($\epsilon\lesssim 1$). The latter may be relevant to disk regions with particle stirring, as dust diffusion limits the midplane dust-to-gas ratio. Furthermore, dissipation, neglected in our main linear calculations, can suppress high-wavenumber classical SI and MRI modes, which could extend the relevance of the BDHI beyond the above regimes. 

Consider a stratified disk with a vertically integrated dust-to-gas ratio $Z = \epsilon \Hd/\Hg \simeq \epsilon \sqrt{\delta/\st}$, where $\Hd$ is the particle scale-height and $\delta= D/C_s\Hg$ is the dimensionless diffusion coefficient \citep{lin21}. For the BDHI-dominated disk presented in \S\ref{simulations} with $\st = 0.1$ and $\delta =2\times 10^{-5}$ (Figure \ref{hallSI_examples}, left panel), we find $Z\simeq 0.003$, which is smaller than the canonical value of 0.01. This suggests that the BDHI could facilitate planetesimal formation in dust-depleted regions of PPDs.

\subsection{Application to protoplanetary disks}

The new BDHI is found under a couple of conditions. First of all, the background magnetic field has a dominant vertical component. Secondly, the magnetic flux is anti-parallel to the disk rotation spin ($\Ha<0$). Third, the Hall effect needs to be of moderate strength (small $|\Ha|$). We discuss the relevance of such conditions in the physical context of PPD formation under the effects of non-ideal MHD.

Several numerical and theoretical works \citep{lee21, zhao21, tsukamoto15} discuss scenarios of non-ideal MHD disk formation under conditions where rotation and magnetic field are either parallel or anti-parallel. The actual behavior and the strength of the Hall effect depend on many factors such as the cosmic ionization rate, dust grain size distribution, gas chemical composition, and temperature \citep{marchand16}. At PPD densities, it is reasonable to assume that the Hall effect is dominated by the electrons, giving a positive Hall coefficient $\eta_{\rm H}$. If the Hall effect is strong enough to have a dynamic effect on the formation process of the disk, then the spin-up of the disk is induced by the Hall drift regardless of the initial rotation of the disk-forming core. The resulting configuration is always anti-parallel, yielding a negative $\Ha$ independently of the initial conditions. Within the disk, there is a background radially outward drift of magnetic flux that is the result of the synergy between the Hall effect and the ambipolar diffusion, guaranteeing that $B_z$ will grow less rapidly with respect to the disk mass \citep{zhao21}. This also means that $\beta_z$ is reasonably high and will only increase with time. Typical values of $\beta_z \gtrsim 10^4$ are easily reached.

Due to the vertical differential rotation, a strong azimuthal component of the magnetic field is generated through magnetic induction. As a consequence, the presumption of a magnetic configuration dominated by the vertical field should be valid only at the very early phase of PPD formation \citep[Strong-B case in][which does not last for long]{lee21}, or not too far away from the midplane, which is more physically relevant. Given that the BDHI proposed in this work happens at $k_zH_\mathrm{g}$ typically larger than order unity for $\beta_z = 10^4$, the development is limited to the midplane where the vertical field is dominant due to symmetry. As the PPD evolves and $\beta_z$ increases, the toroidal field is more developed and the region dominated by the vertical field becomes even narrower around the midplane. Meanwhile, the BDI develops at large wavenumbers and thus can still be a relevant mechanism for instability growth.

For a typical Minimum Mass Solar Nebula (MMSN) surface density profile \citep{hayashi81} and a temperature of a few hundred kelvin around a sub-Solar mass protostar, $\St = 0.1$ corresponds roughly to millimeter-to-centimeter-sized grains within a few tens of AU. Recent observations suggest that dust grains may grow through coagulation up to millimeter sizes in such region \citep{liu21}. According to the giant planet instability models \citep[e.g.][]{desousa20}, this is where the giant planets were believed to have formed. The BDHI proposed in this study might be a viable channel to help the dust grains cross the centimeter barrier in such a regime. 

Furthermore, the BDHI is weak when the dust-to-gas mass ratio $\epsilon$ is close to that of the typical ISM (0.01). The value $\epsilon=0.2$, with which the BDHI growth rate dominates, is mostly relevant for an early disk with moderate dust settling toward the midplane. As new evidence suggests that giant planets might have formed early and close-in \citep[e.g.][]{manara18,morbidelli20}, the newly discovered BDHI could effectively be a plausible mechanism for early planetesimal formation during the class 0/I phases of PPD evolution. 

\subsection{Caveats and outlook}

Our calculations are based on unstratified, local shearing box models with a single species of dust grains. The gas is magnetized but for the most part it is only subject to the Hall effect. How our results carry over to a more realistic setup should be addressed in future work. 

In a global disk, the Hall effect can drive laminar gas accretion flows in the midplane \citep[e.g.][]{bai17}.  This can lead to an `azimuthal drift SI' (AdSI), even in the absence of a radial pressure gradient \citep{Lin-Hsu-2022,Hsu_Lin_2023}, which may facilitate planetesimal formation. We neglected this effect but it may be relevant to Hall-effected disks as these can naturally develop zonal flows that trap dust near pressure maxima \citep{bethune+17,krapp18}. Future work should incorporate such a background accretion flow, for example by applying an appropriate torque onto the gas \citep{mcnally17} and examining how it interacts with the BDHI. 

Recent generalizations of the classical SI to a grain size distribution have shown that when the total dust-to-gas ratio is less than unity, the instability can be significantly weakened compared to the single-size approximation adopted here \citep{Krapp2019,Paardekooper2020,Zhu_Yang_2021}. Since this is the regime where the BDHI is the most relevant, it will also be necessary to generalize the BDHI to polydisperse dust grains. 

 In a realistically stratified disk, the dust-rich midplane has nearly Keplerian rotation while the dust-poor, pressure-supported layers above and below rotate at a sub-Keplerian speed. Such a vertical gradient in the gas and dust rotation velocities, or vertical shear, can drive Kelvin-Helmholtz instabilities \citep{chiang08,barranco09,lee10} and `vertically shearing' SIs \citep{ishitsu09,lin21}, which tend to disrupt the dust layer. Future work will also need to clarify how the BDHI operates in a stratified disk. 

Finally, we have neglected ambipolar diffusion (AD), which is applicable to low gas density regions \citep{Lesur-review}. Without an equilibrium azimuthal field, as considered here, AD behaves similarly to ohmic resistivity \citep{Latter-Kunz2022}. \cite{Lin-Hsu-2022} did not find new phenomena when considering resistive MHD with dust; we therefore do not expect new instabilities when combining AD and dust dynamics in the case. However, in the presence of an equilibrium azimuthal field, AD can destabilize oblique modes with $k_x\neq0$ \citep{kunz04}. It remains to be seen how these AD shear instabilities are affected by dust; and whether or not AD and the background gas drift yield an analogous instability to the BDHI.  

\section{Summary}\label{sec:8}
In this paper, we examine the interplay between non-ideal magnetohydrodynamics (MHD) and dust dynamics in PPDs. We focus on the Hall effect and dust-gas instabilities such as the streaming instability (SI) since they are directly relevant to planet-forming regions of PPDs.  We consider vertical fields anti-parallel to the disk's rotation since otherwise, MHD instabilities dominate the system. 

Our linear analyses confirm, in line with theoretical expectations, variations of the SI that are attributable to the Hall effect. These are associated with resonances between the equilibrium dust-gas relative drift and the phase velocity of whistler or ion-cyclotron waves in the magnetized gas. While these Hall-mediated `resonant drag instabilities' (RDIs) do exist, they have smaller growth rates and span a narrower range of wavenumbers compared to the classical hydrodynamic SI or pure MHD instabilities. This implies that these Hall RDIs will unlikely replace or override the classical SI or MRI. 

\begin{figure}
\centering
\includegraphics[width=1\hsize]{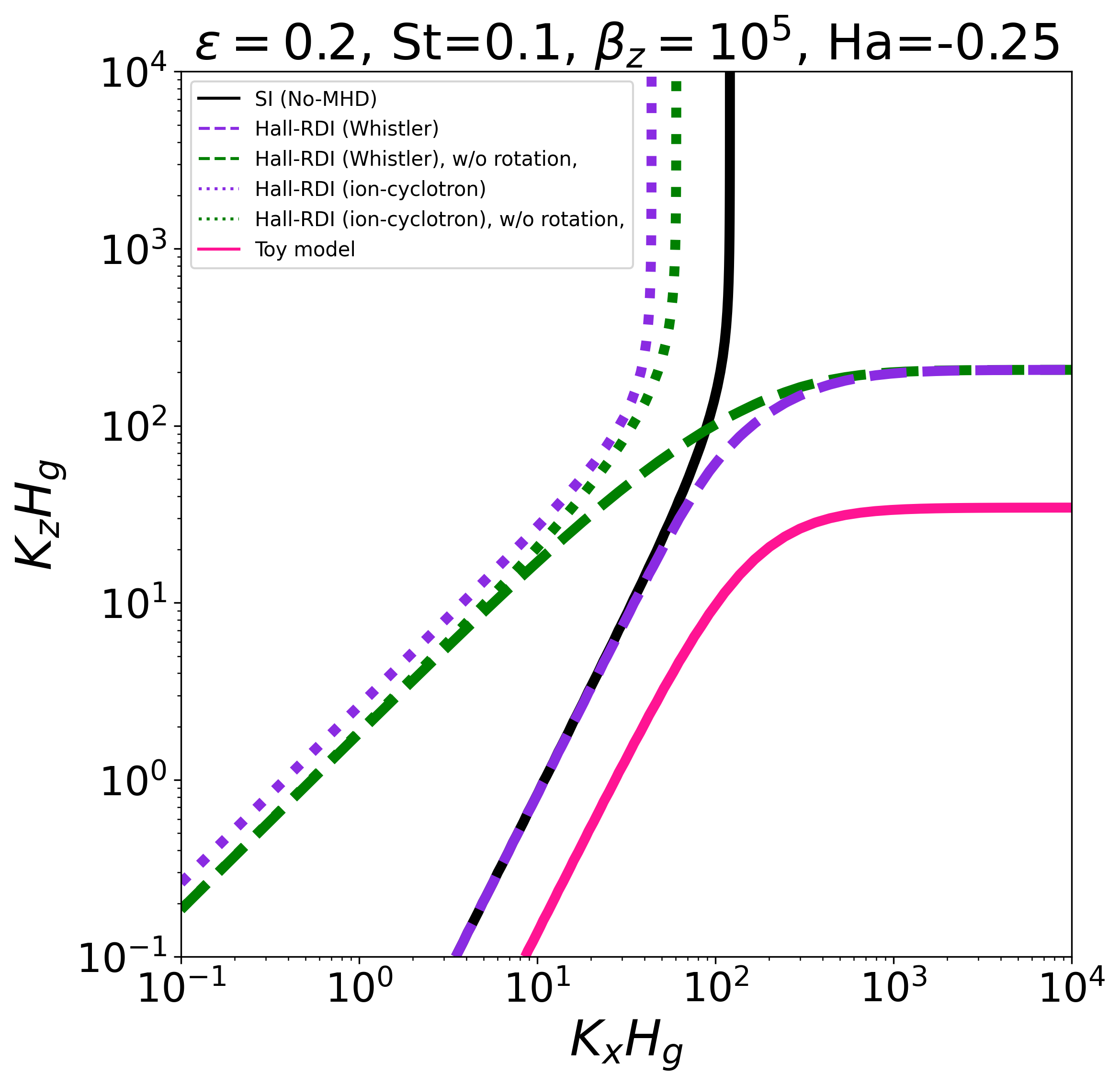}
\caption{Summary of unstable modes for model ``e02s01b5h-025", relevant parameters are marked on the title. The black solid line is the classical SI, the violet dashed line and dotted line are the two resolutions of Hall-RDI,
these there instabilities as shown in Figures \ref{fig:1} and \ref{BDHI_full_novx}. The pink solid line is BDHI, as shown in Figure \ref{fig:full vs toy}. The green dashed line and dotted line are the two resolutions of Hall-RDI without Keplerian rotation.}\label{summary}
\end{figure}

However, we also discover a novel instability attributable to the interaction of the Hall effect with the dust-induced equilibrium gas flow. While unrelated to the SI or RDIs, this new `Background Drift Hall Instability' (BDHI) can dominate in low dust-to-gas ratio environments with weak magnetic fields. We reproduce the BDHI in a single-fluid toy model and briefly explore its nonlinear evolution using spectral simulations. We summarize all the instability modes involved in this study in Figure \ref{summary}. As demonstrated in Figure \ref{summary}, our analyses indicate that in anti-parallel Hall-dusty disks, the composition of instabilities arises from the classical SI, Hall-RDI (whistler and ion-cyclotron), and the newly discovered BDHI.

In a dust-poor disk, we find the BDHI drives much more significant dust enhancements than the classical SI. High-resolution simulations with larger Reynolds numbers than that considered in this work ($\re\sim 10^5$) will be required to address whether or not the BDHI can drive dust densities to the point of gravitational collapse. If so, the BDHI would lower the super-solar metallicity threshold for planetesimal formation as based on the classical SI. 

\section*{Aknowledgement}
This work was initiated at the 2022 ASIAA Summer Student Program. We thank Xue-Ning Bai and Zhaohuan Zhu for inspiring discussions and comments. YW gratefully acknowledges the support of the DUSTBUSTERS RISE project (grant agreement number 823823) for his secondment at the University of Arizona. YW further thanks ASIAA for his two-week visit to Taipei. M-KL is supported by the National Science and Technology Council (grants 111-2112-M-001-062-, 112-2112-M-001-064-, 111-2124-M-002-013-, 112-2124-M-002 -003-) and an Academia Sinica Career Development Award (AS-CDA110-M06). CC acknowledges funding from STFC grant ST/T00049X/1. Y-NL acknowledges the support of the National Science and Technology Council (111-2636-M-003-002) and the Ministry of Education Yushan Young Scholar Fellowship. 

\section*{Data availability}
The data obtained in our simulations can be made available on reasonable request to the corresponding author.

\clearpage
\appendix




\section{Single-fluid, gas-based model}\label{full_single_fluid}
We describe a simplified, single-fluid model of the dusty, magnetized gas. We take the gas to be incompressible with constant density $\rhog$. The gas equations are
\begin{align}
\nabla\cdot\bmv &=\nabla\cdot\bmB = 0, \label{gas_based_gas} \\
\frac{\partial\textbf{\emph{v}}}{\partial t} + \textbf{\emph{v}}\cdot\nabla\textbf{\emph{v}} &= 
    2v_{y}\Omega\hat{\textbf{\emph{x}}} - v_{x}\frac{\Omega}{2}\hat{\textbf{\emph{y}}} - \frac{1}{\rho_{g}}\nabla \Pi + 2\eta r \Omega^2 \xhat
    +\frac{1}{\mu_0\rhog}\bmB\cdot\nabla\bmB + \frac{\epsilon\bmu}{\taus} + \nu_1 \nabla^2\bmv \label{gas_based_vg},\\
      \frac{\partial\textbf{\emph{B}}}{\partial t} & = \bmB\cdot\nabla\bmv - \bmv\cdot\nabla\bmB
      -\frac{3}{2}\Omega B_x\yhat  - \frac{\eta_H}{B_{z0}}\nabla\times\left[\left(\nabla\times\bmB\right)\times\bmB\right] + \chi\nabla^2\bmB ,\label{gas_based_induction}
\end{align}
where
\begin{align}
\Pi \equiv p + \frac{|\bmB|^2}{2\mu_0}
\end{align}
is the total pressure, and recall $\bmu \equiv \bmw - \bmv$ 
is the relative velocity between dust and gas.

In the induction equation (\ref{gas_based_induction}), we simplified the Hall term by normalizing it with the constant vertical field $B_{z0}$, rather than the evolving  magnetic field strength $|\bmB|$ that appears through $\widehat{\bmB}$ 
in the proper treatment  (i.e. Eq. \ref{shear-eq-B}). This modification renders the Hall term a quadratic nonlinearity, which was found to make our spectral simulations more robust. However, it has no effect on linear stability. 
Furthermore, we have included the gas viscosity and magnetic resistivity terms $\propto \nu_1, \chi$ to regularize the system on small scales, which are needed for numerical stability in nonlinear simulations.

\subsection{Relative drift equation}

The evolutionary equation for $\bmu$ is obtained from subtracting the full equations for $\bmw$ and $\bmv$ without viscosity (Eqs. \ref{shear-eq-velocity-dust} and \ref{gas_based_vg} with $\nu_1=0$, respectively). We then add a viscous term, $\nu_2\nabla^2\bmu$, where $\nu_2$ is a second viscosity coefficient, again for numerical stability. We obtain: 
\begin{align}
  \frac{\p \bmu}{\p t} + \left(\bmv\cdot\nabla\right)\bmu + \left(\bmu\cdot\nabla\right)\bmv + \left(\bmu\cdot\nabla\right)\bmu 
  = 2\Omega u_y \xhat - \frac{\Omega}{2}u_x\yhat 
 + \frac{\nabla \Pi}{\rhog} - 2\eta r \Omega^2\xhat - \frac{1}{\mu_0\rhog}\bmB\cdot\nabla\bmB - \frac{(1+\epsilon)} {\taus}\bmu + \nu_2 \nabla ^2 \bmu.\label{gas_based_Dv}
\end{align} 
For tightly coupled grains with small $\taus$, it is possible to solve this equation approximately and obtain an explicit expression for $\bmu$. We describe this procedure in Appendix \ref{TVA}.  

\subsection{Dust-to-gas ratio equation}\label{single_fluid_dg}
For an incompressible gas, the dust mass equation (\ref{shear-eq-density-dust}) can be written as 
\begin{align}
\frac{\p \epsilon}{\p t} + \bmv\cdot\nabla\epsilon = - \nabla\cdot\left(\epsilon\bmu\right) + D\nabla^2\epsilon, \label{gas_based_dg}
\end{align} 
where the dust diffusion term $\propto D$ is again added for numerical stability. An alternative form of this dust-to-gas ratio equation can be obtained by taking the divergence of the gas momentum equation (Eq. \ref{gas_based_vg}), utilizing incompressibility, and eliminating the relative velocity. We find 
\begin{align}
    \frac{\p \epsilon}{\p t} + \bmv\cdot\nabla\epsilon = \taus\nabla\cdot\left(2\Omega v_y\xhat\right)
    -\taus\nabla\cdot\left(\bmv\cdot\nabla\bmv\right) - \frac{\taus}{\rhog}\nabla^2\Pi + \frac{\taus}{\mu_0\rhog}\nabla\cdot\left(\bmB\cdot\nabla\bmB\right) + D \nabla^2 \epsilon,\label{gas_based_dg_alt}
\end{align}
where we have assumed axisymmetry throughout. We use Eq. \ref{gas_based_dg_alt} for  stability analyses as its linearized form is particularly simple (see Appendix \S\ref{single_fluid_linear}).

\subsection{Equilibrium drift}

The equilibrium dust-gas drift is:
\begin{align}
    u_x & = -\frac{2(1+\epsilon)\st}{\Delta^2}\st,\label{eqm_ux}\\
    u_y &= \frac{\st^2}{\Delta^2}\eta r \Omega.
\end{align}
Recall $\Delta^2 = \st^2 + (1+\epsilon)^2$ and the equilibrium gas velocities are given by Eqs. \ref{eqm_vx}---\ref{eqm_vy}.

\subsection{Linearized single-fluid equations}\label{single_fluid_linear}
Linearizing the single-fluid model (Eqs. \ref{gas_based_gas}---\ref{gas_based_induction}, \ref{gas_based_Dv}, \ref{gas_based_dg_alt}) in the same manner as in the main text, we obtain: 
\begin{align}
    0 =& \ikx \delta v_x + \ikz \dd v_z,\label{gas_based_incompress}\\
    \sigma\dd\bmv =& - \ikx v_x \dd \bmv +  2\Omega\dd v_y\xhat - \frac{\Omega}{2}\dd v_x \yhat - \left(\ik\frac{\dd\Pi}{\rhog} - \frac{\ikz B_z}{\mu_0\rhog}\dd\bmB\right) + \frac{1}{\taus}\left(\epsilon\dd\bmu + \bmu \dd\epsilon \right) - \nu_1 k^2\dd \bmv,\label{gas_based_linear_vg} \\
    \sigma \dd \bmu = & -\ikx (v_x + u_x) \dd \bmu - \ikx u_x \dd\bmv  + 2\Omega \dd u_y\xhat - \frac{\Omega}{2}\dd u_x\yhat + \ik\frac{\dd\Pi}{\rhog} - \frac{\ikz B_z}{\mu_0\rhog}\dd\bmB - \frac{1}{\taus}\left[(1+\epsilon)\dd\bmu + \bmu\dd\epsilon\right] - \nu_2 k^2\dd\bmu,\label{gas_based_linear_u} \\ 
    \sigma \dd\bmB =& \ikz B_z\dd\bmv - \ikx v_x \dd \bmB - \frac{3}{2}\Omega\dd B_x\yhat -\etahall k_z^2\dd B_y\xhat + \etahall k^2 \dd B_x \yhat +\etahall k_x k_z \dd B_y\zhat - \chi k^2\dd\bmB,\label{gas_based_linear_induction}\\
    \sigma \dd \epsilon =& 2\ikx \Omega\taus\dd v_y + \taus k^2\frac{\dd\Pi}{\rhog} - \ikx v_x \dd\epsilon - Dk^2\dd\epsilon.\label{gas_based_linear_dg}
\end{align}

\subsection{Implementation in \textsc{dedalus}}\label{full_model_dedalus}

In our \textsc{dedalus} simulations (\S\ref{simulations}), we solve for the gas velocity $\widetilde{\bmv}$ relative to the dust-free, Keplerian shear flow offset by the pressure gradient,
\begin{align}
    \widetilde{\bmv} \equiv \bmv + \eta r \Omega\yhat.
\end{align}
For axisymmetric flow, we can simply replace $\bmv$ by $\widetilde{\bmv}$ in the governing equations and remove the background pressure gradient term $\propto \eta$ in the momentum equation (\ref{gas_based_vg}). 
The equilibrium gas azimuthal velocity is
\begin{align}
    \widetilde{v}_y = \frac{\epsilon(1+\epsilon)}{\Delta^2}\eta r \Omega.
\end{align}


\subsubsection{Induction equation}

To ensure $\nabla\cdot\bmB =0$ numerically, we solve for the magnetic field deviation $\Delta\bmB = \bmB - B_{z0}\zhat$ from the equilibrium vertical field by defining the vector potential $\bm{A}$ such that 
\begin{align}
    \Delta\bmB = \nabla\times\bm{A},
\end{align}
and work in the Coulomb gauge 
\begin{align}
\nabla\cdot\bmA = 0. \label{coulomb_gauge}
\end{align}
The induction equation (\ref{gas_based_induction}) becomes   
\begin{align}
      \frac{\partial\bmA}{\partial t} =& 
      \widetilde{\bmv}\times\bmB - \left(\widetilde{\bmv}\times\bmB\right)_0 
      +\frac{3}{2}\Omega A_y\xhat - \frac{\eta_H}{B_{z0}}\left(\nabla\times\bmB\right)\times\bmB + \chi\nabla^2\bmA +  \nabla\phi,\label{vector_potential_eqn_full}
\end{align}
where the scalar potential $\phi$ enforces the gauge condition \citep[Eq. \ref{coulomb_gauge}, e.g.][]{lecoanet19}. The second term on the right-hand side is the initial value of $\widetilde{\bmv}\times\bmB$. It is added for convenience so that $\bmA=0$ in equilibrium. Its inclusion is not necessary, however, since it only leads to a uniform growth in $\bmA$ over the domain, which does not generate any magnetic fields.  

\subsubsection{Postitive-definite dust-to-gas ratio equation}

To ensure that the dust-to-gas ratio $\epsilon>0$ in the simulations, we instead solve for 
\begin{align}
    Q_\epsilon \equiv \epsilon_0 \ln{\left(\frac{\epsilon}{\epsilon_0}\right)},
\end{align}
where $\epsilon_0$ is the initial dust-to-gas ratio. Then $\epsilon = \epsilon_0\exp{\left(Q_\epsilon/\epsilon_0\right)}>0$ and thus $Q_\epsilon=0$ initially. The equation for $Q_\epsilon$ is 
\begin{align}
    \frac{\p Q_\epsilon}{\p t} = -\left(\widetilde{\bmv} + \bmu\right)\cdot\nabla Q_\epsilon - \epsilon_0\nabla\cdot\bmu + D\left(\nabla^2 Q_\epsilon + \frac{\left|\nabla Q_\epsilon\right|^2}{\epsilon_0}\right).\label{dg_Q_form}
\end{align}
The term $\propto |\nabla Q_\epsilon|^2$ arises from expressing the diffusion term $\propto \nabla^2\epsilon$ (Eq. \ref{gas_based_dg}) in terms of $Q_\epsilon$.

\section{Magnetized terminal velocity approximation}\label{TVA}

Our aim here is to find an approximate expression for the relative velocity, $\bmu$, from Eq. \ref{gas_based_Dv}. Here, we neglect the viscosity term $\propto \nu_2$ as it was only introduced for numerical stability. We assume that the dust fluid reaches terminal velocity such that 
\begin{align}
\frac{\p \bmu}{\p t} + \left(\bmv\cdot\nabla\right)\bmu = 0,
\end{align}
and that relative velocities are small and write 
\begin{align}
\bmu = \bmu^{(0)}\st + \bmu^{(1)}\st^2+\cdots
\end{align}
Inserting this expansion into Eq. \ref{gas_based_Dv} and balancing terms of the same order in $\st$, we find at zeroth order
\begin{align}
    \bmu^{(0)} = \frac{\fgas}{\Omega}\left(\frac{\nabla \Pi}{\rhog} - 2\eta r \Omega^2\xhat - \frac{1}{\mu_0\rhog}\bmB\cdot\nabla\bmB \right),\label{tva_order0}
\end{align}
and at first order in $\st$,
\begin{align}
\bmu^{(1)} = \frac{\fgas}{\Omega}\left[2\Omega u_y^{(0)}\xhat - \frac{\Omega}{2}u_x^{(0)}\yhat  - \bmu^{(0)}\cdot\nabla\bmv\right].
\end{align}
Note that generally $u_y^{(0)}\neq 0$ due to the Lorentz force, unlike the hydrodynamic case under axisymmetry. 


The gas momentum equation (\ref{gas_based_vg}) becomes
\begin{align}
\frac{\partial\textbf{\emph{v}}}{\partial t} + \left[\textbf{\emph{v}} + \st\fdust\bmu^{(0)}\right]\cdot\nabla\textbf{\emph{v}} &= 
    2\Omega v_{y} \hat{\textbf{\emph{x}}} - \frac{\Omega}{2} v_{x}\hat{\textbf{\emph{y}}} 
    -\fgas\left(\frac{1}{\rho_{g}}\nabla \Pi - 2\eta r \Omega^2 \xhat
-\frac{1}{\mu_0\rhog}\bmB\cdot\nabla\bmB\right) + \st\fdust\left[2\Omega u_y^{(0)}\xhat - \frac{\Omega}{2}u_x^{(0)}\yhat\right] + \nu_1 \nabla^2 \bmv . \label{gas_based_mom_tva}
\end{align}
The difference from the original gas momentum equation is the reduction in the pressure and magnetic tension forces by the gas fraction $\fgas$, the dust-modified advection velocity on the LHS (which is the mixture's center of mass velocity to first order in $\st$), and drag forces now appear as the penultimate term on the RHS:

\begin{align}
    \st\fdust\left[2\Omega u_y^{(0)}\xhat - \frac{\Omega}{2}u_x^{(0)}\yhat \right]
    = -\st\fdust\fgas \left[\frac{2}{\mu_0\rhog}\left(\bmB\cdot\nabla B_y\right)\xhat + \frac{1}{2}\left(\frac{1}{\rhog}\frac{\p\Pi}{\p x} - 2 \eta r \Omega^2 - \frac{1}{\mu_0\rhog}\bmB\cdot\nabla B_x\right)\yhat\right].
\end{align}

\subsection{TVA equilibrium}
The TVA equilibrium consists of constant densities and velocities. The latter is given as follows. First, we note from Eq. \ref{tva_order0} that
\begin{align}
    u_x^{(0)} = -2\eta r \Omega\fgas.
\end{align}
The relative drift is then 
\begin{align}
&u_x = u_x^{(0)}\st = -2\eta r \Omega \st \fgas,\\
&u_y = u_y^{(1)}\st^2= -\frac{1}{2}\fgas u_x^{(0)} \st^2 = \eta r \Omega \fgas^2 \st^2.
\end{align}
The gas velocities can be read from the gas momentum equation in equilibrium, giving 
\begin{align}
    &v_x = - \st\fdust u_x^{(0)} = 2\eta r \Omega \fgas \fdust \st,\\
    &v_y = - \eta r \Omega \fgas,
\end{align}
while $u_z = v_z=0$. 

\subsection{Linearized momentum equation in the TVA}
Linearizing Eq. \ref{gas_based_mom_tva} gives: 
\begin{align}
    \sigma\dd\bmv =& 2\Omega\dd v_y\xhat - \frac{\Omega}{2}\dd v_x \yhat - 2\eta r \Omega^2\fgas^2\dd\epsilon \xhat - \fgas\left(\ik\frac{\dd\Pi}{\rhog} - \frac{\ikz B_z}{\mu_0\rhog}\dd\bmB\right) + \st \eta r \Omega^2\fgas^3(1-\epsilon)\dd\epsilon\yhat,\label{tva_linearized_mom} \\
    \phantom{\sigma\dd\bmv=}& - \st\fdust\fgas \left[\frac{2\ikz B_z}{\mu_0\rhog}\dd B_y \xhat + \frac{1}{2}\left(\ikx \frac{\dd\Pi}{\rhog} - \frac{\ikz B_z}{\mu_0\rhog}\dd B_x\right)\yhat\right] - \nu_1 k^2 \dd\bmv, \label{gas_based_linear_vg_TVA},
\end{align} 
where we utilized the results
\begin{align}
    &\dd \fdust = - \dd\fgas = \fgas^2 \dd\epsilon,\\
    &\dd \left(\fdust\fgas\right) = (1-\epsilon)\fgas^3\dd\epsilon. 
\end{align} 
Notice also that the background radial gas flow does not appear in Eq. \ref{tva_linearized_mom}, as it cancels with the background radial dust-gas drift. That is the advection term on the LHS of the gas momentum equation (Eq. \ref{gas_based_vg}) does not enter the linear problem. However, the background radial gas flow still appears in the linearized induction equation (Eq. \ref{gas_based_induction}). This difference from the momentum equation drives a new instability described below.


\section{RDI condition in the absence of rotation}\label{RDI-NoRotation}
The following are the specific forms of solutions for RDI condition when rotation is ignored ($\S$\ref{3.2.2}, or green lines in Figure \ref{summary}):
\begin{align}\label{eq: whistlers-RDI-no-rotation} k_z^{2}=&\frac{\left(V_{A}k_{x}\zeta_{x}\right)^{2}}{V_{A}^{4}-\left(V_{A}k_{x}l_{H}\zeta_{x}\right)^{2}}+\frac{\left(V_{A}k_{x}^{2}l_{H}\zeta_{x}\right)^{2}}{4\left[V_{A}^{4}-\left(V_{A}k_{x}l_{H}\zeta_{x}\right)\right]}+\frac{V_{A}k_{x}^{3}l_{H}\zeta_{x}^{2}\sqrt{4V_{A}^{2}+\left(V_{A}k_{x}l_{H}\right)^{2}+4\zeta_{x}^{2}}}{2\left[V_{A}^{4}-\left(V_{A}k_{x}l_{H}\zeta_{x}\right)\right]},
\end{align}
\begin{align}\label{eq: ion-cyclotron-RDI-no-rotation} k_z^{2}=&\frac{\left(V_{A}k_{x}\zeta_{x}\right)^{2}}{V_{A}^{4}-\left(V_{A}k_{x}l_{H}\zeta_{x}\right)^{2}}+\frac{\left(V_{A}k_{x}^{2}l_{H}\zeta_{x}\right)^{2}}{2\left[V_{A}^{4}-\left(V_{A}k_{x}l_{H}\zeta_{x}\right)\right]}-\frac{V_{A}k_{x}^{3}l_{H}\zeta_{x}^{2}\sqrt{4V_{A}^{2}+\left(V_{A}k_{x}l_{H}\right)^{2}+4\zeta_{x}^{2}}}{2\left[V_{A}^{4}-\left(V_{A}k_{x}l_{H}\zeta_{x}\right)\right]}.
\end{align}

The $``+"$ waves (eq. \ref{eq: whistlers-RDI-no-rotation}) are commonly referred to as whistler or electron-cyclotron modes, whereas those marked with $``-"$ (eq. \ref{eq: ion-cyclotron-RDI-no-rotation}) are identified as ion-cyclotron modes.

\section{Toy model with dissipation }\label{hallSI_diss_toy_model}

For numerical stability, we add gas viscosity and ohmic resistivity terms to the momentum and induction equations in the toy model (\S\ref{bg_drift_hallSI_toy}). We obtain: 
\begin{align}
    &\frac{\partial\textbf{\emph{v}}}{\partial t} + \textbf{\emph{v}}\cdot\nabla\textbf{\emph{v}} = 
    2v_{y}\Omega\hat{\textbf{\emph{x}}} - v_{x}\frac{\Omega}{2}\hat{\textbf{\emph{y}}} - \frac{\fgas}{\rho_{g}}\nabla \Pi 
    +\frac{\fgas}{\mu_0\rhog}\bmB\cdot\nabla\bmB + \nu_1\nabla^2 \bmv \label{toy_model_vg_diss},\\
      \frac{\partial\textbf{\emph{B}}}{\partial t} & = \nabla\times\left(\bm{v}\times\bmB\right) + \nabla\times\left(\bm{v}_\mathrm{ext}\times\Delta\bmB\right)
      -\frac{3}{2}\Omega B_x\yhat  - \left(\frac{\eta_H}{B_{z0}}\right)\nabla\times\left[\left(\nabla\times\bmB\right)\times\bmB\right] + \chi\nabla^2\bmB.\label{toy_model_induction_diss}
\end{align}
In terms of the magnetic vector potential $\bmA$, the induction equation (\ref{toy_model_induction_diss}) becomes   
\begin{align}
      \frac{\partial\bmA}{\partial t} =& 
      \bm{v}\times\bmB + \bm{v}_\mathrm{ext}\times\Delta\bmB 
      +\frac{3}{2}\Omega A_y\xhat - \frac{\eta_H}{B_{z0}}\left(\nabla\times\bmB\right)\times\bmB + \chi\nabla^2\bmA +  \nabla\phi.\label{vector_potential_eqn_toy}
\end{align}
Recall $\Delta\bmB = \nabla\times \bmA$, $\bmB = B_{z0}\zhat + \Delta\bmB$, and $\phi$ is the scalar potential. In \textsc{dedalus} simulations, we solve Eq. \ref{vector_potential_eqn_toy} instead of Eq. \ref{toy_model_induction_diss} to ensure that $\nabla\cdot\bmB = 0$. 

Linearizing Eqs. \ref{toy_model_vg_diss}---\ref{toy_model_induction_diss} about $\bmv = 0$ and $\bmB = B_{z0}\zhat$ yields: 
\begin{align}
    \sigma\dd\bmv =& 2\Omega\dd v_y\xhat - \frac{\Omega}{2}\dd v_x \yhat - \fgas\left(\ik\frac{\dd\Pi}{\rhog} - \frac{\ikz B_{z0}}{\mu_0\rhog}\dd\bmB\right) - \nu_1 k^2\dd\bmv \label{toy_linear_mom_diss},\\
    \sigma \dd\bmB =& \ikz B_{z0}\dd\bmv - \ikx v_{x} \dd \bmB - \frac{3}{2}\Omega\dd B_x\yhat -\etahall k_z^2\dd B_y\xhat + \etahall k^2 \dd B_x \yhat +\etahall k_x k_z \dd B_y\zhat - \chi k^2\dd\bmB.\label{toy_linear_induction_diss}
\end{align}
Note that Eq. \ref{toy_linear_mom_diss}  may also be obtained from Eq. \ref{gas_based_linear_vg_TVA} by setting $\delta\epsilon=\st=0$ where they appear explicitly. 


\bibliography{Hall-SI}{}

\begin{thebibliography}{}
\expandafter\ifx\csname natexlab\endcsname\relax\def\natexlab#1{#1}\fi
\providecommand{\url}[1]{\href{#1}{#1}}
\providecommand{\dodoi}[1]{doi:~\href{http://doi.org/#1}{\nolinkurl{#1}}}
\providecommand{\doeprint}[1]{\href{http://ascl.net/#1}{\nolinkurl{http://ascl.net/#1}}}
\providecommand{\doarXiv}[1]{\href{https://arxiv.org/abs/#1}{\nolinkurl{https://arxiv.org/abs/#1}}}

\bibitem[{{Abod} {et~al.}(2019){Abod}, {Simon}, {Li}, {Armitage}, {Youdin}, \& {Kretke}}]{Abod2019}
{Abod}, C.~P., {Simon}, J.~B., {Li}, R., {et~al.} 2019, \apj, 883, 192, \dodoi{10.3847/1538-4357/ab40a3}

\bibitem[{{Armitage}(2011)}]{armitage11}
{Armitage}, P.~J. 2011, \araa, 49, 195, \dodoi{10.1146/annurev-astro-081710-102521}

\bibitem[{{Armitage}(2015)}]{armitage15}
---. 2015, arXiv e-prints, arXiv:1509.06382, \dodoi{10.48550/arXiv.1509.06382}

\bibitem[{{Bai}(2015)}]{bai15}
{Bai}, X.-N. 2015, \apj, 798, 84, \dodoi{10.1088/0004-637X/798/2/84}

\bibitem[{{Bai}(2017)}]{bai17}
---. 2017, \apj, 845, 75, \dodoi{10.3847/1538-4357/aa7dda}

\bibitem[{{Bai} \& {Stone}(2010{\natexlab{a}})}]{bs2010a}
{Bai}, X.-N., \& {Stone}, J.~M. 2010{\natexlab{a}}, \apj, 722, 1437, \dodoi{10.1088/0004-637X/722/2/1437}

\bibitem[{{Bai} \& {Stone}(2010{\natexlab{b}})}]{bs2010b}
---. 2010{\natexlab{b}}, \apjl, 722, L220, \dodoi{10.1088/2041-8205/722/2/L220}

\bibitem[{{Bai} \& {Stone}(2013)}]{bs13}
---. 2013, \apj, 769, 76, \dodoi{10.1088/0004-637X/769/1/76}

\bibitem[{{Bai} \& {Stone}(2017)}]{bs17}
---. 2017, \apj, 836, 46, \dodoi{10.3847/1538-4357/836/1/46}

\bibitem[{{Balbus} \& {Hawley}(1991)}]{bh91}
{Balbus}, S.~A., \& {Hawley}, J.~F. 1991, \apj, 376, 214, \dodoi{10.1086/170270}

\bibitem[{{Balbus} \& {Terquem}(2001)}]{bt01}
{Balbus}, S.~A., \& {Terquem}, C. 2001, \apj, 552, 235, \dodoi{10.1086/320452}

\bibitem[{{Balsara} {et~al.}(2009){Balsara}, {Tilley}, {Rettig}, \& {Brittain}}]{balsara09}
{Balsara}, D.~S., {Tilley}, D.~A., {Rettig}, T., \& {Brittain}, S.~D. 2009, \mnras, 397, 24, \dodoi{10.1111/j.1365-2966.2009.14606.x}

\bibitem[{{Barranco}(2009)}]{barranco09}
{Barranco}, J.~A. 2009, \apj, 691, 907, \dodoi{10.1088/0004-637X/691/2/907}

\bibitem[{{Ben{\'\i}tez-Llambay} {et~al.}(2019){Ben{\'\i}tez-Llambay}, {Krapp}, \& {Pessah}}]{FARGO3D_multifluid}
{Ben{\'\i}tez-Llambay}, P., {Krapp}, L., \& {Pessah}, M.~E. 2019, \apjs, 241, 25, \dodoi{10.3847/1538-4365/ab0a0e}

\bibitem[{{B{\'e}thune} {et~al.}(2017){B{\'e}thune}, {Lesur}, \& {Ferreira}}]{bethune+17}
{B{\'e}thune}, W., {Lesur}, G., \& {Ferreira}, J. 2017, \aap, 600, A75, \dodoi{10.1051/0004-6361/201630056}

\bibitem[{{Birnstiel} {et~al.}(2010){Birnstiel}, {Dullemond}, \& {Brauer}}]{Birnstiel2010}
{Birnstiel}, T., {Dullemond}, C.~P., \& {Brauer}, F. 2010, \aap, 513, A79, \dodoi{10.1051/0004-6361/200913731}

\bibitem[{{Birnstiel} {et~al.}(2012){Birnstiel}, {Klahr}, \& {Ercolano}}]{Birnstiel2012}
{Birnstiel}, T., {Klahr}, H., \& {Ercolano}, B. 2012, \aap, 539, A148, \dodoi{10.1051/0004-6361/201118136}

\bibitem[{{Blum} \& {Wurm}(2008)}]{Blum2008ARAA}
{Blum}, J., \& {Wurm}, G. 2008, \araa, 46, 21, \dodoi{10.1146/annurev.astro.46.060407.145152}

\bibitem[{{Burns} {et~al.}(2020){Burns}, {Vasil}, {Oishi}, {Lecoanet}, \& {Brown}}]{burns20}
{Burns}, K.~J., {Vasil}, G.~M., {Oishi}, J.~S., {Lecoanet}, D., \& {Brown}, B.~P. 2020, Physical Review Research, 2, 023068, \dodoi{10.1103/PhysRevResearch.2.023068}

\bibitem[{{Chen} \& {Lin}(2020)}]{Chen_Lin_2020}
{Chen}, K., \& {Lin}, M.-K. 2020, \apj, 891, 132, \dodoi{10.3847/1538-4357/ab76ca}

\bibitem[{{Chiang}(2008)}]{chiang08}
{Chiang}, E. 2008, \apj, 675, 1549, \dodoi{10.1086/527354}

\bibitem[{{Chiang} \& {Youdin}(2010)}]{Chiang_Youdin_2010}
{Chiang}, E., \& {Youdin}, A.~N. 2010, Annual Review of Earth and Planetary Sciences, 38, 493, \dodoi{10.1146/annurev-earth-040809-152513}

\bibitem[{{Crutcher}(2012)}]{Crutcher_2012_ARAA}
{Crutcher}, R.~M. 2012, \araa, 50, 29, \dodoi{10.1146/annurev-astro-081811-125514}

\bibitem[{{Cui} \& {Bai}(2021)}]{cui2021}
{Cui}, C., \& {Bai}, X.-N. 2021, \mnras, 507, 1106, \dodoi{10.1093/mnras/stab2220}

\bibitem[{{de Sousa} {et~al.}(2020){de Sousa}, {Morbidelli}, {Raymond}, {Izidoro}, {Gomes}, \& {Vieira Neto}}]{desousa20}
{de Sousa}, R.~R., {Morbidelli}, A., {Raymond}, S.~N., {et~al.} 2020, \icarus, 339, 113605, \dodoi{10.1016/j.icarus.2019.113605}

\bibitem[{{Dittrich} {et~al.}(2013){Dittrich}, {Klahr}, \& {Johansen}}]{dittrich+13}
{Dittrich}, K., {Klahr}, H., \& {Johansen}, A. 2013, \apj, 763, 117, \dodoi{10.1088/0004-637X/763/2/117}

\bibitem[{{Dr{\k{a}}{\.z}kowska} {et~al.}(2023){Dr{\k{a}}{\.z}kowska}, {Bitsch}, {Lambrechts}, {Mulders}, {Harsono}, {Vazan}, {Liu}, {Ormel}, {Kretke}, \& {Morbidelli}}]{Drazkowska-PPVII}
{Dr{\k{a}}{\.z}kowska}, J., {Bitsch}, B., {Lambrechts}, M., {et~al.} 2023, in Astronomical Society of the Pacific Conference Series, Vol. 534, Protostars and Planets VII, ed. S.~{Inutsuka}, Y.~{Aikawa}, T.~{Muto}, K.~{Tomida}, \& M.~{Tamura}, 717, \dodoi{10.48550/arXiv.2203.09759}

\bibitem[{{Fromang} \& {Nelson}(2005)}]{fromang+05}
{Fromang}, S., \& {Nelson}, R.~P. 2005, \mnras, 364, L81, \dodoi{10.1111/j.1745-3933.2005.00109.x}

\bibitem[{{Fromang} \& {Papaloizou}(2006)}]{fromang+06}
{Fromang}, S., \& {Papaloizou}, J. 2006, \aap, 452, 751, \dodoi{10.1051/0004-6361:20054612}

\bibitem[{{Gammie}(1996)}]{gammie96}
{Gammie}, C.~F. 1996, \apj, 457, 355, \dodoi{10.1086/176735}

\bibitem[{{Goldreich} \& {Lynden-Bell}(1965)}]{Goldreich1965}
{Goldreich}, P., \& {Lynden-Bell}, D. 1965, \mnras, 130, 125, \dodoi{10.1093/mnras/130.2.125}

\bibitem[{{Goldreich} \& {Ward}(1973)}]{Goldreich1973}
{Goldreich}, P., \& {Ward}, W.~R. 1973, \apj, 183, 1051, \dodoi{10.1086/152291}

\bibitem[{{Gole} {et~al.}(2020){Gole}, {Simon}, {Li}, {Youdin}, \& {Armitage}}]{Gole2020}
{Gole}, D.~A., {Simon}, J.~B., {Li}, R., {Youdin}, A.~N., \& {Armitage}, P.~J. 2020, \apj, 904, 132, \dodoi{10.3847/1538-4357/abc334}

\bibitem[{{Hayashi}(1981)}]{hayashi81}
{Hayashi}, C. 1981, in Fundamental Problems in the Theory of Stellar Evolution, ed. D.~{Sugimoto}, D.~Q. {Lamb}, \& D.~N. {Schramm}, Vol.~93, 113--126

\bibitem[{{Hopkins} \& {Squire}(2018)}]{hs18}
{Hopkins}, P.~F., \& {Squire}, J. 2018, \mnras, 479, 4681, \dodoi{10.1093/mnras/sty1604}

\bibitem[{{Hsu} \& {Lin}(2022)}]{Hsu_Lin_2023}
{Hsu}, C.-Y., \& {Lin}, M.-K. 2022, \apj, 937, 55, \dodoi{10.3847/1538-4357/ac8df9}

\bibitem[{{Ishitsu} {et~al.}(2009){Ishitsu}, {Inutsuka}, \& {Sekiya}}]{ishitsu09}
{Ishitsu}, N., {Inutsuka}, S.-i., \& {Sekiya}, M. 2009, arXiv e-prints, arXiv:0905.4404, \dodoi{10.48550/arXiv.0905.4404}

\bibitem[{{Jacquet} {et~al.}(2011){Jacquet}, {Balbus}, \& {Latter}}]{Jacquet2011}
{Jacquet}, E., {Balbus}, S., \& {Latter}, H. 2011, \mnras, 415, 3591, \dodoi{10.1111/j.1365-2966.2011.18971.x}

\bibitem[{{Jaupart} \& {Laibe}(2020)}]{Jaupart2020}
{Jaupart}, E., \& {Laibe}, G. 2020, \mnras, 492, 4591, \dodoi{10.1093/mnras/staa057}

\bibitem[{{Johansen} {et~al.}(2014){Johansen}, {Blum}, {Tanaka}, {Ormel}, {Bizzarro}, \& {Rickman}}]{Johansen_2014}
{Johansen}, A., {Blum}, J., {Tanaka}, H., {et~al.} 2014, in Protostars and Planets VI, ed. H.~{Beuther}, R.~S. {Klessen}, C.~P. {Dullemond}, \& T.~{Henning}, 547--570, \dodoi{10.2458/azu_uapress_9780816531240-ch024}

\bibitem[{{Johansen} \& {Klahr}(2005)}]{johansen+05}
{Johansen}, A., \& {Klahr}, H. 2005, \apj, 634, 1353, \dodoi{10.1086/497118}

\bibitem[{{Johansen} {et~al.}(2006){Johansen}, {Klahr}, \& {Mee}}]{johansen+06}
{Johansen}, A., {Klahr}, H., \& {Mee}, A.~J. 2006, \mnras, 370, L71, \dodoi{10.1111/j.1745-3933.2006.00191.x}

\bibitem[{{Johansen} {et~al.}(2007){Johansen}, {Oishi}, {Mac Low}, {Klahr}, {Henning}, \& {Youdin}}]{johansen+07}
{Johansen}, A., {Oishi}, J.~S., {Mac Low}, M.-M., {et~al.} 2007, \nat, 448, 1022, \dodoi{10.1038/nature06086}

\bibitem[{{Johansen} \& {Youdin}(2007)}]{johansen07}
{Johansen}, A., \& {Youdin}, A. 2007, \apj, 662, 627, \dodoi{10.1086/516730}

\bibitem[{{Johansen} {et~al.}(2009){Johansen}, {Youdin}, \& {Klahr}}]{johansen+09}
{Johansen}, A., {Youdin}, A., \& {Klahr}, H. 2009, \apj, 697, 1269, \dodoi{10.1088/0004-637X/697/2/1269}

\bibitem[{{K{\"o}nigl} {et~al.}(2010){K{\"o}nigl}, {Salmeron}, \& {Wardle}}]{konigl+10}
{K{\"o}nigl}, A., {Salmeron}, R., \& {Wardle}, M. 2010, \mnras, 401, 479, \dodoi{10.1111/j.1365-2966.2009.15664.x}

\bibitem[{{Krapp} {et~al.}(2019){Krapp}, {Ben{\'\i}tez-Llambay}, {Gressel}, \& {Pessah}}]{Krapp2019}
{Krapp}, L., {Ben{\'\i}tez-Llambay}, P., {Gressel}, O., \& {Pessah}, M.~E. 2019, \apjl, 878, L30, \dodoi{10.3847/2041-8213/ab2596}

\bibitem[{{Krapp} {et~al.}(2018){Krapp}, {Gressel}, {Ben{\'\i}tez-Llambay}, {Downes}, {Mohandas}, \& {Pessah}}]{krapp18}
{Krapp}, L., {Gressel}, O., {Ben{\'\i}tez-Llambay}, P., {et~al.} 2018, \apj, 865, 105, \dodoi{10.3847/1538-4357/aadcf0}

\bibitem[{{Kunz} \& {Balbus}(2004)}]{kunz04}
{Kunz}, M.~W., \& {Balbus}, S.~A. 2004, \mnras, 348, 355, \dodoi{10.1111/j.1365-2966.2004.07383.x}

\bibitem[{{Kunz} \& {Lesur}(2013)}]{kunz13}
{Kunz}, M.~W., \& {Lesur}, G. 2013, \mnras, 434, 2295, \dodoi{10.1093/mnras/stt1171}

\bibitem[{{Latter} \& {Kunz}(2022)}]{Latter-Kunz2022}
{Latter}, H.~N., \& {Kunz}, M.~W. 2022, \mnras, 511, 1182, \dodoi{10.1093/mnras/stac107}

\bibitem[{Lecoanet {et~al.}(2019)Lecoanet, Vasil, Burns, Brown, \& Oishi}]{lecoanet19}
Lecoanet, D., Vasil, G.~M., Burns, K.~J., Brown, B.~P., \& Oishi, J.~S. 2019, Journal of Computational Physics: X, 3, 100012, \dodoi{https://doi.org/10.1016/j.jcpx.2019.100012}

\bibitem[{{Lee} {et~al.}(2010){Lee}, {Chiang}, {Asay-Davis}, \& {Barranco}}]{lee10}
{Lee}, A.~T., {Chiang}, E., {Asay-Davis}, X., \& {Barranco}, J. 2010, \apj, 718, 1367, \dodoi{10.1088/0004-637X/718/2/1367}

\bibitem[{{Lee} {et~al.}(2021){Lee}, {Charnoz}, \& {Hennebelle}}]{lee21}
{Lee}, Y.-N., {Charnoz}, S., \& {Hennebelle}, P. 2021, \aap, 648, A101, \dodoi{10.1051/0004-6361/202038105}

\bibitem[{{Lesur}(2021)}]{Lesur-review}
{Lesur}, G. 2021, Journal of Plasma Physics, 87, 205870101, \dodoi{10.1017/S0022377820001002}

\bibitem[{{Li} \& {Youdin}(2021)}]{Li2021}
{Li}, R., \& {Youdin}, A.~N. 2021, \apj, 919, 107, \dodoi{10.3847/1538-4357/ac0e9f}

\bibitem[{{Li} {et~al.}(2018){Li}, {Youdin}, \& {Simon}}]{Li2018}
{Li}, R., {Youdin}, A.~N., \& {Simon}, J.~B. 2018, \apj, 862, 14, \dodoi{10.3847/1538-4357/aaca99}

\bibitem[{{Li} {et~al.}(2019){Li}, {Youdin}, \& {Simon}}]{Li2019}
---. 2019, \apj, 885, 69, \dodoi{10.3847/1538-4357/ab480d}

\bibitem[{{Lin}(2021)}]{lin21}
{Lin}, M.-K. 2021, \apj, 907, 64, \dodoi{10.3847/1538-4357/abcd9b}

\bibitem[{{Lin} \& {Hsu}(2022)}]{Lin-Hsu-2022}
{Lin}, M.-K., \& {Hsu}, C.-Y. 2022, \apj, 926, 14, \dodoi{10.3847/1538-4357/ac3bb9}

\bibitem[{{Lin} \& {Youdin}(2017)}]{Lin-Youdin_2017}
{Lin}, M.-K., \& {Youdin}, A.~N. 2017, \apj, 849, 129, \dodoi{10.3847/1538-4357/aa92cd}

\bibitem[{{Lin} {et~al.}(2023){Lin}, {Li}, {Tobin}, {Ohashi}, {J{\o}rgensen}, {Looney}, {Aso}, {Takakuwa}, {Aikawa}, {van't Hoff}, {de Gregorio-Monsalvo}, {Encalada}, {Flores}, {Gavino}, {Han}, {Kido}, {Koch}, {Kwon}, {Lai}, {Lee}, {Lee}, {Phuong}, {Sai Insa Choi}, {Sharma}, {Sheehan}, {Thieme}, {Williams}, {Yamato}, \& {Yen}}]{eDiskII}
{Lin}, Z.-Y.~D., {Li}, Z.-Y., {Tobin}, J.~J., {et~al.} 2023, \apj, 951, 9, \dodoi{10.3847/1538-4357/acd5c9}

\bibitem[{{Liu} {et~al.}(2021){Liu}, {Tsai}, {Chen}, {Liu}, {Zhang}, {Ma}, {Elbakyan}, {Green}, {Hales}, {Liu}, {Takami}, {P{\'e}rez}, {Vorobyov}, \& {Yang}}]{liu21}
{Liu}, H.~B., {Tsai}, A.-L., {Chen}, W.~P., {et~al.} 2021, \apj, 923, 270, \dodoi{10.3847/1538-4357/ac31b9}

\bibitem[{{Manara} {et~al.}(2018){Manara}, {Morbidelli}, \& {Guillot}}]{manara18}
{Manara}, C.~F., {Morbidelli}, A., \& {Guillot}, T. 2018, \aap, 618, L3, \dodoi{10.1051/0004-6361/201834076}

\bibitem[{{Marchand} {et~al.}(2016){Marchand}, {Masson}, {Chabrier}, {Hennebelle}, {Commer{\c{c}}on}, \& {Vaytet}}]{marchand16}
{Marchand}, P., {Masson}, J., {Chabrier}, G., {et~al.} 2016, \aap, 592, A18, \dodoi{10.1051/0004-6361/201526780}

\bibitem[{{McNally} {et~al.}(2021){McNally}, {Lovascio}, \& {Paardekooper}}]{McNally2021}
{McNally}, C.~P., {Lovascio}, F., \& {Paardekooper}, S.-J. 2021, \mnras, 502, 1469, \dodoi{10.1093/mnras/stab112}

\bibitem[{{McNally} {et~al.}(2017){McNally}, {Nelson}, {Paardekooper}, {Gressel}, \& {Lyra}}]{mcnally17}
{McNally}, C.~P., {Nelson}, R.~P., {Paardekooper}, S.-J., {Gressel}, O., \& {Lyra}, W. 2017, \mnras, 472, 1565, \dodoi{10.1093/mnras/stx2136}

\bibitem[{{Morbidelli} \& {Nesvorn{\'y}}(2020)}]{morbidelli20}
{Morbidelli}, A., \& {Nesvorn{\'y}}, D. 2020, in The Trans-Neptunian Solar System, ed. D.~{Prialnik}, M.~A. {Barucci}, \& L.~{Young}, 25--59, \dodoi{10.1016/B978-0-12-816490-7.00002-3}

\bibitem[{{Nayakshin} {et~al.}(2022){Nayakshin}, {Elbakyan}, \& {Rosotti}}]{Nayakshin_2022}
{Nayakshin}, S., {Elbakyan}, V., \& {Rosotti}, G. 2022, \mnras, 512, 6038, \dodoi{10.1093/mnras/stac833}

\bibitem[{{Ohashi} {et~al.}(2023){Ohashi}, {Tobin}, {J{\o}rgensen}, {Takakuwa}, {Sheehan}, {Aikawa}, {Li}, {Looney}, {Williams}, {Aso}, {Sharma}, {Sai Insa Choi}, {Yamato}, {Lee}, {Tomida}, {Yen}, {Encalada}, {Flores}, {Gavino}, {Kido}, {Han}, {Lin}, {Narayanan}, {Phuong}, {Santamar{\'\i}a-Miranda}, {Thieme}, {van't Hoff}, {de Gregorio-Monsalvo}, {Koch}, {Kwon}, {Lai}, {Lee}, {Plunkett}, {Saigo}, {Hirano}, {Lam}, \& {Mori}}]{eDiskI}
{Ohashi}, N., {Tobin}, J.~J., {J{\o}rgensen}, J.~K., {et~al.} 2023, \apj, 951, 8, \dodoi{10.3847/1538-4357/acd384}

\bibitem[{{Paardekooper} {et~al.}(2020){Paardekooper}, {McNally}, \& {Lovascio}}]{Paardekooper2020}
{Paardekooper}, S.-J., {McNally}, C.~P., \& {Lovascio}, F. 2020, \mnras, 499, 4223, \dodoi{10.1093/mnras/staa3162}

\bibitem[{{Pandey} \& {Wardle}(2012)}]{DMRI}
{Pandey}, B.~P., \& {Wardle}, M. 2012, \mnras, 423, 222, \dodoi{10.1111/j.1365-2966.2012.20799.x}

\bibitem[{{Riols} {et~al.}(2020){Riols}, {Lesur}, \& {Menard}}]{riols20}
{Riols}, A., {Lesur}, G., \& {Menard}, F. 2020, \aap, 639, A95, \dodoi{10.1051/0004-6361/201937418}

\bibitem[{{Salmeron} {et~al.}(2011){Salmeron}, {K{\"o}nigl}, \& {Wardle}}]{salmeron+12}
{Salmeron}, R., {K{\"o}nigl}, A., \& {Wardle}, M. 2011, \mnras, 412, 1162, \dodoi{10.1111/j.1365-2966.2010.17974.x}

\bibitem[{{Sano} \& {Stone}(2002{\natexlab{a}})}]{ss22a}
{Sano}, T., \& {Stone}, J.~M. 2002{\natexlab{a}}, \apj, 570, 314, \dodoi{10.1086/339504}

\bibitem[{{Sano} \& {Stone}(2002{\natexlab{b}})}]{ss22b}
---. 2002{\natexlab{b}}, \apj, 577, 534, \dodoi{10.1086/342172}

\bibitem[{{Sch{\"a}fer} {et~al.}(2020){Sch{\"a}fer}, {Johansen}, \& {Banerjee}}]{Schafer2020}
{Sch{\"a}fer}, U., {Johansen}, A., \& {Banerjee}, R. 2020, \aap, 635, A190, \dodoi{10.1051/0004-6361/201937371}

\bibitem[{{Sch{\"a}fer} {et~al.}(2017){Sch{\"a}fer}, {Yang}, \& {Johansen}}]{Schafer2017}
{Sch{\"a}fer}, U., {Yang}, C.-C., \& {Johansen}, A. 2017, \aap, 597, A69, \dodoi{10.1051/0004-6361/201629561}

\bibitem[{{Schaffer} {et~al.}(2021){Schaffer}, {Johansen}, \& {Lambrechts}}]{Schaffer2021}
{Schaffer}, N., {Johansen}, A., \& {Lambrechts}, M. 2021, \aap, 653, A14, \dodoi{10.1051/0004-6361/202140690}

\bibitem[{{Schaffer} {et~al.}(2018){Schaffer}, {Yang}, \& {Johansen}}]{Schaffer2018}
{Schaffer}, N., {Yang}, C.-C., \& {Johansen}, A. 2018, \aap, 618, A75, \dodoi{10.1051/0004-6361/201832783}

\bibitem[{{Shi} \& {Chiang}(2013)}]{Shi2013}
{Shi}, J.-M., \& {Chiang}, E. 2013, \apj, 764, 20, \dodoi{10.1088/0004-637X/764/1/20}

\bibitem[{{Simon} {et~al.}(2016){Simon}, {Armitage}, {Li}, \& {Youdin}}]{Simon2016}
{Simon}, J.~B., {Armitage}, P.~J., {Li}, R., \& {Youdin}, A.~N. 2016, \apj, 822, 55, \dodoi{10.3847/0004-637X/822/1/55}

\bibitem[{{Simon} {et~al.}(2017){Simon}, {Armitage}, {Youdin}, \& {Li}}]{Simon2017}
{Simon}, J.~B., {Armitage}, P.~J., {Youdin}, A.~N., \& {Li}, R. 2017, \apjl, 847, L12, \dodoi{10.3847/2041-8213/aa8c79}

\bibitem[{{Squire} \& {Hopkins}(2018{\natexlab{a}})}]{Squire-Hopkins-2018a}
{Squire}, J., \& {Hopkins}, P.~F. 2018{\natexlab{a}}, \apjl, 856, L15, \dodoi{10.3847/2041-8213/aab54d}

\bibitem[{{Squire} \& {Hopkins}(2018{\natexlab{b}})}]{Squire-Hopkins-2018b}
---. 2018{\natexlab{b}}, \mnras, 477, 5011, \dodoi{10.1093/mnras/sty854}

\bibitem[{{Squire} \& {Hopkins}(2020)}]{Squire-Hopkins2020}
---. 2020, \mnras, 498, 1239, \dodoi{10.1093/mnras/staa2311}

\bibitem[{{Testi} {et~al.}(2014){Testi}, {Birnstiel}, {Ricci}, {Andrews}, {Blum}, {Carpenter}, {Dominik}, {Isella}, {Natta}, {Williams}, \& {Wilner}}]{Testi_PPVI}
{Testi}, L., {Birnstiel}, T., {Ricci}, L., {et~al.} 2014, in Protostars and Planets VI, ed. H.~{Beuther}, R.~S. {Klessen}, C.~P. {Dullemond}, \& T.~{Henning}, 339--361, \dodoi{10.2458/azu_uapress_9780816531240-ch015}

\bibitem[{{Tilley} {et~al.}(2010){Tilley}, {Balsara}, {Brittain}, \& {Rettig}}]{tilley10}
{Tilley}, D.~A., {Balsara}, D.~S., {Brittain}, S.~D., \& {Rettig}, T. 2010, \mnras, 403, 211, \dodoi{10.1111/j.1365-2966.2009.16145.x}

\bibitem[{{Tsukamoto} {et~al.}(2015){Tsukamoto}, {Iwasaki}, {Okuzumi}, {Machida}, \& {Inutsuka}}]{tsukamoto15}
{Tsukamoto}, Y., {Iwasaki}, K., {Okuzumi}, S., {Machida}, M.~N., \& {Inutsuka}, S. 2015, \apjl, 810, L26, \dodoi{10.1088/2041-8205/810/2/L26}

\bibitem[{{Umurhan} {et~al.}(2020){Umurhan}, {Estrada}, \& {Cuzzi}}]{Umurhan2020}
{Umurhan}, O.~M., {Estrada}, P.~R., \& {Cuzzi}, J.~N. 2020, \apj, 895, 4, \dodoi{10.3847/1538-4357/ab899d}

\bibitem[{{Wang} {et~al.}(2019){Wang}, {Bai}, \& {Goodman}}]{wang19}
{Wang}, L., {Bai}, X.-N., \& {Goodman}, J. 2019, \apj, 874, 90, \dodoi{10.3847/1538-4357/ab06fd}

\bibitem[{{Wardle}(1999)}]{wardle99}
{Wardle}, M. 1999, \mnras, 307, 849, \dodoi{10.1046/j.1365-8711.1999.02670.x}

\bibitem[{{Wardle} \& {Koenigl}(1993)}]{wardle93}
{Wardle}, M., \& {Koenigl}, A. 1993, \apj, 410, 218, \dodoi{10.1086/172739}

\bibitem[{{Weidenschilling}(1977)}]{Weidenschilling1977}
{Weidenschilling}, S.~J. 1977, \mnras, 180, 57, \dodoi{10.1093/mnras/180.2.57}

\bibitem[{{Whipple}(1972)}]{Whipple1972}
{Whipple}, F.~L. 1972, in From Plasma to Planet, ed. A.~{Elvius}, 211

\bibitem[{{Williams} \& {Cieza}(2011)}]{Jonathan_2011_ARAA}
{Williams}, J.~P., \& {Cieza}, L.~A. 2011, \araa, 49, 67, \dodoi{10.1146/annurev-astro-081710-102548}

\bibitem[{{Xu} \& {Bai}(2022)}]{xb22}
{Xu}, Z., \& {Bai}, X.-N. 2022, \apj, 924, 3, \dodoi{10.3847/1538-4357/ac31a7}

\bibitem[{{Yang} \& {Johansen}(2014)}]{Yang_Johansen_2014}
{Yang}, C.-C., \& {Johansen}, A. 2014, \apj, 792, 86, \dodoi{10.1088/0004-637X/792/2/86}

\bibitem[{{Yang} {et~al.}(2017){Yang}, {Johansen}, \& {Carrera}}]{Yang2017}
{Yang}, C.-C., {Johansen}, A., \& {Carrera}, D. 2017, \aap, 606, A80, \dodoi{10.1051/0004-6361/201630106}

\bibitem[{{Yang} {et~al.}(2018){Yang}, {Mac Low}, \& {Johansen}}]{yang+18}
{Yang}, C.-C., {Mac Low}, M.-M., \& {Johansen}, A. 2018, \apj, 868, 27, \dodoi{10.3847/1538-4357/aae7d4}

\bibitem[{{Youdin} \& {Johansen}(2007)}]{Youdin-Johansen2007}
{Youdin}, A., \& {Johansen}, A. 2007, \apj, 662, 613, \dodoi{10.1086/516729}

\bibitem[{{Youdin} \& {Goodman}(2005)}]{Youdin-Goodman2005}
{Youdin}, A.~N., \& {Goodman}, J. 2005, \apj, 620, 459, \dodoi{10.1086/426895}

\bibitem[{{Youdin} \& {Shu}(2002)}]{Youdin_Shu_2002}
{Youdin}, A.~N., \& {Shu}, F.~H. 2002, \apj, 580, 494, \dodoi{10.1086/343109}

\bibitem[{{Zhao} {et~al.}(2021){Zhao}, {Caselli}, {Li}, {Krasnopolsky}, {Shang}, \& {Lam}}]{zhao21}
{Zhao}, B., {Caselli}, P., {Li}, Z.-Y., {et~al.} 2021, \mnras, 505, 5142, \dodoi{10.1093/mnras/stab1295}

\bibitem[{{Zhu} \& {Yang}(2021)}]{Zhu_Yang_2021}
{Zhu}, Z., \& {Yang}, C.-C. 2021, \mnras, 501, 467, \dodoi{10.1093/mnras/staa3628}

\end{thebibliography}
\bibliographystyle{aasjournal}

\end{document}